\documentstyle[12pt,axodraw,epsfig]{article}
\begin{document}



\def\CompHEP{{CompHEP}}
\def\VEGAS{{\it Vegas}}
\newcommand{\C}{{{\it C}}}
\def\FORTRAN{{\it Fortran}}
\def\fortran{{\it Fortran}}
\def\Fortran{{\it Fortran}}
\def\REDUCE{{\it Reduce}}

\def\GRACE{{\it GRACE}}
\def\FeynArt{{\it FeynArts/FeynCalc}}
\def\HELAS{{\it HELAS}}
\def\MADGRAPH{{\it MADGRAPH}}
\def\CTEq{{\it CTEQ4m}}
\def\CTEQ{{\it CTEQ}}
\def\MRS{{\it MRS}}
\def\MATHEMATICA{{\it Mathematica}}
\def\Esc{{\it Esc}}
\def\PgUp{{\it PgUp}}
\def\PgDn{{\it PgDn}}
\def\Enter{{\it Enter}}

\def\Home{{\it Home}}
\def\End{{\it End}}
\def\Insert{{\it Insert}}
\def\Delete{{\it Delete}}
\def\Backspace{{\it Backspace}}
\def\Arrows{{\it Arrows}}

\def\itemm{\item[\bf --]}

\language=0

\title{\CompHEP -  a package for evaluation of  Feynman diagrams   
and integration over  multi-particle phase space}
\author{A.Pukhov\thanks{contact persons: Alexander Pukhov:
pukhov@theory.npi.msu.su  and Viacheslav Ilyin: ilyin@theory.npi.msu.su} 
\and{E.Boos}\and{M.Dubinin}\and{V.Edneral}\and{V.Ilyin}
\and{D.Kovalenko}\and{A.Kryukov}\and{V.Savrin}\\
\it Skobeltsyn Institute of Nuclear Physics, Moscow State University
\and{S.Shichanin} \\
\it Institute for High Energy Physics, Protvino, Russia
\and{A.Semenov}\\
\it Joint Institute for Nuclear  Research, Dubna, Russia \\ }

\date{{\bf User's manual for version 33}\\ ~~ \\ Preprint INP-MSU  98-41/542}

\maketitle

\newpage
\tableofcontents

\newpage

\section {Preface} 
    \subsection{Introduction} 
             \label{introduction}       
 \CompHEP~ is a package for automatic calculations of elementary particle
decay and collision properties in the lowest order of perturbation theory
(the tree approximation).  The main idea prescribed into the \CompHEP~ is to
make available passing on from the Lagrangian to the final distributions
effectively with a high level of automation.  
Other packages  created to solve  a similar problem
are  
\FeynArt\cite{FeynArt}, \GRACE\cite{GRACE}, \HELAS\cite{HELAS}, 
\MADGRAPH\cite{MADGRAPH}. See also the review \cite{Harlander}.

  \CompHEP~ is a menu-driven system with the context help. The notations
used in \CompHEP~ are very similar to those used in particle physics. 

  The present version has 4 built-in physical models. Two of them are the
versions of the Standard Model (SU(3)xSU(2)xU(1)) in the unitary and
t'Hooft - Feynman gauges. The user can change particle interaction and model
parameters. It is also possible to create new  models of particle
interaction.

  In the present version polarizations are not taken into account.
Averaging over initial and summing over final polarizations are performed
automatically. 
   
  The \CompHEP~ package consists of two parts: symbolic and numerical. The
symbolic part is written in the \C~  programming language. 
It produces \Fortran~ and \C~  codes for a squared matrix
element, and they  are used in the numerical calculation  later on.
There are two versions
of the numerical part: one is written in \Fortran~ and another one is done  in
\C. The facilities of both versions are almost equal. The \C~  version has
more comfortable interface but it does not possess an option to 
generate events and does not  perform calculations with a  quadruple  precision.

\vskip 0.2cm  
\noindent
The symbolic part of \CompHEP~ lets the user:

\begin{itemize}    
\item  select a process by specifying incoming and outgoing particles for
the decays of $1 \rightarrow 2, \ldots ,1 \rightarrow 5$ types and the collisions of $2
   \rightarrow 2, \ldots , 2 \rightarrow 4$ types;
   
\item   generate  Feynman diagrams,  display them, and create the
corresponding \LaTeX~ output;

\item   exclude some diagrams;

\item   generate and display squared Feynman diagrams;

\item   calculate analytical expressions corresponding to squared diagrams
   by using the fast built-in symbolic calculator;
          
\item   save symbolic results corresponding to the squared diagrams
   calculated in the \REDUCE~ and \MATHEMATICA~
codes for  further
   symbolic manipulations;

\item   generate the optimized \Fortran~ and \C~  codes for the squared matrix
   elements for further numerical calculations;

\item   launch the built-in interpreter for numerical calculations;  
\end{itemize}

\noindent
The numerical part of \CompHEP~ offers to:

\begin{itemize}
\item  convolute  the squared matrix element  with structure functions and
beam spectra. \CTEQ~ and \MRS~ parton
   distribution functions, the  ISR and  Beamstrahlung spectra of electrons,
   the laser photon spectrum,
   and the Weizsaecker-Williams photon structure functions are
   available;

\item   modify physical parameters (total energy, charges, masses etc.)
   involved in the process;

\item   select the scale parameter for evaluation of the QCD coupling constant and
   parton structure functions;

\item   introduce various kinematic cuts.

\item   define the kinematic scheme (phase space parameterization) for
   effective Monte Carlo integration;

\item introduce a phase space mapping in order to 
smooth sharp peaks of a squared matrix element and structure functions;

\item   perform  a   Monte Carlo phase space  integration by \VEGAS;

\item   store values of the calculated matrix element  in a file for subsequent
   event generation;

\item   generate events; 

\item   display distributions in various kinematic variables;

\item    create the graphical and \LaTeX~ outputs for histograms.
\end{itemize}

See the review \cite{Results}  about physical results  produced by means of \CompHEP.

    \subsection{History and contributions}     
The \CompHEP~ project  was founded in 1989 by the group of 
 physicists and programmers of D.V. Skobeltsyn Institute of Nuclear
Physics of
Moscow State University.  The project was initiated by Edward Boos,
 Viacheslav Ilyin  and  Victor Savrin. The primary formulation of  
physical problems  for the project was done by E.~Boos, Mikhail Dubinin, and
Dmitri Slavnov.
The first software working group was organized and managed by V.~Ilyin.

    The main author of the \CompHEP~ software is Alexander Pukhov.
He has developed almost all algorithms and data-representation 
structures  of the package. Namely, the structure  of physical
model database, the algorithm for generation of Feynman diagrams, 
the algorithm for evaluation of squared matrix elements, the structure 
of output codes for different programming languages, 
the algorithm for optimization of numerical codes, the algorithm for 
phase-space integration with smoothing of propagator peaks. He also has
created  the specialized   symbolic manipulation package for \CompHEP.

  The first version of the package appeared in 1989 \cite{First,AIHENP1}.
It was written  in the {\it Turbo Pascal} programming language for 
the {\it  MS-DOS}  operation system. The program produced a code
for calculation of squared  diagrams, written in  the \REDUCE~
symbolic manipulation language.    Routines for evaluation of the color
factors
 were written by Alexander Kryukov. The \REDUCE~ code-generation 
routines were written by Alexander Taranov and A.~Pukhov.

The authors of \CompHEP~ were being lead by an idea to create a user
friendly software. So they paid a special attention
to the interface and data-representation facilities.
The general part of graphical interface was  designed  by A.~Pukhov.
 The routines  for graphic representation of diagrams
were written by Victor Edneral.  The context-sensitive help facility
was designed by Sergey Shichanin. The program for the \CompHEP~ database
table manipulation was done by A.~Kryukov. Later on 
the plot drawing and \LaTeX~ output were designed  by
V.~Edneral, A.~Pukhov and S.~Shichanin.

The  \CompHEP~ symbolic answers were compared by Mikhail Dubinin
with a large number of  known symbolic 
expressions for  differential  and integral cross sections 
\cite{sCompare}. Starting from this point 
the physicists of \CompHEP~ group  E.~Boos, M.~Dubinin, and
V.~Ilyin applied \CompHEP~ for studying  new physics signals and relevant 
backgrounds.
 
 In 1991 \CompHEP~ got the built-in symbolic 
manipulation package created by A.~Pukhov and the \FORTRAN~
code output written by A.~Pukhov and S.~Shichanin \cite{AIHENP2}.

In 1992 Andrei Davydychev  proposed to use the t'Hooft-Feynman
gauge for evaluation of squared  diagrams. This idea was
realized by A.~Pukhov.
 It has opened the real possibility 
to  calculate any 2~\verb|->|~4 process with the help of the \CompHEP~
package. In the same  year V.~Ilyin and S.~Shichanin designed
a \FORTRAN~ program of  phase-space integration  for  2~\verb|->|~3
processes
and A.~Pukhov developed the numerical interpreter for 2~\verb|->|~2 
processes. These achievements were announced in \cite{AIHENP2}.

The numerical results by \CompHEP~ for a large  set of processes  were
compared  with the {\it GRACE} package \cite{nCompare}. E.~Boos,
M~.Dubinin,  V~.Ilyin and S~.Shichanin  performed  some other cross-checks
of  the \CompHEP~ package.  Later on   A.~Belyaev, E.~Boos and L.~Dudko 
compared \CompHEP~  with the FNAL program
{\it VECBOS} \cite{Vecbos}  and
extensive cross-checks of numerical results for $e^+,e^- \to 4\quad fermions$
set of processes have been performed by M.~Dubinin \cite{YellowPreprint}.

In 1993 the symbolic part of \CompHEP~ was rewritten in the 
{\cal C}  programming language by V.~Edneral \cite{PtoC}.
 The realization 
of the {\it Turbo Pascal} graphic routines by means of X11 tools
was done by Andrey Semenov \cite{X11}. 
It has opened  the way to create the version for  UNIX 
workstations  \cite{PtoC, AIHENP3, Seoul}.

   The main problem remaining in this version 
was the  phase-space  integration. \CompHEP~  created 
the \FORTRAN~ code for  squared matrix elements  with a high 
level of  automation. Generally  the matrix elements have a lot 
of singularities caused by the propagators of virtual particles.
   In order to succeed in  the Monte Carlo  phase-space
integration of singular matrix elements the user was forced every time to
modify the program of  phase-space parameterization.  

For  automation of this operation step   A.~Pukhov proposed
a general approach to the  generation  of  multi-particle kinematics
and to the regularization of  matrix elements. In 1995-96  this scheme 
was realized by A.~Pukhov, Dmitri Kovalenko and V.~Ilyin \cite{AIHENP4,
AIHENP5, kinemat}. At the same time V.~Ilyin  wrote  new 
\FORTRAN~ interface
programs like  {\it menus} and together with  A.~Kryukov
  embedded the parton structure functions
in \CompHEP. Later on the structure-function package was improved  by
A.~Pukhov to include the regularization of integration over   Feynman
parameters. As a result   we get a version which 
provides the user with a possibility   of automatic evaluation  starting from 
the input of Lagrangian  and finishing with distributions in 
physical parameters.  The corresponding service for a histogram filling
also was done by A.~Pukhov. The list of needed distributions 
was compiled by E.~Boos.

During this work it was realized that  the \Fortran~ programming 
language is not convenient for  the future development of the numerical
part of \CompHEP.   In 1997 the \C~ code output for the numerical
calculation
was designed and  the \Fortran~ program for the numerical evaluation
was rewritten in \C~ as well. This work was done by A.~Pukhov. 
In the same year A.~Kryukov  imported  \CompHEP~ onto  the MS-Windows95/NT
platform.

The development of \CompHEP~ was being under a continuous 
pressure of physicists' requests. 
The works  of  E.~Boos, 
M.~Dubinin, V.~Ilyin, V.~Savrin and S.~Shichanin, who first used  \CompHEP~
for physical calculations, at the same time
were defining a direction of the package development. They
also contribute to  and  are responsible for debugging the package.

 It must be especially  noted the role of E.~Boos and V.~Ilyin 
in  the popularization of  \CompHEP~  for  the scientific community.

The manual was written by A.~Pukhov, V.~Savrin, and S.~Shichanin.

    \subsection{Acknowledgments }        The \CompHEP~ project was supported by\\

    Russian State Program on High Energy Physics;
 
    RFBR (93-02-14428, 96-02-19773a, 96-02-18635, 98-02-17699);

    ISF (M9B000, M9B300);

    INTAS (1010-CT93-0024, 93-1180, 93-1180ext,96-0842);
 
    Japanese  Society for the Promotion of Science;
 
    Japanese companies KASUMI Co, Ltd. and SECOM Co, Ltd.;
  
    Royal Society of London (UK);

    Organizing Committee of the AIHENP series of International Workshops.\\

    During a long time the \CompHEP~  group had not got a possibility to
develop the project on UNIX workstations in Russia. The adaptation  of the
package for different UNIX  platforms was  done 
 during the visits of the group members to  various
 universities and scientific centers of the world. We very much
appreciate this support and are grateful for cooperation to our foreign
colleagues   
Y.~Shimizu, (KEK, Japan),  
H.S.~Song  (Seoul National University, Korea),
O.~Eboli (University of Sao Paulo, Brazil), 
H.-J.~Schreiber (DESY, Germany).

We acknowledge beneficial discussions with 
F.~Cuypers, I.~Ginzburg, F.~Gutbrod, B.~Mele, 
M. Sachwitz, W.~von~Schlippe, P.~Osland,
and members of Minami-Tateya group (KEK),
and their benevolent attitude to our project during many years.
In this context we are especially grateful to D.~Perret-Gallix.

  We  also express our gratitude to our colleagues A.~Taranov,
P.~Baikov, H.~Eck, L.~Gladilin, P.~Silaev, S.~Ostapchenko who 
contributed to the development of \CompHEP~ software as well as
to A. Davydychev, A. Rodionov and D. Slavnov for some helpful
ideas.
 
  We thank S.~Ambrosanio  and A.~Belyaev for their numerous
reports on \CompHEP~ bugs.

\newpage
 \section{Installation procedure}
             \label{install}
  \subsection{\CompHEP~ Web page}
             \label{Web}                

\CompHEP~ codes and manual are disposed on the following Web site\\
\hspace*{2cm}{\tt http://theory.npi.msu.su/\verb|~|comphep}\\
Also there is a mirror in DESY\\
\hspace*{2cm}{\tt http://www.ifh.de/\verb|~|pukhov}
    
  \subsection{License}
             \label{license}            \centerline{\bf        Non-profit Use License Agreement }
\vskip 0.2cm

This Agreement is to be held between the Authors of the \CompHEP~
program and a Party which acquires the program. On acquiring the
program the Party agrees to be bound by terms of this Agreement.

\begin{enumerate}
\item This License entitles the Licensee (one person) and the
Licensee's research group to obtain a copy of the source or
executable code of \CompHEP~ and to use the acquired program for
academic research or other non-profit purposes within the research
group; or, it entitles the Licensee (a company, organization or
computing center) to install the program and allow an access to the
executable code to the members of the Licensee for academic research
or other non-profit use. 
\item No user or site will re-distribute the
source code or executable code to a third party in the 
modified form. Any re-distribution  must be accompanied with the 
current license.
\item Publications which result from using the program will contain
references to the articles describing \CompHEP. See the necessary
references on the \CompHEP~ Web  page.

\item This License does not permit any commercial (profit-making or
proprietary) use or re-licensing or re-distributions. Persons
interested in a for-profit use should contact the Authors.

\item The Authors of \CompHEP~ do not guarantee that the program is free
of errors or meets its specification and cannot be held responsible
for loss or consequential damage as a result of using it.

\end{enumerate}

  \subsection{How to get the codes}   
             \label{get-code}             If you agree with the license above, you may get \CompHEP~ code for the 
version 33 on  the \CompHEP~ Web page. The name of received file should be 

\vskip 0.2cm
{\tt comphep\_33.\#.tar.gz }
\vskip 0.2cm

\noindent
 where \# denotes a  number of release. Unpack this file by 

\vskip 0.2cm
 {\tt        gzip -d comphep\_33.\#.tar.gz }
\vskip 0.2cm
{\tt        tar -xf comphep\_33.\#.tar }
\vskip 0.2cm

\noindent
 As a result a directory  comphep\_33.\# should be created.

 This directory contains  source codes  of the \CompHEP~ package for UNIX
platforms. After compilation of these codes  the \CompHEP~ binary
executable files  appear in the same directory.      
 We shall refer to it below as a 
\CompHEP~ root directory.

  \subsection{Compilation procedure}  
             \label{compilation}          In order to compile the \CompHEP~ source code you need a \C~ compiler
with the X11 graphics library. If you would like to use \Fortran~  for
numerical evaluation you need also a \Fortran~ compiler. The
compilation of the \CompHEP~ \C~ source code is launched by the

\vskip 0.2cm
{\tt     ./create\_c  }
\vskip 0.2cm
\noindent
command started from the \CompHEP~ root directory.

  If the  \C~ compiler is detected and the \C~ sources are compiled
successfully  you will see the message:

\vskip 0.2cm
{\tt
   "C-part of code  has been successfully compiled and linked." 
   
   "CompHEP could be started."
}
\vskip 0.2cm

Otherwise  the corresponding error message comes out.
See the next section for a discussion of possible problems.

If the  \Fortran~ compiler is available and you would like to use
\Fortran~ for numerical calculation  launch the 

\vskip 0.2cm
{\tt   ./create\_f}
\vskip 0.2cm
\noindent
command. If the  \Fortran~ compiler is detected and the \Fortran~  sources are
also compiled successfully,  the message appears:

\vskip 0.2cm
{\tt     "Fortran code has been successfully compiled."}
\vskip 0.2cm
\noindent
See  the next section in the case of problem with  compilation.

  The files in the \CompHEP~ root directory will be
used only for reading and execution during the user session. 
It provides us with a possibility to have one \CompHEP~ root
directory for several users. To set the appropriate files modes
start the 

\vskip 0.2cm
{\tt  ./set\_mod }
\vskip 0.2cm
\noindent
command.

Each user has to create his own directory to work with \CompHEP. 
See an instruction for the user installation  in the next section.
However for an  express check of compiled  version you 
could go to the {\it ./test} directory and launch {\it ./comphep} there.

\subsubsection*{Possible problems} 
           
  Due to its first instruction the {\it create\_c} command tests the
existence of the {\it CC} file in the \CompHEP~ root directory.
If this file exists, the C compiler name and its options are read  from
this file. Otherwise, as we usually have at the first
start of {\it create\_c}, this file  is created and  contains  default 
parameters.

After that the {\it create\_c} program tests the necessary
compiler options. 
For this goal it generates various  programs  with the same  name
{\it test.c}  and tries to
compile and link them.  If  compilation is not satisfactory,
the
{\it create\_c} command finishes with the corresponding error
message and asks you to rewrite the command file {\it CC} in
order to fit  the requirement. The current test file {\it test.c}
 is saved. So in the case of such an error you could update your
{\it CC} file and start {\it create\_c} again. 
   The options listed below must be supported by \C~ compiler:
\begin{itemize}
\itemm     the  ANSI mode of function prototypes;
\itemm     the  signed  char type;
\itemm     an  access to the X11 include and library
files.
\end{itemize}
\noindent
An example of contents of the {\it CC} file for Linux platform 
is presented below:
\begin{center}
{\tt gcc -fsigned-char -I/usr/X11R6/include -L/usr/X11R6/lib}
\end{center}

   The work of {\it create\_f} is similar to that of {\it create\_c}
described above.   The \Fortran~ compiler must possess long character 
variables (e.g. CHARACTER*5000).  The  name of  \FORTRAN~ compiler and options  of compilation 
are stored in the {\it F77}  file.

If all tests are passed on successfully the message appears:
\begin{center}
{\tt
C (Fortran) compiler OK. Starting \CompHEP~ source code compilation.\\
}
\end{center}
If your UNIX platform is one of the listed:
Linux, IRIX, IRIX64, HP-UX, AIX, OSF1, the necessary options are
known from the beginning except of the  path to {\it X11}.
Anyway, in the  case of some problem on this step you can send a request
for help to the \CompHEP~ authors. 

 A correction of the {\it CC} and {\it F77} files could  be used to tune compiler 
options  according to  your UNIX platform. In this case you should 
create {\it CC} or {\it F77} by  starting {\it makeCC} or  {\it
makeF77}, perform 
tunings in the {\it CC} or {\it F77} files and launch {\it create\_c} or  {\it
create\_f} after that.
For example, you could switch on an optimization flag. 

\subsubsection*{\C~ compiler tuning}
       
  There are two  {\it macro}   definitions for the \C~ code, which could
be useful to tune:

  The first one is {\it STRSIZ}. This is a maximum size of strings in the \CompHEP~
models.  By default {\it STRSIZ=2048}. If you would like to embed in \CompHEP~
some new interaction with a very cumbersome vertex you have to 
increase this size by the

\vskip 0.2cm
{\tt  -DSTRSIZ=<new value>}
\vskip 0.2cm
\noindent
option.

  The second {\it macro} defines a type of integer numbers which are used in
symbolic calculations by \CompHEP. By default \CompHEP~ uses the 'long'
type.
\vskip 0.2cm
{\it -DNUM\_LONG\_LONG} forces the compiler to use  {\it 'long long'}. 
 This type is not the ANSI standard  and, perhaps, it is  not
 supported in your case. Different realizations of the {\it 'long~long'} type
use different formats for reading and writing such numbers. The user may 
specify  this format  defining the {\it 'NUM\_STR'}. By default \CompHEP~ uses

{\tt \#define  NUM\_STR "lld"}  

\vskip 0.2cm
{\it -DNUM\_DOUBLE} option forces \CompHEP~ to emulate integer numbers via
 {\it 'double'} float ones. 

  The {\it makeCC} program inserts an option which guarantees 8-byte size
integer calculation in \CompHEP. If the size of the standard `long'
type is not enough,  then the 'double' one is used.

  \subsection{User installation and start of the \CompHEP~ session}     
             \label{u-install}          
The \CompHEP~ root directory is not intended to start  a session.
For this  purpose  the user has to prepare a special  working directory. 
A few working directories may be created for  various tasks and by different
users. 

    In order to provide the programs with an
access to the  \CompHEP~ files and commands  the environment variable
COMPHEP  should contain the corresponding path. We recommend to define the 
COMPHEP variable in the user startup file.
 The name of the startup file and the syntax
of the assignment  instruction depends on  the command interpreter. 
For example, in the case
of  {\it tcsh} the  name of  startup file is {\it \verb|~|/.tcshrc} or
 {\it \verb|~|/.cshrc} and  the assignment  is realized by the instruction\\
\hspace*{2cm} {\tt setenv COMPHEP <Path to CompHEP directory>}

Let you have  defined the COMPHEP variable and  create a directory, 
say, {\it WORK} for the \CompHEP~ session. To prepare this directory for 
processing  call the program {\it install} from within this directory:\\
\hspace*{2cm} {\tt \$COMPHEP/install }\\
As a result  the  following sub-directories and files
should appear in your {\it WORK} directory:\\
\hspace*{2cm} {\tt models/~~~   tmp/~~~~~~~  results/}\\
\hspace*{2cm} {\tt comphep~~~   comphep.ini}

The directory   {\it models} is used for  files which describe models 
of particle interactions. The directory  {\it tmp} is created for temporal
files.  The directory {\it results} is  assigned for a \CompHEP~ output.

 To start a \CompHEP~ session  you should issue a command\\
\hspace*{2cm}{\tt  ./comphep}

 The {\it comphep.ini} file allows  to choose an 
appropriate text font for the \CompHEP~ window, select the  color or colorless
mode for the window and also switch on/off the sound signal. The syntax of  
this file is self-explanatory.

  \subsection{Installation under MS Windows9x/NT}
             \label{w-install}          The \CompHEP~ Windows9x/NT version is distributed
in a compiled form due to the absence of standard C/Fortran compiler
under the Windows9x/NT. Thus the program does not provide the user with
compiling and linking options within the working session. The user
should use the "Numerical interpreter" option to get numerical results.

The installation file {\tt
bNNi.zip} (where {\tt NN} denotes a number of the release) is available
to copy from the \CompHEP~ Web-page.
The installation procedure is the following: 

\begin{enumerate}

\item[1)] create an installation directory and copy the distributive {\tt
bNNi.zip} to this directory;

\item[2)] unpack the distributive file {\tt bNNi.zip} by the command

{\tt unzip bNNi.zip}

As a result a set of files and subdirectories should be created. 
This set corresponds to the user working directory in the UNIX release
(see
section \ref{u-install}).
Executable and other files corresponding to those
in  the CompHEP
root directory of the UNIX version (see section \ref{get-code}) are stored
in the subdirectory {\tt bin}.

\end{enumerate}

To start a \CompHEP~ session  the user should launch the command\\
\hspace*{2cm}{\tt comphep.pif} \\
in the installation directory. 

Contact person for the  Windows9x/NT version of \CompHEP~ is A.P. Kryukov
(e-mail: kryukov@theory.npi.msu.su).

\newpage
\section {User guide}  \label{guide}
  \subsection{Elements of the user interface }
             \label{interface}          \CompHEP~ software is written in \C~ and \Fortran.
The symbolic part is done in \C. The numerical one was written in \Fortran~
but later on was converted to \C. However both versions of numerical part are
now available. The \C~  programs  use a graphical  window interface based on
the   X11 or MS-Windows facilities,
whereas the \Fortran~ ones  use only a standard input/output. 
So in the latter case the service   is  more ugly.

\subsubsection{Graphical interface} \label{X11interface}

  There are the following elements of the user interface in the \CompHEP~
package: {\it On-line Help, Menu, Message, String Editor, Table Editor,
Diagram Viewer} and {\it Plot Viewer}. You can control them using the
\Arrows~ keys, \Enter, \Esc\footnote{Use the  {\tt 'Ctrl ['} sequence if the \Esc~ key is  absent on your
keyboard}, \Backspace,
\PgUp/\PgDn\footnote {On some keyboards there are {\it Prev/Next} instead of 
\PgUp/\PgDn}  
keys and the mouse click. \CompHEP~ is sensitive to the left mouse button release.

  You can toggle on/off colors  and a sound for the \CompHEP~ session as well
as choose the most appropriate font for the \CompHEP~ window.  Just edit
{\it comphep.ini}  which appears as a result of user installation (Section
\ref{u-install}). We hope the syntax  is  obvious.

\paragraph  {1. On-line Help.} 

At almost every point when \CompHEP~ is waiting for your input, you can
press the {\it F1} key to get a context sensitive help information. If the screen 
height is not enough to display the full help message, you will see the 
\PgDn~ mark in the right-bottom corner of the help window. To get the next 
page of the message press the \PgDn~ key or click the mark. To close the 
help window  press the \Esc~ key or click the asterisk in the top-left corner 
of the help border.

\paragraph { 2. Menu.} 

 The   menu program displays a list of menu functions. One of
them is highlighted. See a typical example of  menu in Fig.\ref{screen_menu}.  
Use the arrow keys or a mouse click to highlight
a desired function. Press the \Enter~ key or click on the highlighted
function to activate it.

   If the  menu is  too large you will
see only a part of it. Use the \PgDn/\PgUp~ buttons or click on the
corresponding marks in the menu corners to scroll. 

   In order to get back to the previous menu level press the \Esc~ key or
click the asterisk in the top-left corner of the menu border. 

   The menu program is also sensitive to the functional keys {\it F1, F2,
..,F9}. The list of 
active functional keys depends on the program point and is displayed on the 
bottom line of the screen. Generally the functional keys  activate the 
following  programs:\\
\begin{tabbing}
F1-~ \= Help~~~~~~~\=: \= displays a help message about the highlighted  menu
function;\\
F2- \>  Manual     \>:\>  displays an  information about service facilities.\\
F3- \> Models     \>:\>   displays  contents of the current model of particle 
interactions.\\ 
F4- \> Diagrams \>:\> browses the generated Feynman diagrams.\\
F6- \> Results  \>:\> views and  deletes \CompHEP~ output files.\\ 
F9-\> Quit     \>:\> quits the \CompHEP~ session.\\ 
\end{tabbing}
   To call one of these programs just press the functional key  or click
on the corresponding symbol on the bottom line of the screen. The digit keys 
act as the functional keys. For example, {\it '3'} acts as {\it F3}.

\paragraph{ 3. Message.}
 
    \CompHEP~ writes informative and dialogue  messages during the session. 
The informative messages finish with the {\it "Press~any~key"} string. You can
continue
your work either following this instruction or clicking the mouse on the
message area. In the second case the  message has a label {\it(Y/N?)}. You
should  press  the {\it Y} or {\it N} key or just click on them in the
message window to answer `Yes' or `No'.

\paragraph{4. String Editor.} 

    If you would like to enter a new process or change a  parameter 
value,  \CompHEP~ calls the String Editor.  As a rule, previous
information about this item is available and the kept string is
displayed.  If you would like to edit the original string, use the
{\it left/right arrow} keys or the mouse click to put the cursor on the
desired position. Otherwise, if the fist input character is a printing
symbol, the original string will be deleted.  The \Delete~ key works as
the \Backspace~ key and removes 
a character left to the cursor.   To terminate the input you can press  \Enter~ 
to accept the resulting string or the \Esc~ key to cancel it.

\paragraph{5. Table Editor.} 

    \CompHEP~ uses {\it tables} to store the information about parameters, particles,
vertices, cuts and distributions. For all these cases any unit 
(displayed as a table line) consists  of several fields (table columns). 
The program  Table Editor is  invented to provide the user a possibility 
to  view and change  the  table contents.   In some program 
points  the Table Editor is  used to browse a table contents without 
a permission to change data.

\begin{figure}
\begin{picture}(410,230)
\put(0,0){\epsfig{file=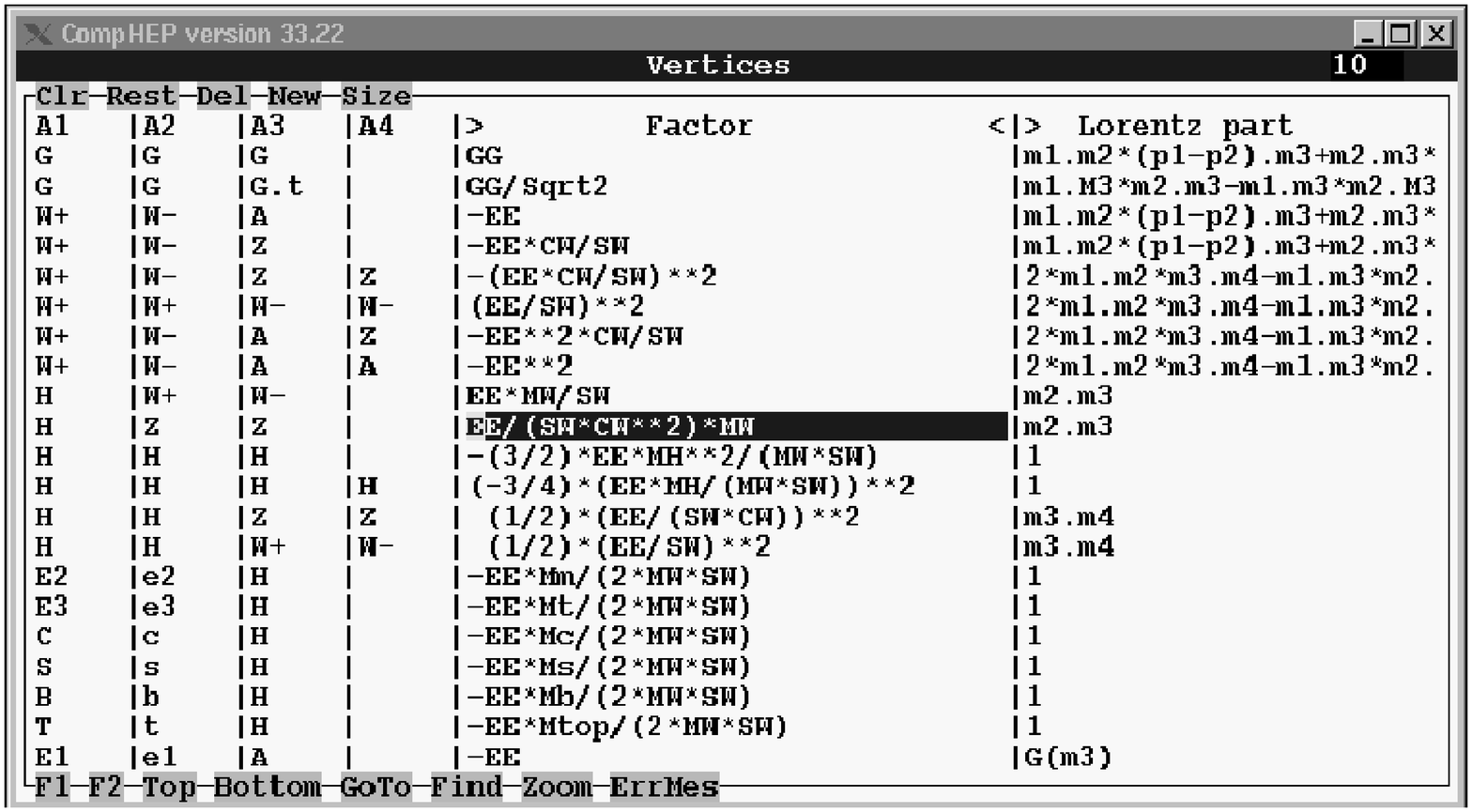,width=410pt, height=230pt}}
\end{picture}

\caption{ Example of the  \CompHEP~  tables}

\label{screen_table}

\end{figure}

    Table is displayed on the screen as it follows (see Fig.\ref{screen_table}).
The top line of the window  contains a  title of the table. Below there are  
table columns 
surrounded by a frame box.  The columns are separated by vertical lines. The first 
horizontal line contains column names. 
One cell (a line - column intersection) is highlighted. If the table is open
for  changes, the   highlighted  cell  contains the cursor. The ordering 
number of the corresponding  line  is displayed in the 
top-right corner   of the window.

To change position of the  cursor and  the   highlighted  cell one  can use 
the arrow keys, the {\it Tab} key  and the mouse click. If one types any printing symbol it 
will be inserted into the table at the cursor position. 
  The  \PgUp, \PgDn~  keys are used to scroll the table. 
The {\it F1} and {\it F2} functional  keys provide   
information about  the meaning of table fields  and about  facilities
of the Table Editor.
 To exit the table  one has to press the \Esc~ key.

There are some auxiliary commands which help the user to operate the 
tables. These commands can be realized by means of {\it Control } 
symbols or by mouse click on the command label displayed on the table
border:  
 
{\it Top (\^~T)} moves the cursor ( highlighted cell) to the top line of the 
table.

{\it Bottom (\^~B)} moves the cursor ( highlighted cell) to the bottom line of
the  table.

{\it GoTo (\^~G)}  moves the cursor to the line directed by the user.

{\it Find (\^~F)}  searches the string directed by the user.

{\it Find again (\^~A)} repeats previous command {\it Find}.
 
{\it Zoom (\^~Z)} key switches on the zoom mode to view/edit  
contents of  the  highlighted  cell. In this case
\CompHEP~ opens a special window and  the field text wraps this window.
  To terminate the {\it Zoom} mode  one has either to press \Enter~ to
accept changes  or \Esc~ to forget them.

{\it ErrMess (\^~E)}   redisplays an  error message   concerning the contents
 of one of the  tables 
 which has been previously generated by \CompHEP.
 
The above commands are available in both modes of the Table Editor.
The labels of these commands are disposed on the bottom border  of the 
table. The following  commands 
 are available only  if the table is open
for  changes:

{\it Clr (\^~C)} clears the contents of the current field right to the cursor
position.

{\it Rest (\^~R)} restores the contents of the current field which existed
before last entering  the corresponding cell.

{\it Del (\^~D)}   cuts the current line from the table and
put it into  the buffer.

{\it New (\^~N)} creates a new line and fill it with the buffer contents.
Also you can  press the \Enter~ key to create a new line.

{\it Size (\^~S)} allows  the user to change the width of current field.
This command is active only if the  cursor is disposed in the column 
whose name  is surrounded by the   '\verb|>|','\verb|<|' symbols.

The labels of these commands are disposed on the top border of the window.

\begin{figure}            
\begin{picture}(410,230)
\put(0,0){\epsfig{file=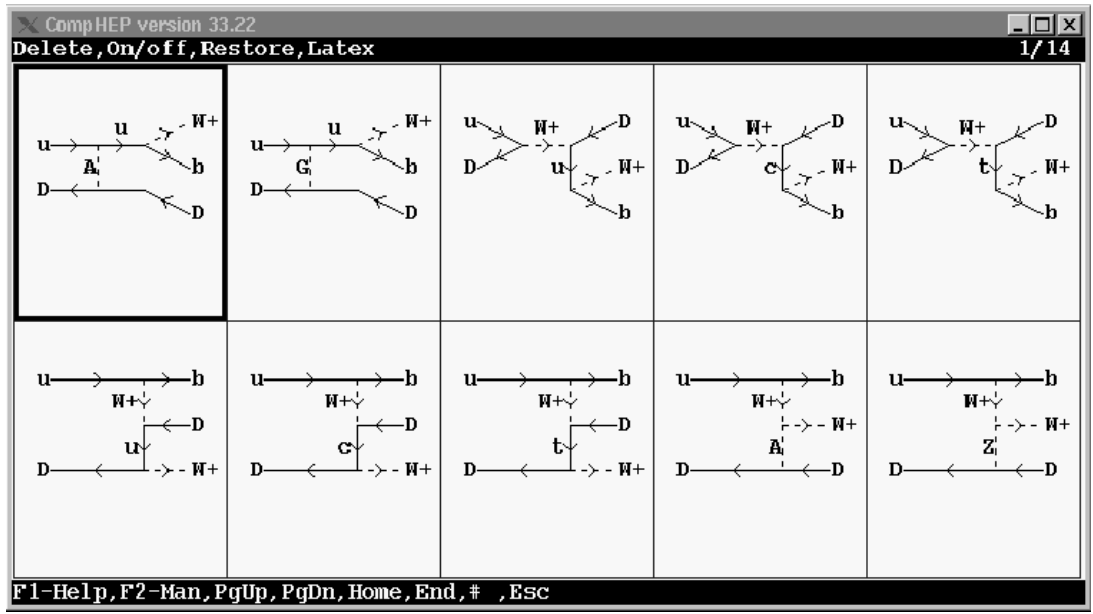,width=410pt, height=230pt}}
\end{picture}

\caption{ Example of the \CompHEP~ diagram images}

\label{screen_diag}

\end{figure}

\paragraph{ 6.  Diagram Viewer.} 
   This program was designed to display  several Feynman 
diagrams on the   screen. The Viewer splits the screen into   rectangle 
cells and puts the 
diagram images in these cells one by one. One cell is marked by  surrounding
box frame. The total number of diagrams and the ordinal number of the marked
one are  displayed in 
the right-top corner of the screen.  See an example  in Fig.\ref{screen_diag}.
 
The number of diagrams which can be displayed simultaneously
depends on the  window size. 
  If you would like to see more diagrams on one screen, increase the window  
using the window manager.

    You can use the \PgDn/\PgUp~ keys to scroll diagram set. The
\mbox{\Home/\End~} 
 keys display  the beginning/end of the set. To display a diagram with some
ordinal number  you should press the {\it '\#'} key and after that the needed 
number.  You can move the position of the surrounding box 
 by the \Arrows~ keys or by the mouse click.     To finish work with the Diagram Viewer
 press the \Esc~ key.

The labels  for the above  commands  are displayed on the bottom border of the
window and  you may use the mouse click to activate one of them. 

 The Diagram Viewer may have some optional functions which depend  
on the context. The  labels 
for these functions are shown on the top border of the
window. One of them  generates a file with  graphical diagram image in  
the \LaTeX~ format. 
 Press   the 
{\it F1} key to get an information about these commands.
You may use the mouse click  on the label  or its first symbol
to activate  the function.   

\begin{figure}            
\begin{picture}(410,230)
\put(0,0){\epsfig{file=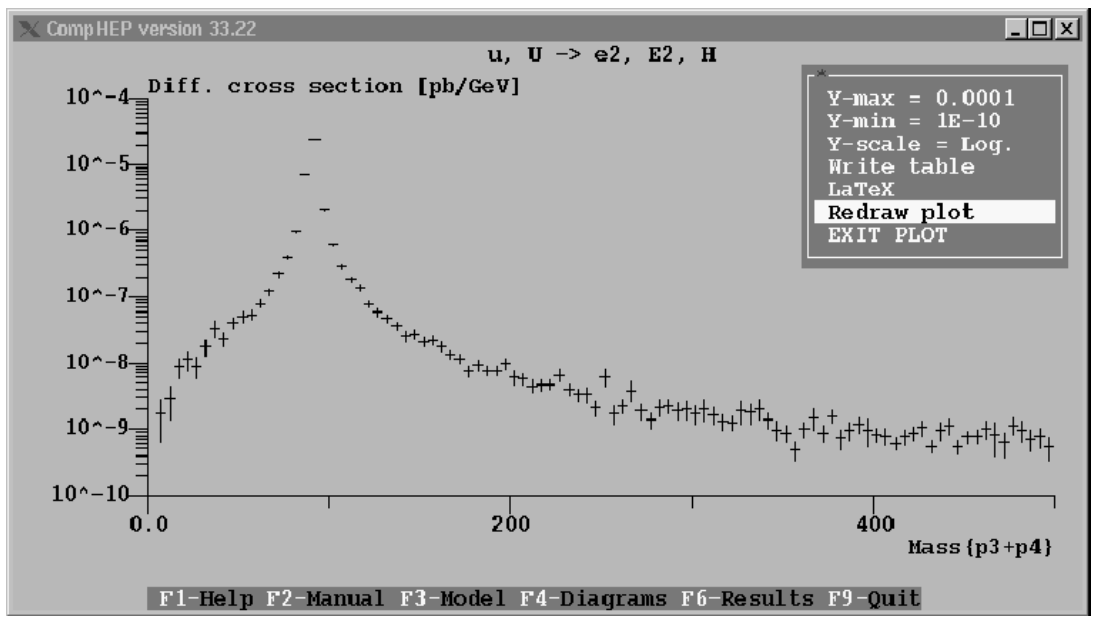,width=410pt, height=230pt}}
\end{picture}

\caption{ Example of the plot image }

\label{plot-image}

\end{figure}

\begin{figure}            
\begin{picture}(410,230)
\put(0,0){\epsfig{file=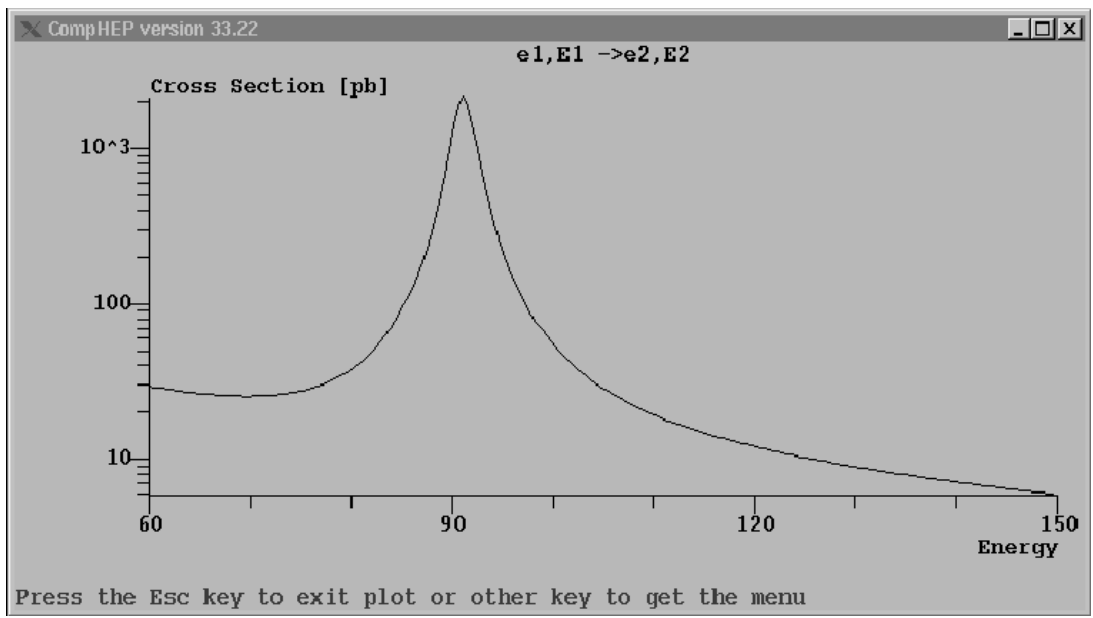,width=410pt, height=230pt}}
\end{picture}

\caption{ Plot for a 2-$>$2 process}

\label{results_2-2}

\end{figure}

\paragraph{ 7. Plot Viewer.}
This program is designed to display smooth curves and histograms.
Examples are presented in Fig.\ref{results_2-2} and  Fig.\ref{plot-image}. 

Being launched the {\it Plot Viewer} displays a picture and waits for the 
keyboard signal as it is shown in Fig.\ref{results_2-2}. The program ends
with pressing the \Esc~ key. 
If some other key is pressed then  the user gets
a menu which you can see in  Fig.\ref{plot-image}. This menu allows to  change 
the limits  of
vertical axis and its scale. Note that the logarithmic scale is available 
only if the lower limit is positive and the ratio of upper and lower limits 
is more then ten. To re-display the plot  choose the {\it 'Redraw plot'} option.  

The menu also provides the user with a possibility to save a graphical plot
image as a \LaTeX~ file and as a numerical table. The name of the 
corresponding  file is displayed on the screen just after  writing the file.
The numerical table created in this way can  be displayed on the screen 
later on by the command\\
\hspace*{2cm} {\tt \$COMPHEP/tab\_view < table\_file}\\
This {\it tab\_view} is just the same {\it Plot Viewer} but compiled as a stand-along
program.
 
The {\it 'Exit Plot'} menu function completes the {\it Plot Viewer} session.

\subsubsection  {Interface for a text-screen mode}
         
 Only the standard input/output \Fortran~ facilities are used for this part of
\CompHEP.  So the service is  primitive. The user input
here is a string terminated by the \Enter~ key.

\paragraph{1. Menu.}

  When you work with the menu the numerated list of menu functions is displayed.
 Possible inputs are: 
 
 \verb|H<number>| - to get help information about the
$\verb|<number>|^{th}$ menu function;
        
 \verb|<number>|   - to execute the corresponding menu function;

 \verb|X|   - to quit  the current menu.

The lower case characters {\it h} and {\it x} may be used also as 
key characters of the above commands. 
You can see in  Fig.\ref{screen_menu_f} how this menu looks like
on the display.

\paragraph{2. Tables.}

   Tables are used to enter and change  information about physical cuts
and a phase space regularization.   When you use this program a numerated 
list of  elements is displayed. Elements which have a  number  older than 9 are 
numerated by letters (A,B,C,D ...).
 Under this list you see a prompt for commands:\\
\hspace*{2cm}{\tt   N - new, \verb|D<set>| - delete, C<n> - change, X - quit.} 

\verb|N|   is  used  to insert a new table line. The subsequent input
           depends on the table.

\verb|D<set>|  is  used to delete records enumerated in \verb|<set>|. 
For example, \verb|D13A| will delete  the first, third and tenth  records.

\verb|C<n>|    allows you to change contents of the $\mbox{\verb|<n>|}^{th}$ item.

\verb|X|  is used to quit.  

The lower case characters {\it n, d, c} and {\it x} may be used also as 
key characters of the above commands.

\paragraph{3. Viewer.}
 
 When you browse the help information or  some file contents 
the following commands may be used:

to look at the next page      type   the {\it Spacebar} key;

to look at the preceding page      type  `b';

to exit from the viewer         type  `x'.

 \subsection{Menu system for symbolic calculation} 
             \label{symb-menu}          
\begin{figure} 
\begin{picture}(450,350)
\put(0,0){\epsfig{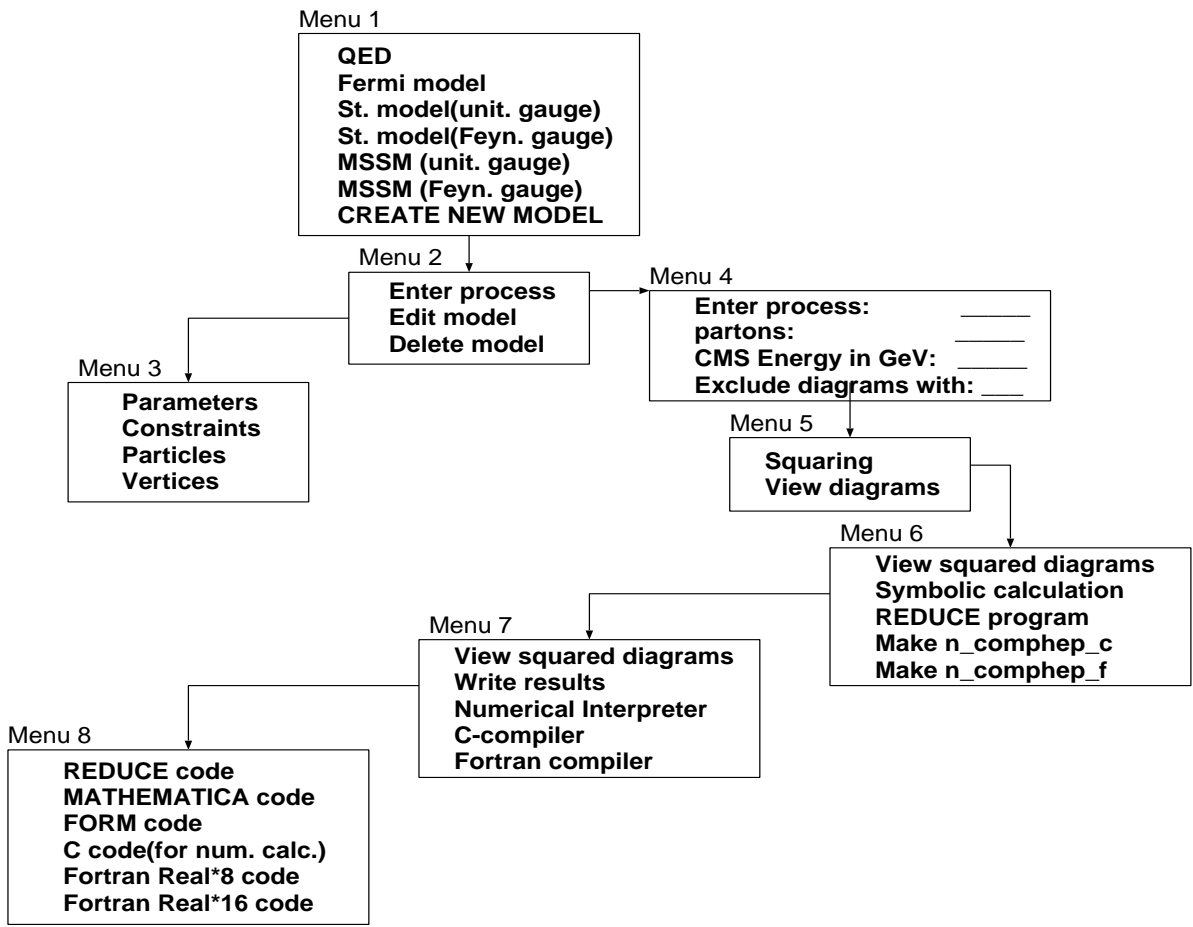}}
\end{picture}

\caption{ Menu scheme for the  symbolic session }

\label{s_chain}

\end{figure}

The scheme of menus  for the symbolic calculation session 
is presented  in Fig.\ref{s_chain}.

\subsubsection{Choice of  the model and work with it} 

\paragraph{ Menu 1. } 

This menu  contains a list of available models.  
It offers you  an option to select a model  of 
elementary particle interaction for  subsequent work. 

There are four  models built in \CompHEP:
quantum electrodynamics, the model of electroweak four-fermion
interaction, and two versions  of the Standard Model. 
  The QED model is included  as an example of realization of the simplest 
particle interaction scheme in \CompHEP. The 
four-fermion interaction  model gives an example of realization of the 
four-fermion interaction in the  \CompHEP~ notations. 
The Standard Model
is presented in two gauges: the unitary  one and the t'Hooft-Feynman one.
  We recommend  to choose the latter 
for  calculations because the ultra-violet cancellations
between diagrams caused by  gauge invariance 
are  absent in this case.  See the discussion in   Section \ref{squaring}.
The Standard Model in the unitary gauge  is used for verification 
of  gauge invariance. 
 
 The bottom  menu function  provides you with  a possibility to 
include a new model into the \CompHEP~ list.
New model is created as a copy of one of existing
models.  On the next menu level you can change this copy.
 If you  choose  the "New Model" menu function  you will  be prompted 
for  a new model name  and   a template source.  \CompHEP~
adds the underscore symbol '\_' in front of the  name of new model.
It serves to distinguish  user's models   from the built-in 
ones.  
To choose a template the list of  existing models appears.

\paragraph{  Menu 2.}

 The first function of this menu lets you  enter the physical process which 
you  wish to deal with. A format of process specification 
is explained below. You can also use the context help facility pressing
the {\it F1} key on any  step of the input.

  Before entering a process you  may  also edit the model
contents by means of  the {\it Edit Model} menu.
Later on you will be able only to browse  the
model contents (by  pressing {\it F3}), but not to
change it. 

If the currently used model is a user-created one,
the menu function {\it Delete model}  removes this model and 
\CompHEP~ returns to {\it Menu~1}.
In the  case of a built-in model   {\it Delete model} 
restores the default version of  model instead. Before the deletion 
or restoration the corresponding
warning appears and you can cancel the operation. 

\paragraph{ Menu 3.}

 Information about a model is  stored in four tables.  
Generally they are text files 
which are  disposed in  user's {\it models}
directory  and may be corrected by an ordinary text editor. 
But we strongly recommend to use \CompHEP~ facilities to edit these files
because in this case \CompHEP~ can control possible mistakes 
of the user input.

  \CompHEP~  displays  a menu of model tables. 
By choosing a position of this menu 
 you can   edit the corresponding  part of the model.  

The {\it Parameters}, {\it Constraints}, {\it Particles} and 
 {\it Vertices}  menu   functions let you browse  and edit
correspondingly:

1.  independent parameters of the chosen model;

2.  parameters depending on the basic ones;

3.  list of particles and their properties;

4.  vertices of interaction.

See  Section \ref{models} for the format of these tables 
and  also  Section \ref{X11interface}  for  the 
explanation of  facilities of the table editor.

\CompHEP~  verifies  the  model when  you try to leave this menu 
after some changes made in one of the tables.
If some error is detected the corresponding  message appears  
and no exit from the menu occurs.
This message  contains the diagnostics, the table name,  and  the
number of line where the error has been detected. You can  recall this
message later on within  the table editor by pressing  the {\it Ctrl E} key.

The check stops  when   the first  error is detected.
You can fix the error and  try to leave the {\it Edit Model} menu
once more.

 When you enter the {\it Edit~Model} menu  the  current  version of the  
model  is saved  and you have a possibility  to return to this version 
forgetting  your corrections. Just answer   {\it N}  the
question \\
\begin{center}
\begin{tabular}{|c|}
\hline
    Save correction ? \\ Y/N? \\
\hline 
\end{tabular}
\end{center}
which appears every time when you try to leave this menu after 
 correcting the model.      

  There is a sequence of   points which are being checked:
\begin{itemize}
\item  correctness of identifiers and numbers;
\item  declaration of an identifier before its use in the 
expression;
\item  declaration of a particle  before its use  in the  
vertex;
\item  correctness of algebraic expressions;
\item  compatibility of Lorentz indices;
\item  existence of the conjugated vertex.
\end{itemize}

See Section \ref{models} for a full list of requirements on the model.

\subsubsection{Input of the process}

\begin{figure}            
\begin{picture}(410,230)
\put(0,0){\epsfig{file=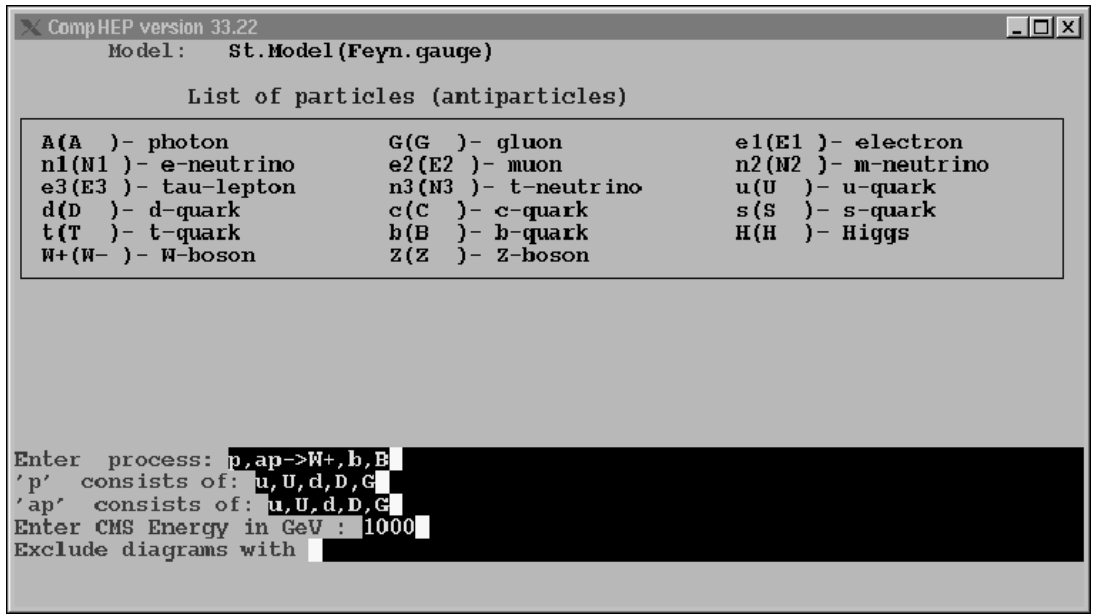,width=410pt, height=230pt}}
\end{picture}

\caption{ Example of the  process input }

\label{screen_ent}

\end{figure}

 After  activating   the  {\it Enter process}  function of {\it Menu~3}
the list of particles together with their   
notation conventions is displayed. The notation of  anti-particle is shown in
     parentheses after  that of particle. In the 
    case of  the Standard Model  the corresponding screen is 
shown in  Fig.\ref{screen_ent}.
 If the list is too long one may use the \PgUp~ and \PgDn~ buttons
to scroll it.

In the bottom part of the screen the prompt '{\it Enter process:}' appears.
      The  syntax for the input is: \\
\hspace*{2cm} {\tt P1[,P2] \verb|->|  P3,P4 [,...,[N*x]]} 

These  'P1'..'P4'  are  particle names, {\it N} is a  number of inclusive 
particles. The sets of {\it in}- and 
 {\it out}-particles are separated by the  arrow \verb|'->'| formed of two 
printing symbols. The
particles inside of each set are separated by commas.
The total number of particles should not exceed 6. 

  For example,  the input  $u,U $ \verb|->| $ G,G$ 
     denotes the process of annihilation of the $u$-quark and $\bar{u}$-quark 
     into two gluons. 

     One can also construct inclusive processes. For example, the input\\
\hspace*{2cm} {\tt         u,U \verb|->| G,G,2*x }\\
     is a request to construct all processes of 
        annihilation of the $u$-quark and $\bar{u}$-quark
     into two gluons accompanied by  two arbitrary particles. 

         If the program finds an unknown name  among the
     in-particles it will try to  consider it  as a name of composite 
     particle  and will ask you  about its parton contents. For
     instance, after the input\\
\hspace*{2cm} {\tt            e,p \verb|->| 3*x } \\
      the question appears:\\
     \hspace*{2cm}  {\tt  Is 'p' a composite  particle Y/N  ? } \\
     If you choose {\it 'Y'} you  will be prompted to specify the parton
     structure of  {\it 'p'}.  A possible input is\\  
      \hspace*{2cm}{\tt      'p' consists of: $ u,U,d,D,G $ }

     If one  enters a collision process,  the information on 
     total energy of colliding particles in the  center-of-mass system 
     is demanded:\\
     \hspace*{2cm} {\tt  Enter Sqrt(S) in GeV: {\it 300 } }

     \CompHEP~ generates only those channels  where 
the total mass of incoming particles and the total mass of
outgoing particles are smaller than  {\it Sqrt(S)}. 
  
     On the next step of input you are prompted  to exclude
diagrams with  specified virtual particles. 
The input should be\\
    \hspace*{2cm}{\tt Exclude diagrams with : $ P1>n1$ [$,P2>n2$ ...] }\\
where $P1,P2,...$ are particle names, $n1,n2,..$ are the quantity  limits.
Such an input means that diagrams where the number of virtual particles 
$P_i$ is more than   $n_i$ will not be constructed. For example:\\
  \hspace*{2cm}{\tt  Exclude diagrams with : $ W+>1 $ }\\
means that  only those diagrams  will be generated  which contain
less than two virtual W-bosons.

 Several restrictions separated by commas are allowed. If one 
has a restriction for a particle, the restriction for the corresponding 
anti-particle is not needed. If one would like to forbid the appearance of
some virtual particle $P$ at all, the input "$P>0$" may be 
shortened to  the "$P$".

This option may be  used to exclude diagrams which
are suppressed due to a large virtual  particle mass, or a small coupling
constant, or for  some other reasons.   Use the empty input to get a
full set of diagrams.

Use the \Esc~ key to return to the previous level of input and
the {\it F1} key to get the online help.

After the  input is completed \CompHEP~  starts the Feynman  diagram
generation. If the number of generated diagrams 
is zero  the corresponding warning appears and you  return 
to the beginning of process input, otherwise  the next menu appears.

\subsubsection{Squaring of diagrams and symbolic calculation}
\paragraph { Menu 5.}

This menu appears on the  screen just after  construction of   
Feynman  diagrams and together with the  information about numbers of diagrams and 
subprocesses generated.

The first function of this menu ({\it Squaring})  is the instruction
for \CompHEP~ to create squared diagrams. \CompHEP~ uses the  squared diagram
technique  for evaluation of squared matrix  elements. See Section
\ref{squaring} for details.

The  {\it View diagrams}  function gives you a tool 
to view a graphic  representation  of  generated  Feynman diagrams,
to  remove some  diagrams before the squaring and to create 
the \LaTeX~ output for undeleted diagrams. 

If a few   subprocesses have been  generated then the subprocess menu  appears
after an activation of the {\it View diagrams}  menu function. There is a
possibility  to remove all diagrams in the  highlighted subprocess by
pressing the {\it F7} button. In its turn  the {\it F8} key  restores all
diagrams of  the highlighted subprocess deleted before. 

When you  choose a subprocess the diagram graphic viewer is 
launched. See  Section \ref{X11interface} for details
or use the $F1$ and   $F2$
functional keys for online help.

Here we would like to note some peculiarities  of constructed diagrams.
\begin{itemize}
\item Incoming particles are drawn on the left side of diagrams,
whereas the outgoing ones are shown on the right. 
\item  Particles with  spins $ 0,\,\frac{1}{2}\,,1 $ are represented
by the dotted, solid, and dashed line correspondingly.
\item Charged particles are  represented by arrow lines. 
The  arrow indicates  the direction of particle (not anti-particle)
propagation.
\item  Incoming and outgoing particles are labeled by 
relevant names in the end of the line. Virtual particles are labeled
by their names at the mid-line. If a particle is not self-conjugated,
the particle's name is used for the labeling (not the anti-particle's one).   
\item In the case of collision process a first colliding particle is
disposed in the top part of picture, whereas the second one is drawn 
in the bottom part.  
\item \CompHEP~ produces only one representative of a  set of diagrams 
which  can be transformed one to another by replacing  the identical
outgoing particles.  For example, \CompHEP~ creates only one diagram for 
the $e1,E1$~\verb|->|~$A,A$  process, whereas all textbooks present two diagrams
in this case. In other words, momenta are not assigned to the outgoing 
particles  on the step of diagrams generation.

\item \CompHEP~ does not generate  diagrams   with  Faddeev-Popov and
Goldstone  ghosts\footnote{In the following text we use  {\it Goldstone  ghost}
instead of commonly used {\it Goldstone boson}. For our convenience we 
name both of these two kinds of fields and the auxiliary tensor field as 
{\it ghosts}.}. Such diagrams are restored on the step of 
squared  diagram evaluation. Any ghost has a real particle as a prototype
(see Section  \ref{ghostFields}). As a rule, a diagram with ghosts has got
the  parent diagram where ghosts are replaced by real particles. 
During the 
calculation of these parent diagrams the  contributions of the corresponding
ghost diagrams  are also  calculated and added to the contribution of the  parent 
one (see Section \ref{squaring}). There exist some exceptions 
from this rule. For example, the 
Standard Model contains the vertex with four Goldstone bosons associated 
with the $Z$-boson, however the $Z^4$ vertex is absent in the theory. 
  To provide a possibility to take into account the ghost diagrams with a
four-Goldstone  vertex \CompHEP~ generates  false diagrams with $Z^4$
interaction.

\item Vertices with a complicated color structure, for example, 
the four-gluon one, are implemented  by means of  unphysical tensor  field.
This field is treated by \CompHEP~ as a special gluon  ghost  and
does not appear in the constructed diagrams. 
A contribution of the four-gluon vertex is restored when
squared diagrams are  evaluated, in the same manner as the contribution of
Faddeev-Popov and Goldstone  ghosts.   
See Sections \ref{ghostFields}, \ref{lagrangian}, \ref{examples} for  further 
explanations.   

\end{itemize}

\paragraph{ Menu  6.} \label{sq_diagrams_menu}

The {\it View squared diagrams}   function is similar to 
the {\it View diagram} one of the previous menu
but is applied to the set of squared diagrams. 
Each squared diagram is a graphic representation 
of  $A B^{*}$ 
contribution in   the  squared matrix element,
where $A$ and $B$ are some amplitudes corresponding to 
Feynman diagrams. 

Let us summarize   some features  of \CompHEP~ squared diagrams 
generation:
\begin{itemize}  

\item   \CompHEP~ never constructs both   
$A  B^{*}$ and   $B  A^{*}$ diagrams  but only one of them.
Instead, for simplicity,  \CompHEP~ calculates $2*Re(A  B^{*})$
on the symbolic and numerical levels. 

\item  \CompHEP~  constructs only one representative  of a set of
diagrams which can be transformed one to another by permutation of 
identical outgoing particles. The needed symmetrization  for 
such particles is performed later on,  at the step of
numerical or symbolic summation.

\item  Each \CompHEP~ diagram represents a set of squared 
diagrams where some physical particles are replaced by their  ghosts
in all possible ways according to the  existing vertices of the  model.
To see the ghost  squared diagrams for each displayed one 
just  press  the {\it 'G'} key.
\end{itemize} 

If you   browse the squared diagrams after usage of the 
 {\it Symbolic calculation}  function  you  will see that each of 
the squared diagrams is marked by one of the following  labels 
 {\it CALC, ZERO, Out of memory, Del}. They mean that the
diagram has been  successfully calculated; gives a zero contribution;
cannot be calculated; or has been  deleted,  correspondingly.

The {\it Symbolic calculation}  function starts symbolic 
evaluation of the generated  squared diagrams. This  evaluation
is performed by the built-in symbolic calculator created
specially in the framework of the  \CompHEP~ project.

During the calculation the information about a number of 
calculated diagrams and about a step of evaluation of the current
diagram is shown.

The {\it Reduce program}  function creates  a code of  symbolic 
evaluation in the format of 
\REDUCE~ language \cite{REDUCE} separately  for each squared diagram. 
These codes  are not used for the further \CompHEP~ processing, but 
they can be useful for  cross-checking  the  \CompHEP~ software.
On  one hand, one  can  investigate the  \REDUCE~ code to get
conviction that it correctly calculates a contribution of the 
squared diagram. 
On the  other hand, you may compare the  result of the \REDUCE~ evaluation of
diagram code with the result of built-in symbolic calculator. 
There are some tools created for this purpose. See Section \ref{symb-calc-check}
for details.

General purpose of a package like \CompHEP~ is to create the
corresponding \C~ or \FORTRAN~ source code for  further numerical 
processing and compile  this code  using the corresponding 
system facilities. User's control is not necessary  for this  step as well as
for  the step of symbolic calculations. 
\CompHEP~ provides the user with a possibility to perform the above steps 
in a non-interactive mode. To start the non-interactive session  one could activate 
the {\it Make~n\_comphep\_c} or {\it Make~n\_comphep\_f}
menu function. Then  the current interactive session ends and a
new  batch process starts. See  Section 
\ref{blind} for details of the non-interactive  calculations in \CompHEP. 

As the outcome, the executable file {\it n\_comphep\_c} or {\it n\_comphep\_f}
is created for the further numerical precessing. You may find it in the 
{\it results} directory. For the time of batch
calculation the {\it LOCK} file appears in the user working 
directory to prevent a double start of the \CompHEP~ package within the same 
place of the file system.

 As a rule the  System Administrator  requires to start a  large job
in a special batch regime. The execution of  one of the {\it Make n\_comphep\_\* } 
functions  may require large computer resources for some processes
with large number of diagrams. In this case you could correct your
{\it comphep}   command file according to the System Administrator rules.
The corresponding  calls are disposed after labels 24) and 25) in this 
file.

\subsubsection{Output of results and launching of the numerical calculation} 

\paragraph{ Menu  7.} 
The  {\it View~squared~diagrams}  function is identical to 
the  same  function of the previous menu. Following  a logic  of the
program you can  reach this menu only if all squared diagrams are 
marked either as deleted or as calculated. If you undelete some
diagrams  then you will be returned to the previous menu.
 
The {\it Write~results} function lets  you   save the
results of symbolic calculation in terms of one of computer languages 
for further numerical or symbolic evaluation. See  {\it Menu~8}  for details.
 The files are saved in the user's {\it results} sub-directory.

The next three menu functions   realize  numerical calculations 
in  \CompHEP, particularly Monte Carlo sessions. The  first one calls the built-in numerical
interpreter of the  obtained symbolic expression. This program has 
almost the same possibilities as the numerical code produced by the
$C$-compiler, but works slower than the compiled option.

The menu function {\it C-compiler}/{\it Fortran~ compiler}
calls the corresponding compiler to create and launch an executable file 
{\it n\_comphep\_c}/{\it n\_comphep\_f } in the directory {\it results}.
It is assumed that the corresponding source code was created earlier
using the {\it Write results} menu function.

\paragraph { Menu 8.}

This menu provides the user a possibility to save the obtained 
symbolic results in different formats. The \REDUCE~  and \MATHEMATICA~ 
outputs may be used for further symbolic manipulation. 
For example, you can get  symbolic representation of a
  sum of all diagrams, perform some substitutions or  
expansions, evaluate the symbolic expression  for the total
cross-section or width  of the process. As a rule such manipulations are
possible and reasonable for the processes with a small number of final particles.
The structure of the symbolic output and  some programs for manipulation 
with the  \REDUCE~ output are described  in  Section \ref{output}.

  Otherwise the \C~ and \FORTRAN~ outputs are used  to get numerical results.
They may be compiled and linked with the library  of  
\CompHEP~ routines for phase space integration.
 The produced executable program allows one to obtain numerical values of the
total cross-section or width, distributions and an event flow, taking into
account a variety of cuts.
 In order to  create an executable file  for numerical
calculations based on the \C/\FORTRAN~ output  
use  functions of  {\it Menu~6} or 
launch the {\it \$COMPHEP/make\_\_n\_comphep} program as it has been explained 
in Section \ref{file-commands}.

\subsubsection{Non-interactive session} \label{blind}

   \CompHEP~ was created as a program for calculation in the interactive
regime. But in practice  long-time calculations are typical. Due to this 
circumstance we have implemented the batch (or, say, `blind') calculation 
mode into \CompHEP.

If one launches  {\it s\_comphep} with an option "-blind", e.g.\\
\hspace*{2cm} {\tt   \$COMPHEP/s\_comphep -blind STRING }\\
then \CompHEP~ does not open the X11 window and  reads  the control characters 
from  {\it STRING}.
For imitating functional keys the following control characters are
defined:\\
\hspace*{2cm} '\{':  Escape key\\
\hspace*{2cm} ']':   Right Arrow\\ 
\hspace*{2cm} '[':   Left Arrow key\\
\hspace*{2cm} '\}':  Enter key


  So, the user has a possibility to concatenate into  {\it STRING}
a  list of commands  which simulate the keyboard and execute 
{\it s\_comphep} in the non-interactive mode.

The work of {\it s\_comphep} in the blind mode is simplified a little. 
It automatically passes by all  warnings and  messages  answering 'Y' (Yes)
any question itself. The  \CompHEP~ title screen  is missed in 
this mode and the  simulation of keyboard hit, which is used to leave the
first  screen, is not needed.

An   example of the blind session is 
realized in  your {\it comphep} command  file.  
When  you activate the  {\it Make n\_comphep\_c}  function of {\it  Menu~6}
then  {\it s\_comphep} ends with an exit code 24 and \\
\hspace*{2cm}{\tt   \$COMPHEP/s\_comphep -blind "]\}\}]]]\}9" }\\
is launched. The above command performs  symbolical calculations  
and writes down the C-code for the squared matrix element
into the directory {\it results}. Note that the last symbol '9' simulates 
the work of the {\it F9} functional key which in its turn completes the session
(See Section \ref{X11interface}).

Another utilization of the {\it blind} mode is a check of \CompHEP~ for some 
huge set of processes (Section  \ref{symb-calc-check}).

 \subsection{Numerical calculation by \CompHEP}  
             \label{num-menu}           
 In the framework of  \CompHEP~ package  there are four ways to 
realize numerical calculations, namely using:
\begin{itemize}
\item[1)]  built-in interpreter  
\item[2)]  compiled \C~ output
\item[3)]  compiled  {\it double precision} (REAL*8)   \FORTRAN~ output
\item[4)]  compiled  {\it quadruple precision} (REAL*16) \FORTRAN~ output.   
\end{itemize}
For the  first two items  calculations are performed 
with a {\it double} precision.  The interpreter works 
slower  compared with the stand-alone compiled executables, but it
does not need such steps as the storage and  compilation of symbolic
output. So it is reasonable to use the interpreter either in the case of
simple process or if you have no compiler at all. 

The general  functionality scheme is the same for all four cases.
The main differences  are: 
\begin{itemize}
\item[-] for the first two items the user interface is more comfortable
because it is based on graphic terminal facilities; 
\item[-]  results of calculation produced by the interpreter are
displayed on the screen only, whereas in other cases they are  stored onto a 
disk either;
\item[-]  {\it batch} calculations (set of evaluations launched by one
job task) are not available for the interpreter; 
\item[-] for the current version only  \FORTRAN~ programs provide
the user with a possibility to generate an event flow.
 
\end{itemize}
 
The interpreter is launched  from within the symbolic
session.
Compiled  programs can be  created and  started under the  symbolic session 
(see {\it Menu~7} in Fig.\ref{s_chain})
or created by means of Section \ref{file-commands} commands and
launched as an independent process.

\subsubsection{Sketch of the menu system}

\begin{figure}[htb]        
\begin{picture}(410,230)
\put(0,0){\epsfig{file=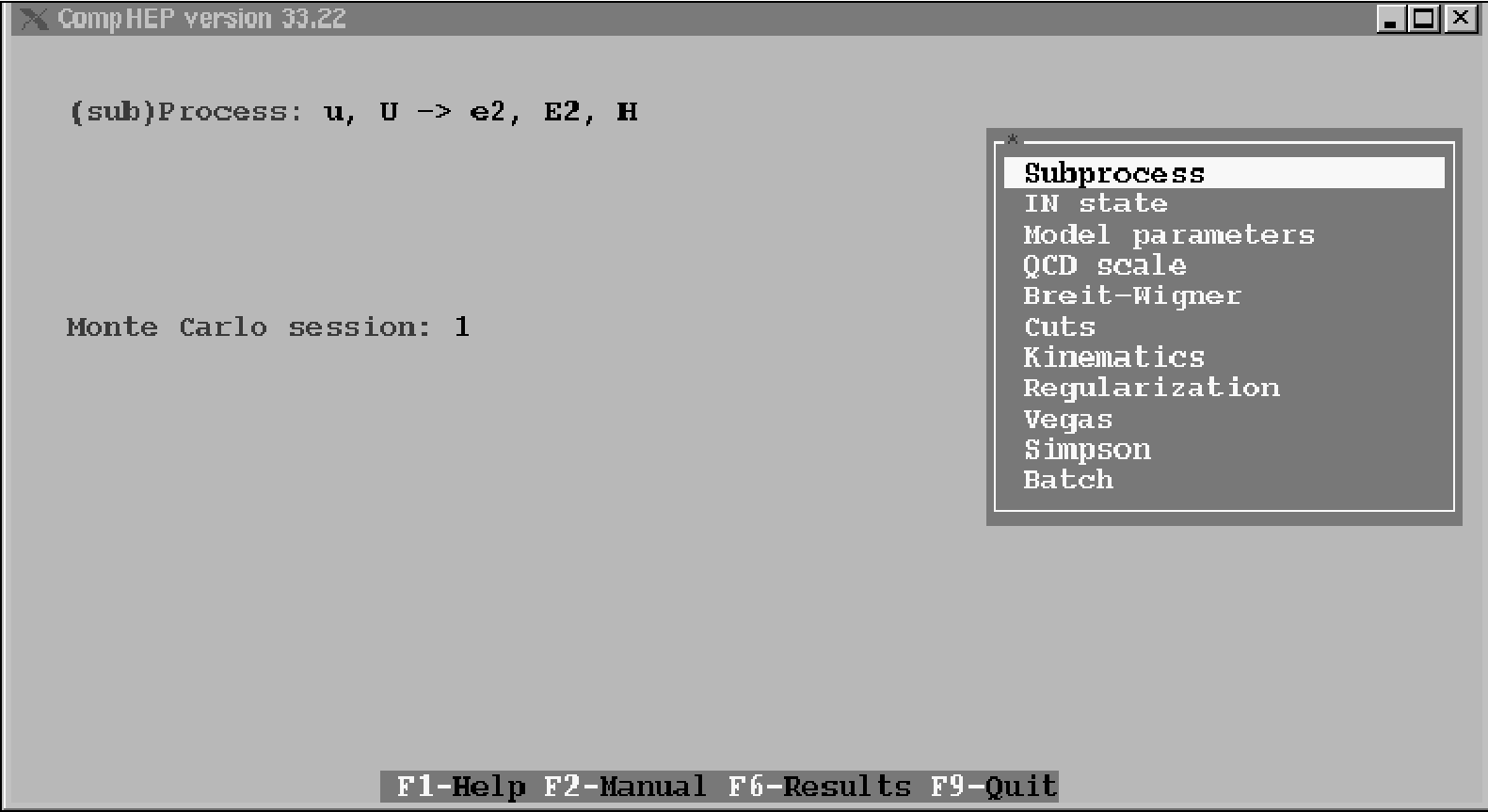,width=410pt, height=230pt}}
\end{picture}
\caption{ Example of the \CompHEP~ menu}
\label{screen_menu}
\end{figure}

\begin{figure}            

\begin{picture}(450,350)
\put(0,0){\epsfig{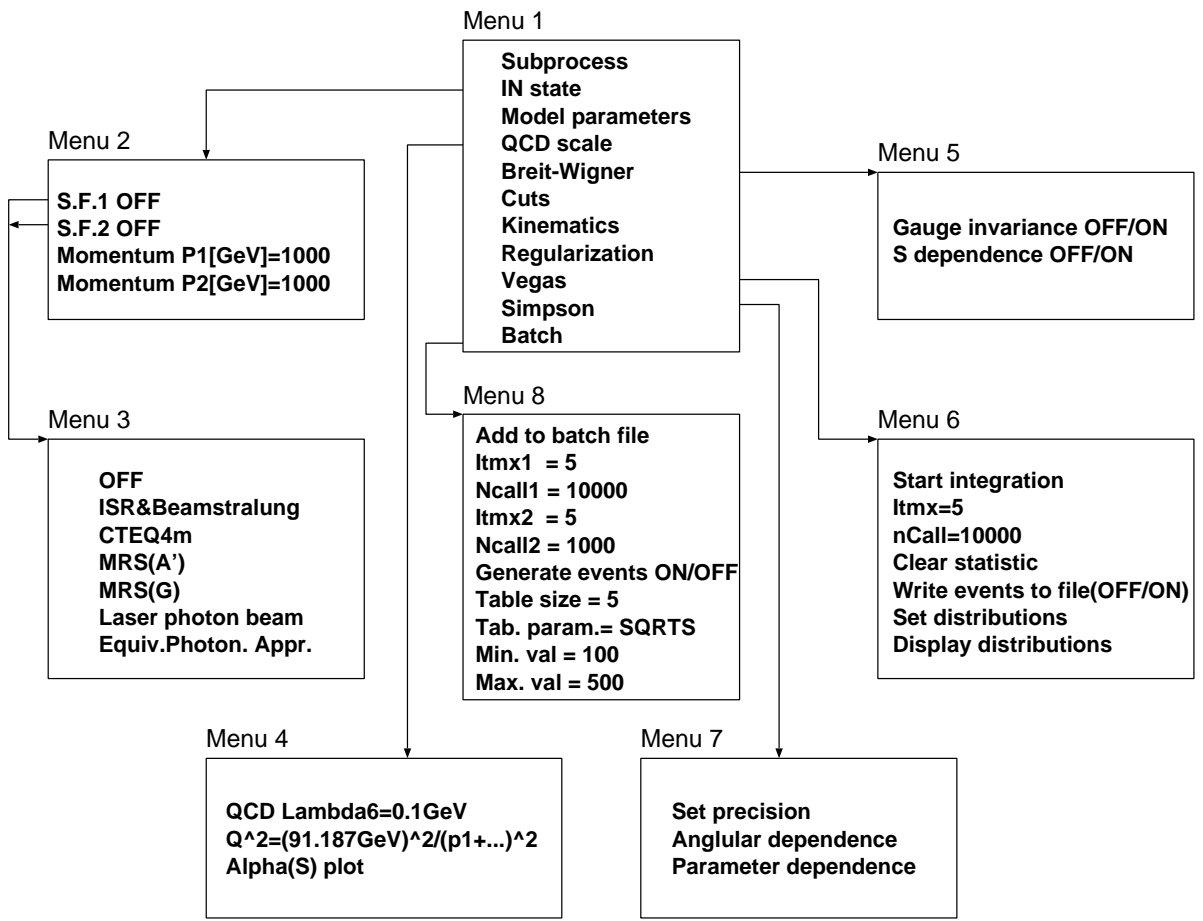}}
\end{picture}

\caption{ Menu scheme for the  numerical session }

\label{n_chain}

\end{figure}

A general scheme of   \CompHEP~  numerical calculations 
is presented in Fig.\ref{n_chain}. 
The \CompHEP~ screen with  the   menu  number 1  is presented in
Figs. \ref{screen_menu} and \ref{screen_menu_f}.  
  Three functions
of this menu  launch   integration programs. One of them is 
 Monte Carlo program \VEGAS~   \cite{Lepage,NumRecFORT}.
  The other one is the ordinary {\it Simpson} 
integration routine which can operate only for the simplest 
2~\verb|->|~2 type of reactions.  The {\it Batch} menu function is used 
to organize a set of \VEGAS~ evaluations in the non-interactive mode (see
Section \ref{batch}).  
 Other menu items are needed to set  an environment for integration.
Below we  briefly   describe all of them. 
More  detailed explanation is given in the following sections.

By means of the {\it  Subprocess} function the user may  select 
one subprocess for a  further processing. This function is active 
only if  several subprocesses have been evaluated  symbolically before.
The name of  current subprocess
is displayed on the screen (Fig.\ref{screen_menu} and
Fig.\ref{screen_menu_f}).
 Subprocess summation is not implemented in the current 
version.

The {\it In\_state}  function gives the user an option
to choose  structure functions for incoming particles and
assign some values for their momenta. The  {\it GeV} units
are assumed for the momenta.
  This menu function is available for  collision processes 
only.

The {\it Model\_parameters}  function allows the user to assign a 
new value to any  independent parameter involved in the
evaluation.  After  an activation of this
function  the corresponding menu of parameters with information
about their  values is displayed.  As a result of  changing  
the  independent parameters 
some constrained parameter can  get non-evaluated as a 
result of a negative square root argument or division by zero in the
relevant  expressions. In this
case an error  message appears on the screen and the user
is required to choose an admissible set of parameter values. 

  Although the user may assign some value to the  QCD strong coupling
parameter {\it GG} it does not influence on results of calculation.
This parameter will be reset according to the rules defined by the 
{\it QCD~scale} menu function.   

  \CompHEP~ substitutes a Breit-Wigner exact propagator for the  s-channel
virtual particles.
 This trick is
needed to avoid  the integral divergence near the mass pole
and   is motivated by high order corrections.
The way of implementation of the  Breit-Wigner propagator  is driven by  
the {\it Breit-Wigner} menu function. See Section (\ref{Breit-Wigner}) 
for details.

The {\it Cuts} menu function provides the user with a possibility  to 
cut a  phase space volume. A set of physical variables which 
could be used to construct a cut function is described in Section \ref{functions}.
Details of  the {\it Cuts} menu processing are given in Section \ref{cuts}.

  The {\it Kinematics} and {\it Regularization}  menu functions are used
to construct an appropriate mapping of phase space onto  the  \VEGAS~ integration 
volume. This is  important for successful Monte Carlo integration.

\subsubsection{Numeration of Monte Carlo sessions}
\CompHEP~  numerical calculations are organized as
a sequence of Monte Carlo integration sessions. Each session has its ordinal
number. The current session number is
displayed. 

All results of calculations are written into the
{\it results} directory.  
To distinguish files produced in different
sessions we include a session number {\it N} in the output file names.
 To distinguish files produced by {\it n\_comphep\_f}
from those produced by {\it n\_comphep\_c} we use prefixes {\it 'c\_'} and {\it
'f\_'} as required.  
   The results  of integration and all settings   are stored in the protocol 
file {\it f\_prt\_}{\tt N} ({\it c\_prt\_}{\tt N}).  Events generated 
by {\it n\_comphep\_f}
are stacked in the file {\it events\_}{\tt N}.
 A current version of the  {\it n\_comphep\_c}
program does not save events.

  The latest settings are saved in the file {\it f\_session.dat
(c\_session.dat)}
when
{\it n\_comphep} is quitted.  On a next launch this file is scanned to restore
the
last settings.

\subsubsection  {QCD scale}

  \CompHEP~ substitutes the QCD strong coupling constant as a
function of the  ${\Lambda}^{(6)}$ parameter and the  $Q^2$ scale value.
It is calculated in the next-to-next-to-leading order (NLLO)
and  depends on a number of  quark flavors ({\it nf}) with masses less
than Q (see Eq.9.5a of \cite{ParticlesFields}). 
Relevant constants  $\Lambda^{(nf)}$ are recalculated
from  ${\Lambda}^{(6)}$  according to the  matching equation (9.7) 
of \cite{ParticlesFields}.
     
If the user is going to change  a numerical value of  $\Lambda^{(6)}$ 
or the $Q^2$ scale definition he  should  call the necessary function of
{\it Menu~4}. For the    $Q^2$ choice he will be prompted 
what  kind of scale he would like to implement, namely, the {\it constant}
scale or  the {\it running} one. In the first case  a subsequent input 
of number in {\it GeV} units is expected. In the second case \CompHEP~ 
provides the  user with  a possibility to define the  $Q^2$ scale as a
squared sum of particle momenta.  Digits on  input will be
transformed into  ordinal momentum numbers. For example: the input '13' will
be treated  as  $Q^2=|(p1-p3)^2|$.

   If $Q^2$  is defined as a constant, the corresponding value
of $\alpha = GG^2 / ( 4 \pi )$ is displayed on the screen.
 The  $\alpha(s)$ plot  for a user-defined region of {\it s}   is  available
also by means of the corresponding  function of {\it Menu~4}. 
 The same $Q^2$ scale is used as an argument
of parton distribution functions.

\subsubsection {Breit-Wigner propagator}
\label{Breit-Wigner}
The propagator of virtual particle
has a pole at $p^2 = m^2$:
$$\frac{1}{p^2 -m^2}\;\;.$$
If the pole is situated inside the phase space volume 
it leads to a non-integrable
singularity. The general solution of this 
problem is an account  of special set of  high order corrections
\cite{BD}. They  transform the propagator to the  Breit-Wigner form
 $$ \frac{1}{p^2 -m^2 -i \cdot \Gamma(p^2) \cdot m}\;\;,$$
where the value $\Gamma(m^2)$  is the particle width (reversed mean life
time).  

First  problem  which  appears in a way of implementation of this
expression is a choice of $\Gamma(p^2)$ dependence. 
The  $\Gamma$  value is 
essential near    the pole point $p^2 = m^2$. Thus, for the
first approximation we can put $\Gamma(p^2)=\Gamma(m^2)=const$.
It corresponds to the position {\it OFF} of {\it 'S~dependence'} switch.
In some papers it is declared that  $\Gamma(p^2)=\Gamma(m^2) \cdot
\sqrt{p^2}/m$  describes the pole shape better. This choice  corresponds to 
the position {\it ON} of  {\it 'S~dependence'} switch.

The second  and even  more important problem is a gauge symmetry breaking.
Generally we  have  this symmetry in any order of perturbation theory
but  the intervention of a part of  higher order terms  to the lowest order
expression via the Breit-Wigner propagator can break it.  

The gauge symmetry
is responsible for some cancellation of  diagram contributions
(see Section \ref{cancellations}), and its  violation, in turn, 
prevents the cancellations 
and can lead to a completely wrong result. 
The user could  solve this problem by setting  the {\it Gauge~invariance}   
menu switch to the position {\sc ON}. 
In this case the contribution of a
diagram which does not contain the Breit-Wigner propagator is
multiplied by  factor \cite{BW-factor,BW-factor2}  
\begin{equation}
\frac{(p^2-m^2)^2}{(p^2 -m^2)^2 +  (width(p^2) \cdot m)^2}\;\;.
\label{width_factor}
\end{equation}
This trick corresponds to the symbolic summation of all diagram
contributions at a common denominator expression and to
a subsequent substitution of the width term  into the factored
denominator. The  trick
allows to keep all gauge-motivated cancellations. As a defect of the trick it
must be mentioned that the factor (\ref{width_factor}) kills a contribution
of non-resonant diagrams in the resonance point \cite{BadGIWdth}. 
If the  particle
width is very small such an approximation is reasonable, but in some
 cases it can also lead to an error.

\subsubsection{Phase space functions} \label{functions}

There is a special set of phase space functions
 which  may be used to construct cuts and  distributions in the framework of 
 \CompHEP.
  The general notation looks like \\
  \hspace*{2cm} {\it (Key~Character)(momentum~set)}.\\
A set of momenta  is represented as a set of digits. Any digit corresponds 
to a momentum number. For  example, {\it C13} means a cosine of the
angle between  momenta ${\bf p_1}$ and ${\bf p_3} $. 
The momenta are assigned to   particles according to their sequence in the
process name.

Below we list the available  key characters and explain the meaning of 
the corresponding  physical functions.

\begin{itemize}
\item [A] - angle  in degree units. 
\item[C]  - cosine of angle.   
\item[J]  - jet cone angle. The jet cone angle ${\it J_{ij}}$ is defined 
as $\sqrt{\Delta y^2+\Delta\varphi^2}$, where  
$\Delta y$ is the pseudo-rapidity difference and $\Delta \varphi$  is the azimuth angle
difference for momenta ${\bf p_i}$ and ${\bf p_j}$. 

\item[E]  - energy of  the particle set.
\item[M]  - mass of the  particle set. 
\item[P] - cosine in  the rest frame of pair.  ${\it Pij}$ is
defined as follows: we imply a boost in
the direction of ${\bf p_i + p_j}$ to  get the
rest frame of the pair. Then  {\it P} is a cosine of angle between 
the transformed ${\bf  p_i}$  and  the direction of boost.

\item[T]  - transverse momentum   of the  particle set.  
\item[S]  - squared invariant mass of the particle set.
\item[Y] -  rapidity of the particle set. 
\item[U] -  user's implemented  function. The character string following 
 {\it U}  is passed on to the user   \FORTRAN/\C~ function   {\it
usrfun(str)} which, as it is assumed,  calculates a corresponding  value. See
Section \ref{user-prg} for  further explanations.

\end{itemize}

Zero components of all  momenta  are positive, and the conservation law 
means that the sum of momenta of incoming particles is equal to that 
of outgoing particles. Relative momenta signs   for the {\it S}
 function are substituted automatically.

There  are some natural requirements on the momentum set, namely:
 a)~the momentum sets following    {\it A, C, J, P}   must consist 
of two elements; b)~the momenta following   {\it M, P, T}   must belong 
to  outgoing  particles only; c)~any momentum can appear only once
in  one set.

\subsubsection{Cuts} \label{cuts}

The {\it Cuts} menu function opens a table for cuts definition.
This table consists of three columns. 
The  phase space function  in notations of Section \ref{functions}  must be 
 written   in the first  column. 
The second and the third columns  define the  minimal  and maximal
limits  for this function.  If one of them  is empty   the corresponding
limit is not applied.

   Input   formats for limits  are  different 
for the \FORTRAN~ and \C~  realizations.
  In the case of  
{\it n\_comphep\_c} and  {\it interpreter} sessions the user may type
in the corresponding fields any algebraic formula
which contains numbers and identifiers enumerated in the {\it 
Model~parameters} menu. Parentheses "()" and operations $ "+,-,/,*,**,sqrt()"$
also are permitted.

In the case of {\it n\_comphep\_f } session only numbers are expected 
in the  corresponding columns. To escape an input in $GeV^2$ units 
in the case of squared  momenta function
\begin{center} $ V_{min} <  S(momenta~set) < V_{max} $, \end{center} 
the  {\it n\_comphep\_f } program
 asks the user to put  $V_{min}/\sqrt{|V_{min}|}$ and 
  $V_{max}/\sqrt{|V_{max}|}$   in the relevant cells.

  It is very important to note that only the cuts on masses and squared
momenta 
(the {\it S} and {\it M} cases)  are realized via a reduction of the phase space.
Other ones are realized by  multiplication of squared
matrix element by a step function.

\subsubsection{Kinematics} \label{Kinematics}

The {\it Kinematics} function  allows to  display and change  a general 
scheme of the phase space parameterization.
  A kinematics scheme in  \CompHEP~ is
defined by a set of subsequent   1~\verb|->|~2 decays
of the incoming state into the  outgoing particle sets. 

  After activation of this menu function the current 
decay scheme will be displayed and a dialogue message 
will propose the user to accept it or put in another one.

 During the input \CompHEP~  consequently  asks the user  to define
a  first outgoing cluster of  virtual decay. The user input must be 
a sequence of particle (momentum) numbers without any separating symbols.
It is  assumed that 
particles are numerated in the same order as they are written down in 
 the (sub)process name.

  An example of  kinematics 
definition for 2~\verb|->|~4 process is given by the following lines: \\
\hspace*{2cm} 12~~\verb|->|~{\bf 3},456\\
\hspace*{2cm}456~\verb|->|~{\bf 56},3\\
\hspace*{2cm}56~~ \verb|->|~5,6\\
Here the user input is indicated in bold characters. 

In the framework of such a decay scheme the multi-particle phase space
is parameterized by masses of sub-clusters and  by two-dimensional  spherical
angles of  1~\verb|->|~2 decays  \cite{B4,kinematics}.
 A choice of kinematics influences on the phase space mapping
 and by this way it influences  
on convergence of  Monte Carlo integration. See Section \ref{kinematics}.

\subsubsection {Regularization} \label{regul_menu}

  In   general case a squared matrix element is
too singular for direct Monte Carlo integration.
  Singularities of the matrix element are caused
by poles of virtual particle propagators
and can have one of the following forms
\begin{eqnarray}
\label{pole_1}
1/(p^2 - m^2)\\
\label{pole_2}
1/(p^2 - m^2)^2\\
\label{pole_w}
1/( (p^2-m^2)^2 + (m\cdot\Gamma)^2)
\end{eqnarray}
where $m$, $\Gamma$, and $p$  are the  mass, 
 width, and  momentum of virtual particle.

The {\it Regularization} menu function allows the user to
 point out  dangerous denominators for automatic 
smoothing  the  sharp peaks of the squared matrix
element.
The regularization  table contains  four fields: {\it momenta, mass, width
} and {\it power}.

 Momentum of virtual particle is a sum of momenta of incoming and
outgoing particles. Just type the ordering numbers of these
momenta  in the {\it Momentum} field. The sign is substituted automatically.  
For example, for a collision process 
{\bf 12}  is treated as  $(p1+p2)$ and
{\bf 134} is treated as  $(p1-p3-p4)$.

  {\it Mass} and {\it Width} describe a position of the pole.
In the case of \FORTRAN~ program the corresponding  numerical
values must be written down in these fields. In the case of {\it C} program
 the user can write down some algebraic expressions which contain the identifiers
enumerated in the {\it Model parameters} menu.  For  t-channel propagators
(both incoming and outgoing momenta contribute to  $P$)   only a zero  value 
in the {\it Width} field is permitted
because \CompHEP~ ignores particle width term for such propagators.

  The {\it Power} field  defines an exponent of the propagator.
 Acceptable values are 1 and 2. Of course, in a squared
matrix element any propagator appears to the power of 2. But
sometimes as a result of gauge cancellations the exponent can be
efficiently  decreased to 1. If the {\it Width} field is not
equal to 0 such a cancellation is not expected and \CompHEP~
will use value 2 for the exponent ignoring the user input.

The work of regularization program is sensitive to the
{\it S} and {\it M} types of cuts (see Section \ref{cuts}) and 
is not  sensitive to other ones. Consequently, if you would like to 
smooth some singularity due to the pole  inside the phase space, you should 
apply the {\it S}  or {\it M}  cut to exclude the pole point 
from the consideration.

The algorithm of regularization is explained in Section 
\ref{kinematics}.

\subsubsection{VEGAS menu}

 See Section \ref{vegas} or \cite{Lepage,NumRecFORT} for  explanation
of the \VEGAS~ algorithm. The work of \VEGAS~ program is driven by {\it
Menu~6} of Fig.\ref{n_chain}.

The first position of this menu starts the process of  Monte Carlo
integration. It consists of {\it Itmx} runs with subsequent
adaptation of the  Monte Carlo weight function to the integrand. 
For each run \VEGAS~ calls the integrand about  {\it Ncall} times.
The user may set the {\it Itmx} and {\it Ncall} parameters applying  the next 
two items of the menu. 
The results of calculations are displayed  on the screen and
 written to the protocol file.

 If the required  precision of calculation 
is not reached the user may call the {\it Start integration} menu 
function once more to increase the Monte Carlo statistic.  
But if the {\it Clear~statistic} menu function 
is called, then the next  Monte Carlo session
will be started from the beginning.    
The  Monte Carlo weight function is saved between runs anyway.

In the case of \FORTRAN~ {\it n\_comphep} program the user can  write
weighted events generated by \VEGAS~ into a file. The maximum
and average values of event weights are displayed on the screen and written 
into the protocol file.  
The user must take into account that 
any new call of the {\it Start integration} menu function will 
rewrite this file.     

In the case of \C~ version of the  {\it n\_comphep} program the user   
has a possibility to fill in some histograms during  the Monte Carlo
integration. In this case he needs to activate the {\it Set Distribution}
function. This menu function opens a table of distributions 
which can be filled in by the user. The first column defines a phase
space  function  which the user would like to scan. The set of available 
functions and the input format are  defined in Section \ref{functions}.
The next two  columns define a region of variation of the corresponding 
phase space function. There could be either a number or an algebraic
expression containing the identifiers listed in the  {\it Model~parameters} 
menu.

  The user can browse the distributions using the {\it
Display~distributions}
menu function. This menu function asks the user to choose one  
distribution and a number of bins. After that the distribution appears. 
 A call for {\it Clear~statistic} menu function  clears  the stored
distributions.

\subsubsection{Batch calculations} \label{batch}

   \CompHEP~ was created as a program for calculation in an interactive
mode. But in practice   long-time calculations are typical and it is 
reasonable to perform them in a batch queue regime. According
to this 
requirement we have implemented the batch (or, say, `blind') calculation 
mode in \CompHEP.

  During the work with the {\it n\_comphep\_f(c) } program  in the interactive 
mode you have a possibility to store all current settings and task
parameters 
in the  {\it f(c)\_batch.dat} file. The function {\it Batch} of the main menu 
is used 
to define such a task and save it in a file. It opens the  {\it  Menu~8} 
(see Fig.\ref{n_chain}). Its first function  appends  parameters of the  current 
session to the {\it f(c)\_batch.dat} file.  It is possible 
to join several tasks in one {\it f(c)\_batch.dat} file.

  For the batch mode you are provided with a possibility to perform two
sessions
of \VEGAS~ calculation with automatic call for  the {\it Clear~statistic} 
function between the sessions. These sessions are driven by the 
{\it Itmx1, nCall1, Itmx2, nCall2} parameters which in their turn
may be defined by means of functions of {\it Menu~8}. The 
meanings of these parameters are similar to the same name parameters of 
the \VEGAS~ menu.  If you put one of these
parameters equal to zero then only one session will be launched.

If you set the  {\it Generate events} switch   {\tt ON}, then the 
squared matrix elements calculated 
during the last \VEGAS~ session  will be saved in 
a file for further processing (see Section \ref{events}).

There is a  possibility to organize  cycle calculation within one task.
The  {\it Table size}  function  defines a number of steps for
the cycle  calculation.  The user may choose a parameter to scan by means of 
the     {\it Table param}  function. The last two menu functions
allow to set the extreme values for the scanning parameter.

  Every time when you  start {\it n\_comphep\_f}~ it checks a presence of 
the {\it f\_batch.dat} file.
If it exists you will be prompted:\\
\hspace*{2cm}{\tt Would you like to use the BATCH mode?} 

If you answer {\it 'y'}  then {\it n\_comphep} reads the tasks from the 
{\it f\_batch.dat } file and performs requested calculations. 

So, if the {\it f\_batch.dat} file exists the batch calculations can
be launched by
the pipe command:\\
\hspace*{2cm}{\tt echo y |n\_comphep\_f}

The batch calculation in the case of {\it n\_comphep\_c}
is started by \\
\hspace*{2cm} {\tt    n\_comphep\_c -batch}

   \CompHEP~ has an internal counter for Monte Carlo sessions
which accrues after each Monte Carlo session. All output files 
have this session number as a part of their names. Hence, 
outputs from different tasks are written down in  different files and 
all results are  safely remembered after the {\it n\_comphep} run.
In the case of cycle calculation the additional 
joint file {\it f(c)\_table\_{\tt N}\_{\tt M}} with results of calculation 
will be created. Here  {\tt N} and {\tt M} are the first and the last 
number of session for the current cycle.

\subsubsection{Integration by means of the Simpson method} \label{simpson}
In the case of 2\verb|->|2 type of process we provide the user with a possibility
to calculate the cross-section using the  adaptive Simpson  algorithm.
It allows to get the results instantly and with a high precision.

Note that some user's settings are ignored when you choose this 
option. Namely, the structure functions are switched off, the center-of-mass
rapidity  is assigned to zero, regularizations are also ignored   and 
only  angle and cosine cuts for incoming and outgoing particles
are taken into account.

When the {\it Simpson} menu function is activated  \CompHEP~ tries        
to evaluate the cross-section. Sometimes it is not possible, for example, 
if the corresponding value is  infinite, as we have in the case of
 $e^+,e^+ \to e^+,e^+$ reaction. When program 
detects a pole inside of the integration region it produces an error message like

\begin{center}
\begin{tabular}{|c|}
\hline
    Division by zero  \\ Press any key \\
\hline 
\end{tabular}
\end{center}
    
In the  above case  you should  remove the region of small angle 
scattering by applying the corresponding cuts and  after that activate 
the {\it Simpson}  function once more. 

If the cross-section is successfully integrated the result
is displayed on the screen. {\it Menu~6} which is also  displayed on the 
screen  (see Fig.\ref{n_chain}) has the following  options:

a) to set relative precision of integral evaluation. A default value is
$10^{-4}$;
 
b) to display dependence of the differential cross-section on the cosine 
of the scattering angle;

c) to calculate and display on the screen the  total cross-section
or the forward-backward asymmetry  as a function of 
any  parameter involved in the calculation.

\subsubsection{Process of two particle  decay} \label{2decay}

\begin{figure}            
\begin{picture}(410,230)
\put(0,0){\epsfig{file=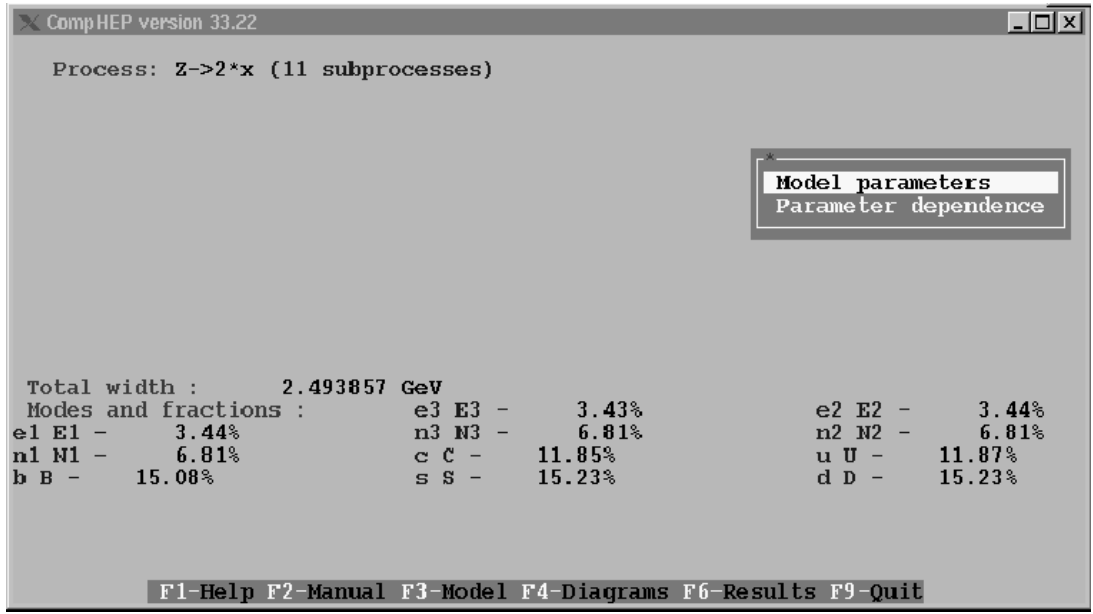,width=410pt, height=230pt}}
\end{picture}

\caption{ Representation of results for  1-$>$2 processes   }

\label{results_1-2}

\end{figure}

In the case of  1\verb|->|2  decay process the menu scheme described  above
is not used  because in this simple case the phase space integration 
is not needed. \CompHEP~ writes on the screen the calculated particle width.
If a multi-channel process like Z\verb|->|2*x  has been introduced then 
branchings for all decay modes are displayed (Fig.\ref{results_1-2}). 
The menu on the screen contains only one function which provides the user with 
a possibility  to calculate the total particle 
width as a function of one of the model parameters.

 \subsection{\CompHEP~ event generator}
             \label{events}             \subsubsection{Concept of the event generator}

  The procedure of event generation consists of two steps.
  During the evaluation of cross section  by  \VEGAS~   \CompHEP~
writes down the calculated squared matrix elements into a binary file
{\it events\_}{\sc N}, where {\sc N} is the  session number.
These files are compact owing to occupation of approximately four
bytes per one event.  All \CompHEP~  settings 
and a current state of the random number generator are stored there too.

On the next step  stand-along command
{\it genEvents}  restores the event flow based on this file.
It repeats the \VEGAS~ session,  but instead of the real matrix element evaluation
 it reads the corresponding  values from the file.
  This trick allows one to repeat with a high speed the calculation of
particle momenta and event weights produced earlier during the
original Monte Carlo integration.

Note that the \CompHEP~  generates the partonic level  events.
 \CompHEP~ passes 
the task of fragmentation  to other programs like PYTHIA \cite{PYTHIA}.

\subsubsection{The {\it genEvents} command}

To restore events from the {\it events\_}{\sc N} file the user
should use the \\ {\it \$COPMHEP/genEvents} command. 
In being launched it  is waiting for   user
commands. These instructions are  read from  the standard input 
and  the outcome is directed  to the standard output.
  Any instruction should  be 
terminated by the end-of-line symbol (the \Enter~ key).

  The first instruction  should contain an event file name,
say,  {\it event\_1}.
If such a file exists and has an appropriate format,  {\it genEvents}
writes the process name, e.g. \\
\hspace*{2cm} {\tt PROCESS: e1,E1 \verb|->| e2,E2} \\  
Then the line(s) with a physical cut definition could follow.
The format of these lines is: \\
\hspace*{2cm}{\tt [function identifier]  [min.limit] [max.limit]}\\
See Section \ref{functions} for a format of the function identifier. The limits 
are just numbers. For the  case of the {\it S} function instead of
the values of squared  momentum  {\it Val}  the user has to put in  
the quantity $ Val/\sqrt{|Val|} $, similar  to the case of cut definition
in Section \ref{cuts}.

The last line of input may be empty or  contain a desired distribution 
specification. For the first case the weighted event flow will be created.
 Otherwise  the corresponding 
distribution  will be produced. The format of input which specifies a 
distribution
is the same as that for cuts, but contains an addition field which
defines the number of bins:\\
\hspace*{0cm}{\tt [function identifier]  [min.limit] [max.limit] [number of bins]}\\
The number of bins has not to exceed 300.

The {\it  genEvents}  command verifies  the input. But the user-friendly 
interface is not implemented here. For the case of wrong input
the error message appears and the program terminates.  So, in order to escape the 
repetition of typing we recommend to   store the input in a file
and pass it to {\it  genEvents} by \\
\hspace*{2cm}{\tt \$COMPHEP/genEvents \verb|<| user\_input\_file  }\\ or\\
\hspace*{2cm}{\tt  cat user\_input\_file | \$COMPHEP/genEvents }

Let us give an example of  {\it user\_input\_file}:\\

\begin{tabular}{l}
event\_1\\
A13 5 175\\ 
A15 5 175\\
M35 1 100 50
\end{tabular}\\

 It is an instruction to  build a 50 bin  distribution for the invariant mass of 
$3^{rd}$  and $5^{th}$ particles under  condition of  a 5 degree cut for the 
angle between the momenta of these particles and the collision axis.

To get  a matching  of particles  with  their momenta
 see the process string in  the corresponding 
protocol file {\it f\_prt\_}{\sc N}.

The output for the  distribution has got the following formats.\\
1. The process string:\\
\hspace*{1.5cm} {\tt PROCESS: e1 , E1  \verb|->|  e2 , E2 }\\
2. The {\it name} of the X-axis in double quotas, limits for the axis and
a number of bins:\\
\hspace*{1.5cm} {\tt  "Cosine(p1,p3)"  from $-1.0$ to   $1.00$ $  N\_bins=  10 $}\\
3. The name of the Y-axis:\\
\hspace*{1.5cm} {\tt Diff. cross section [pb]}\\
4. The sequence of $N\_bins$  lines for the averaged differential cross section
for each bin
 and its statistical uncertainty separated by "+/-":\\
\hspace*{1.5cm}  {\tt  ..............}\\    
\hspace*{1.5cm} {\tt  8.0978E+02  +/-   9.2786E+00}\\
\hspace*{1.5cm} {\tt ...............}

The following normalization of data is assumed: 
a sum of all values of the first column
multiplied by the bin size is equal to the cross section.

The graphical image of the distribution may be displayed 
by the \\ \mbox{\$COMPHEP/tab\_view} command.
One of the ways to use it is the long pipe:
\begin{center}
{\tt cat user\_input\_file | \$COMPHEP/genEvents |\$COMPHEP/tab\_view }
\end{center}
The options of the {\it tab\_view} program are described in Section
\ref{X11interface}. See Fig.\ref{plot-image} as an  example of plot
representation. 
One more  possibility  is to save the  intermediate result and then use
it:\\
\hspace*{2cm}{\tt   cat user\_input\_file | \$COMPHEP/genEvents > hist\_file}\\
\hspace*{2cm}{\tt    \$COMPHEP/view\_tab < hist\_file.}

To generate the event flow  the user has to  enter  an empty 
line instead of the distribution definition. After that the {\it genEvents}
program writes down the weighted events in the standard output.
The first six lines of output contain  general information about the
process, namely the process name, masses of particles,  structure function
records, the QCD scale, and   values of center-of-mass energy and rapidity. For example,
\begin{verbatim}
 PROCESS: e1 , E1    ->  e2 , E2
 MASSES:   0.000000E+00  0.000000E+00  1.057000E-01  1.057000E-01
 StructFun1: OFF
 StructFun2: OFF
 QCD constant scale: 9.12E+01, Lambda6=1.18E-01 GeV
 SQRT(S):   9.000E+01
 Rapidity:   0.000E+00
\end{verbatim}   
 The sign of  rapidity is defined in such a way  
that its change towards a positive value  corresponds to an  acceleration  
of the first  incoming particle. 

The seventh  line  contains the  column titles. The following lines 
contain columns of  numerical information about events.
In the first column the event  weights are written down. They are normalized so
that the sum of all numbers of the first column gives the
calculated cross section or decay width. 
  The other columns contain the components of three-momenta  of particles. 
The corresponding momentum and component numbers are described 
by the  seventh (title) line of the output. For example, {\it P2\_3} means
the third component of the second momentum. 
Note that the $3^{rd}$ component is assigned to the collision axis.
The first incoming particle has this component positive while the second
one has it negative. There are no columns for the momenta of incoming 
particles which are equal to zero by definition, for example, for the
$1^{st}$ and $2^{nd}$ components of momenta of colliding particles.

The  event flow output is normally very large, so we recommend to
redirect  the output to the subsequent  program for some processing  
\begin{center}
{\tt    cat user\_input\_file | \$COMPHEP/genEvents | userProcessing}
\end{center} 
A few such programs were created in the framework of \CompHEP~ project.
They are   disposed in the \$COMPHEP~ directory.

I. {\tt unweight.} It transforms
weighted events into unweighted ones. It needs a parameter $maxw$ which
specifies 
the maximum value of weights. Indeed this program  transforms the floating
number weight  into the  integer number one according  to  the following expression
$$  floor(\frac{weight}{maxw}) + p\left(\frac{weight}{maxw} -
  floor(\frac{weight}{maxw})\right)\;\;, $$
where $floor(x)$  function  rounds  $x$ downwards to the nearest
integer and
 $p(x)$ is the random function which equals 1 with probability $x$ and
zero otherwise. If the integer weight does not equal  zero then this event 
is written down into the output in the same format as the unweighted event
flow. The true maximum weight may be found in the protocol file.

II. {\tt filterD2.}  This program generates decays of one of outgoing 
particles into two others. It needs 6 arguments:\\
1) name of disintegrated particle;\\
2) name of the first decay particle;\\
3) numerical value of the mass of the first particle;\\
4) name of the second decay particle;\\
5) numerical value of the  mass of the second particle;\\
6) branching fraction for this decay mode.
For example, \\
\begin{center}
{\tt filterD2 Z e2 0.1 E2 0.1  0.034}
\end{center}
will generate  decays of the Z boson into  muons.
This program reads  event flow from the standard input and 
writes down generated events into the standard output. It can be used just 
after {\tt genEvents} and after {\tt unweight}.

III.  {\tt filterDN} This function can generate more complicated 
sub-decays, for example, \verb|1->3|  ones. It is assumed that before
its usage  the  events of  corresponding decay have been 
generated by \CompHEP,  transformed into the unweighted format,   and   stored in 
some file. The name of this file must be passed to {\tt filterDN} as the
first  argument. Numerical value of the corresponding   branching fraction
must be passed as the second argument.

 You can obtain a  wrong result  using some part of the whole 
event sample because  \VEGAS~ generates strongly correlated events 
according to the {\it Stratified Sampling} algorithm (see Section \ref{vegas}).
If events are written down in some file they may be mixed up by means of  
the {\tt \$COMPHEP/randomize} program. The first argument of this function
is a name of the file where events are stored. The randomized event flow
 is directed to the standard output.

The programs mentioned above, namely  {\tt unweight, filterD2, filterDN, 
randomize},  can have three optional  arguments. They
must be some integers which  are  used to re-initialize the starting point of
the random number
generator \footnote{ We use standard random number generator {\it drand48}.
}.
 
The generated event flow can be transformed to the table of distribution of
some physical variable  by means of the {\tt mk\_tab} program. This program
needs four arguments, namely, a name of variable in format of 
Section \ref{functions}, minimum and maximum limits of this variable 
and a number of bins, which cannot exceed 300.  The  {\tt mk\_tab} program 
reads the event flow in the standard input and writes the  generated table  down
to the standard output. This table can be transformed into a plot by means
of the {\tt tab\_view} utility.

  \subsection{\CompHEP~ files  and commands}
             \label{file-commands}      
\subsubsection{Files and commands in the working directory}

   We shall use an alias name  {\it WORK} for  the directory which has been created by the user
during  the \CompHEP~ installation (Section \ref{u-install}). It contains the following 
sub-directories and files:\\
\hspace*{2cm}{\tt models/ results/    tmp/ }\\
\hspace*{2cm}{\tt comphep   comphep.ini }

  The {\it models} directory contains files which describe physical models. 
 When you 
modify a model or create a new one  all  changes
are kept in this directory. 

  The directory {\it results} keeps a \CompHEP~ output which is produced as
a result of symbolic or numerical calculations. 

  The directory {\it tmp} is used for temporary files. The file {\it tmp/safe} 
keeps  user settings between the sessions. If this file exists the symbolic 
session  starts its work with reading this file to restore the user settings
of the previous session.
A bug presenting in this file could be a reason for a fatal error in the beginning
of the \CompHEP~ session. In the case of such an error we  recommend to remove  
the {\it tmp/safe} file. 
 
  All these directories are necessary for  \CompHEP~ work. Absence
of one of them will lead to a fatal error.

  You can toggle on/off the color and the sound for a \CompHEP~ session as well
as choose the most appropriate font for your X-terminal. To do this  just edit
the {\it comphep.ini}  file. We hope the syntax there is
self-explanatory enough.

 The command \\
\hspace*{2cm}{\tt       ./comphep  }\\
starts a \CompHEP~ session.  This is a shell script which calls
commands disposed in the  \CompHEP~ root directory. It is assumed that the
{\it \$COMPHEP} environment variable (see Section \ref{u-install}) points to this directory.

\subsubsection{Scheme of calls in the  \CompHEP~ session}

  The command {\it ./comphep} being issued from within the {\it WORK} directory launches 
the 
\begin{center} \tt \$COMPHEP/s\_comphep \end{center}
command. The latter is the main \CompHEP~ program. Prefix {\it 's\_'}
denotes symbolic.  It performs symbolic calculations and generates the \C~ and
\Fortran~ codes of calculated Feynman diagrams. 

Below we would like to describe  commands  for compilation  and numerical
module launching. The command names  in the \C~ and \Fortran~ cases are distinguished  by 
suffixes  {\it 'c'} and {\it 'f'} correspondingly. We shall use the '*' symbol 
as a common notation for these {\it 'c'} and {\it 'f'}. The general scheme of calls
can be presented by the following diagram

\noindent
{\tt
s\_comphep $\rightarrow$ launch\_n\_comphep~* $\rightarrow$ 
\begin{tabular}{|l}
make\_\_n\_comphep~* $\rightarrow$ \begin{tabular}{|l}
                                     nCompil\_*\\ 
                                     ld\_*\\
                                   \end{tabular}           \\
n\_comphep\_*\\
\end{tabular}
}

The binary executable {\it n\_comphep\_*} is created in the {\it WORK/results} 
sub-directory as a result of  compilation. Other commands are
disposed in the {\it \$COMPHEP} directory. They  are shell scripts
except of  binary {\it s\_comphep}. It is assumed that {\it s\_comphep}
and  {\it launch\_n\_comphep}  are started  from  the {\it WORK} directory,
whereas  other  commands   are started from within the {\it WORK/results} 
sub-directory.

The    program {\it nCompil\_* }  compiles  the source
code produced prior by {\it s\_comphep}.  After  successful compilation it
creates a library of object files   and then removes  the source and 
the object files. 

Another command {\it ld\_*} 
calls a  linker to create the  executable file
{\it n\_comphep\_*}  for  numerical job.

 
The  command  {\it make\_\_n\_comphep } with parameter 
'c' or 'f' is  used to start {\it nCompil\_*} and {\it ld\_*}
subsequently. 

In its turn,   {\it make\_\_n\_comphep} is started  by 
{\it launch\_n\_comphep } 
in a  special window. In the  case of successful compilation  the window is closed 
and  just created {\it n\_comphep\_*} is started in 
another window.

  If you copy  {\it \$COMPHEP/make\_\_n\_comphep}  into  your  {\it WORK/} 
directory, then   {\it launch\_n\_comphep} will call for your own version. 
It gives you a possibility to modify a compilation procedure if necessary.

The  {\it s\_comphep} command  starts  {\it launch\_n\_comphep}
via the functions 
{\tt
  \begin{center}
    \begin{tabular}{|c|}
      \hline
       ............\\ C-compiler\\ Fortran compiler \\
      \hline
    \end{tabular}
  \end{center}
}
\noindent
of {\it Menu~6} (Fig.\ref{s_chain})
and thus the chain of above  calls  is realized.  

If you have prepared  the \C~ or \Fortran~ codes you could compile 
them outside of the symbolic session by  the {\it make\_\_n\_comphep }  
command  with the {\it c/f} argument  started from  within the {\it results} directory.

Let us say some  words about other programs stored in the {\it \$COMPHEP} 
directory.
 The program {\it genEvents} is used to prepare an event flow and
 histograms using the information saved by Monte Carlo session. 
The programs {\it unweight}, {\it randomize}, {\it filter2D}, 
{\it filterND}, and {\it mk\_tab} transform the event flow
generated by {\it genEvents}. See Section \ref{events} for details. 
The program {\it tab\_view}   provides the user with a possibility of
visual presentation of distributions. See Section \ref{interface}. 

\subsubsection{LOCK files}

  {\it  WORK/comphep} creates a file {\it LOCK} in the {\it WORK/} 
directory.  
Presence of the {\it LOCK} file prevents a double launching of {\it WORK/comphep}
from within the same place of the file system.  {\it LOCK} is automatically
removed in the end of the {\it WORK/comphep} session.
  In the same manner {\it \$COMPHEP/launch\_n\_comphep} creates a file 
{\it c\_LOCK} or {\it f\_LOCK} in the {\it WORK/results} sub-directory. 

If your session has been
canceled abnormally you have to remove these {\it LOCK,  f\_LOCK, c\_LOCK} 
files manually before launching the next session.

 \subsection{User programs in \CompHEP}  \label{user-prg}
                                        \subsubsection{Concept of user program implementation}

We provide the user with a possibility  
to attach his own codes to the {\it n\_comphep\_f(c)} and {\it genEvents}
programs.  In this way you are able to\\
1) expand the  set of phase space functions  for  cuts and
histograms\\ 
2) implement  new structure functions.

\CompHEP~ archive file {\it num\_f(c).a} contains some patch programs. 
When you create your version of these programs and pass the names 
of its object files to the linker, they will be embedded 
into the executable file. 

In the case of {\it n\_comphep\_f(c)} command the following
instruction does this job:\\  
\hspace*{2cm}{\tt \$COMPHEP/make\_\_n\_comphep  f(c) userObjectFiles.o}

It is assumed that the above command  is started from within the
user's directory 
{\it results}  which contains  sources or their archives.

The {\it genEvents} file is disposed in the {\it \$COMPHEP} directory 
and generally cannot be modified by the user. But one could create 
his own version of this program. In this case use \\     

{\small \tt f77 -o userEvents  \$COMPHEP/events.o userObjFile.o \$COMPHEP/num\_f.a}

\subsubsection[\Fortran~ case]{\Fortran~ case\footnote{ 
Below we insert  parameter  type  definitions into  headers of the \Fortran~ 
routines. It is done for brevity only  and contradicts to the \Fortran~  style}} 

  The user  phase  space function  is evaluated by\\
\hspace*{2cm}{\tt   REAL*8 FUNCTION USRFUN(CHARACTER *9 TEXT) }

The patch is disposed in the file {\it \$COMPHEP/f\_source/usr/u\_var.f}. 

     This function is called by {\it n\_comphep\_f} and 
{\it \$COMPHEP/genEvents} programs  if the phase space
function whose name begins with the 'U' character is used (see Section
\ref{functions}). The characters following   'U' are passed to
the {\it USRFUN} routine as its argument.

To  create this function one  needs to  know the particle momenta.
They are  stored in
\begin{center} {\tt  COMMON/PVECT/ REAL*8 P(0:3,100) }\end{center}
 The first argument of array {\it P} is a Lorentz momentum component.
$P(0,N)$ is the energy of the $N^{th}$ particle. 
We have $P(0,N)>0$ as for 
incoming as for outgoing particles. 
In the case of  collision process  $ P(3,N)$ is a projection of the
particle space momentum onto the collision axis. The direction of this axis is
chosen so that   $P(3,1) > 0$ and $P(3,2) < 0$.

The second argument of this array is a  momentum  ordering  number. 
The first  momentum numbers  are assigned to incoming particles and the 
subsequent ones  are   assigned to   outgoing particles. 
The  numbers of incoming and outgoing particles and 
a  correspondence between  the   particle name and the  momentum number 
may be determined by means of service functions described in 
Section \ref{out-fort}.

The work of user structure function is controlled by a set of routines.
The patches for them 
are disposed in {\it \$COMPHEP/f\_source/usr/sf\_prv.f}. 
All of them should  be realized by the user:

1.    {\it LOGICAL FUNCTION  P\_PRV(CHARACTER*6 P\_NAME) }\\     
      returns  {\it TRUE} if the user structure function can be 
applied to the {\it P\_NAME} particle.
 It is used to create the  menu of possible structure functions 
      for the  particle {\it P\_NAME}. 

2.    {\it SUBROUTINE M\_PRV(I)}\\
       is called for the user input of the structure function parameters
      just after the user structure function  has been chosen. 
       It is assumed that this routine 
saves values of the entered parameters in some COMMON  for
subsequent usage  in  N\_PRV described below.   The argument {\it 'I'} here 
and in the following  means an  incoming  parton  number. It can be 1 or 2.

3.    {\it CHARACTER*60 FUNCTION N\_PRV(I)}\\
      returns the structure function name. If the structure
      function has a few tuning parameters then  the values of these 
      parameters must be included in the returned  name. \CompHEP~ uses the structure 
      function name to keep  information about parameters.

4.   {\it  LOGICAL FUNCTION R\_PRV(I,CHARACTER*60 NAME, REAL*8 CMASS, REAL*8 BE)}\\
      checks {\it NAME}. If  {\it NAME}  has the same format as the one  
produced by {\it N\_PRV()},  {\it R\_PRV()} should  read the parameter values,
 store them in some global variables for subsequent usage,  and 
      return TRUE. Otherwise FALSE is returned.
     {\it  REAL*8 CMASS} and {\it  REAL*8 BE} are outgoing parameters.
     The first one must be equal to the mass on the composite particle 
     which constituents are described by user's structure function. 
     The second one  informs 
     \CompHEP~ that the structure function is singular as

             $  (1-X)^{(BE-1)} $,  where $0<BE\leq 1$.\\
In the nonsingular case  $BE=1$ must be returned.
              
5.  {\it REAL*8 FUNCTION C\_PRV(I,REAL*8 X)}\\
       returns the value of structure function divided by $BE*(1-X)^{(BE-1)}$. 
    Here  $X$ is the Feynman scaling variable. 
It is assumed that possible structure function parameters having been stored 
by {\it R\_PRV} are used for evaluation.  The structure function is
normalized so that 
$$ be(1-x)^{be-1}c\_prv(i,x)dx$$ 
is a probability to find the $i^{th}$  parton with fractions 
$[x, x+dx]$ of initial momentum.

In the simplest case, if the user would like to implement a new 
structure function without singularity and extra parameters, he can just take a
copy of the  {\it \$COMPHEP/f\_source/usr/sf\_prv.f} file, improve  {\it
P\_PRV} so that it returns $'.TRUE.'$ anyway, and  rewrite  originally trivial
{\it C\_PRV}.

\subsubsection{\C~ case}

  The user   phase  space function has a prototype \\
\hspace*{2cm}{\tt   double usrfun(char * name) }\\
The patch is disposed in the file {\it \$COMPHEP/c\_source/num/userFun.c}. 
     This function is called by {\it n\_comphep\_c}   programs  if the phase 
space function whose name is started from the 'U' character is used (see
Section \ref{functions}). The characters following  'U' are passed to
the {\it usrfun} routine as its argument.

 Particle momenta which are needed to evaluate  {\it usrfun} 
are stored in the global variable \\
\hspace*{2cm}  {\tt double pvect[400] } \\
The  $m^{th}$ component of $k^{th}$ momentum occupies the $m+4*(k-1)$ position
in {\it pvect}. We assume that 'm' varies from 0 to 3 and the 'k' momentum  
counter  starts from 1. The agreement about momentum signs is the same 
as in the \Fortran~ case. The  number of incoming and outgoing particles and
the correspondence between  the   particle name and  momentum number
can be determined by means of service functions described in
Section \ref{out-c}.

The work of user {\it structure function} is driven by a  set of routines
similar to those of the \Fortran~ case.
All of them have to be realized by the user if he would like to implement 
a particular structure function. Patches for these routines are disposed in \\
\hspace*{2cm}{\it \$COMPHEP/c\_source/num/strfun/sf\_prv.c}.\\
 This file contains:

1) {\it int p\_prv(char *particleName)}\\
      returns  {\it 1} if the user structure function can be 
implemented to the \mbox{\it  particleName} particle. Otherwise 0 should 
be returned.
 It is used to create a menu of possible structure functions;

2)    {\it void m\_prv(int i) }\\
      is called for the user input of the structure function parameters
      just after the particular structure function  is chosen. It is assumed
that the input is saved in some static variables.
The argument {\it 'i'} here and below
means an  incoming  parton  number. It could be 1 or 2;

3)    {\it void n\_prv(int i, char * funcName )}\\
      returns the structure function name. If the structure
      function has a few tuning parameters then  values of these 
      parameters have to  be included in {\it funcName}. \CompHEP~ uses the structure 
      function name to keep  information about parameters;

4)   {\it int r\_prv(int i, char *funcName)}\\
      checks {\it funcName}. If  {\it funcName}  has an appropriate  format,
  {\it r\_prv} must read  the value of
parameters, store them in some global variables  and 
      returns 1. Otherwise 0 is returned;

5)  {\it int mass\_prv(int i)}\\
    returns the mass of the composite particle which constituents 
    are described by this structure function.

6)    {\it double be\_prv(int i) }\\  
     performs needed  initializations  before structure function evaluation.
     It returns parameter $be$ which  informs 
\CompHEP~ that the structure function is singular as

             $$  (1-x)^{(be -1)}; $$

7)  {\it double c\_prv(int i, double x)}\\
    returns the value of structure function divided by $be(1-x)^{(be-1)}$, 
      where $x$ is the Feynman scaling variable. 
It is assumed that possible turning  parameters have been stored 
by {\it  r\_prv} in some static variables. 

In the simplest case, if the user would like to implement a new 
structure function without singularity and extra parameters, he may  take a
copy of the {\it \$COMPHEP/c\_source/num/strfun/sf\_prv.c} file, improve  {\it p\_prv}
 so that it returns $1$ in any case, and  rewrite the  originally  trivial
{\it c\_prv} function.

\newpage
\section{Implementation of models of particle interactions}
 \subsection{Definition of a model in \CompHEP}
             \label{models}
Description of  particle interaction model  in \CompHEP~ 
consists of four parts. They are  {\it parameters}, {\it constrains}, {\it
particles}, and {\it vertices}.  

  \subsubsection{Independent parameters of the model}
             \label{parameters}           The Table {\it Parameters} consists of three fields:

\begin{enumerate}
\item {\it Name} for an identifier of the parameter. It may
contain up to 6 characters. The first character
must be a letter, others may be either letters or
digits; 
  
  \CompHEP~ identifiers are sensitive to the case of characters. Also names
of different identifiers must be different after rewriting them
in the low case. For example, if the identifier "Me" is
used in the table for the electron mass, then
the forms "ME", "mE","me" are forbidden as for new 
definitions as for identification of the electron mass.

  There are some reserved names which cannot be used
here:
\begin{itemize}
\item {\tt i}    is reserved for imaginary unity;
\item {\tt Sqrt2} is reserved for $\sqrt{2}$;
\item {\tt p1,p2,p3,\ldots} are reserved for particle momenta;
\item {\tt m1,\ldots,M1,\ldots} are reserved for Lorentz indices of 
                particles;
\item {\bf G5} is used for the $\gamma^5$ Dirac matrix;
\end{itemize}
    
\item {\it Value}   for a numerical value of  parameter in some power of 
{\it GeV} units;

\item {\it Comment} for brief description of a parameter.
\end{enumerate}             

  \subsubsection{Constraints between the parameters} 
             \label{constraints}         The Table {\it Constraints} consists of three fields:
\begin{enumerate}
\item {\it Name} for the constrained parameter. The requirements 
for this field are the same as for the name of independent parameres 
(see above);

\item {\it Expression} which  must be an algebraic formula
composed of: 
\begin{itemize}
\item integer numbers, 
\item identifiers enumerated in the {\it Parameters} Table, 
\item identifiers defined above in this Table, 
\item parentheses (), arithmetic operators  +, -, /, *, **, 
  and the $sqrt()$ function.
\end{itemize}
For raising to a power the second operand must be
an integer;

\item {\it Comment} for brief  description of parameters.
\end{enumerate}

  \subsubsection{Description of particles} 
             \label{particles}            Each  row in the {\it Particles} Table describes a  particle -- 
anti-particle pair. The rows consist of 10 fields:

\begin{itemize} 
\item[1.] {\it Full name} for a full name of particle. Just for clear orientation, not
processed  anywhere;

\item[2\&3.] {\it "A"} and  {\it "A+"}  containing designations of the particle and
anti-particle, respectively. Any character is allowed. The name 
may contain one symbol, two symbols or  three symbols 
started from '$\tilde{\;}$'.  For a completely neutral particle the
{\it "A"} and {\it "A+"} fields must be identical;

\item[4.] {\it 2*Spin} for a doubled particle spin: {\it 0} for scalar, {\it 1}
 for spinor  and {\it 2} for vector particles. Neutral spinor particle 
is teated as a Majorana one.

\item[5.] {\it Mass} for  a mass identifier or symbol '{\it 0}'. In the first case
its value must be defined in the {\it Parameters} or {\it Constraints} 
table.  If this field contains zero,
then \CompHEP~ considers this  particle as massless;

\item[6.] {\it Width} for a particle decay width. It must contain  an identifier defined 
in the first two tables  or '{\it 0}'; 

\item[7.] {\it Color} for a dimension of the color SU(3) group representation.  You have
to choose among {\it 1}, {\it 3}, {\it 8}. {\it Unity} corresponds to a colorless 
particle. {\it Three} corresponds to a color triplet (fundamental representation). 
In this case the anti-particle {\it "A+"} is transformed by conjugated $\bar{3}$
representation. {\it Eight} corresponds to a color octet (adjoint
representation);

\item[8.] {\it  Aux} for  an auxiliary  field which allows to modify  particle
propagators. If the {\it Aux} field is
empty the standard expressions for propagators are substituted:
\begin{enumerate}
   \item[(a)] spin 0 case:

      $$ <0|T[A(p_1) , A^+(p_2)]|0> = ScPr(p_1,p_2,M) = \frac{\delta(p_1 + p_2)}
{(2\pi)^4i(M^2 - p_1^2)}\;\;;$$

   \item[(b)]  spin 1/2 case:
 
      $$<0|T[A(p_1)  , A^+(p_2) \; \gamma_0  ]|0> =  (\not p_1 + M) \,
ScPr(p_1,p_2,M)\;\;,$$
      where 
       $$\not p = p^{\mu}\gamma_{\mu}\;\;.$$
      Here "$A^+$" is  the Hermitian conjugation of "$A$"; thus the
      Dirac $\gamma_0$ matrix is needed to make the Dirac conjugate
      field;

   \item[(c)]  spin 1 case:

      \begin{eqnarray}
      <0|T[A^{m_1}(p_1),  \; (A^{m_2})^+(p_2)]0> &=& -(g^{m_1 m_2} + p_1^{m_1} \; p_2^{m_2} / M^2)
\nonumber \\ &\times& ScPr(p_1,p_2,M)\;\;.
      \label{mvPropagator}
      \end{eqnarray}

      Zero mass vector particle must be marked as a {\it gauge} one using the
      {\it Aux} field (see below).
       
\end{enumerate}

Possible objects for the {\it Aux} field are:

\begin{itemize}
\item [{\it 'l','L'}] is permitted for massless fermions ( 2*spin = 1 ) only.
The propagator is changed  to

         $$\frac{  \not p_1  (1 + \gamma_5)}{2} \; ScPr(p_1,p_2,M)\;\;. $$

  This is a way to introduce a left-handed fermion as a neutrino or
a massless polarized fermion.  When \CompHEP~ performs the averaging
over incoming particle polarizations it takes into account that
there is only one polarization state;

\item [{\it 'r','R'}]  is permitted for massless fermion ( 2*spin=1 ) 
particle only.  They change the propagator for
                             
          $$ \frac{\not p_1  (1 - \gamma_5)}{2} \; ScPr(p_1,p_2,M)\;\;.$$

This is a way to introduce right-handed fermions;

\item [{\it '*'}] is permitted for massive particle only. In this case 
$ScPr(p_1,p_2,M)$ is replaced to  
$$ \frac{\delta(p_1+p_2)}{ (2 \pi)^4 i \;M^2}\;.$$
In the case of vector particle we also remove $p_1^{m1}p_2^{m2}$ 
term in the propagator  numerator (\ref{mvPropagator}). 
 
Such particles cannot appear as
incoming or outgoing  ones. They  are used to describe a point-like
interaction as one has in the electroweak 4-fermion interaction model;

\item [{\it 'G','g'}]  is permitted for vector ( 2*spin=2 ) particle. 
In this case the propagator of vector particle accepts the   
Feynman form

              $$ -g^{m_1m_2} \, ScPr(p_1,p_2,M)\;\;;$$

\end{itemize}       

\item[9\&10.] {\it LaTeX(A)} and {\it LaTeX(A+)} for 
particle and anti-particle designations in the \LaTeX~ format. They are 
substituted in the \LaTeX~ image of Feynman diagrams generated by \CompHEP.
The names are used in the mathematical mode.
\end{itemize}

  \subsubsection{Ghost fields in \CompHEP} 
                                        
  \CompHEP~ constructs a list of quantum fields according to 
the Table of particles described above. Besides of names enumerated
in this table  \CompHEP~   generates  auxiliary fields, 
for example, the Faddeev-Popov ghosts \cite{BD}. We use the name
 {\it ghost} for all of them.

  The ghost fields do not  correspond  to physical degrees of freedom,
but each of them has some real particle as a prototype. The names of ghost 
fields are  constructed by \CompHEP~  from  the prototype particle name  followed
by a suffix which specify a type of ghost. The particle name and 
the suffix are separated by the dot symbol. It is assumed that the ghost 
fields explicitly  appear in the vertices of interactions together with
real particle fields and  thus contribute to the  particle interaction.
Below we list \CompHEP~ ghost fields.

\paragraph{ Faddeev-Popov ghost and anti-ghost.}
They  are  generated for any gauge vector particle, in other words,
for those  particles which have the  mark {\it 'g'} in the {\it 'Aux'} column 
of the {\it Particles} Table described in the   previous section.
  The names of Faddeev-Popov ghosts and anti-ghosts  are constructed by means of suffixes {\it 'c'} 
and {\it 'C'} respectively.  For example,  {\it G.c, G.C} are  gluon
ghosts and   {\it W+.c, W+.C, W-.c, W-.C}  are W-boson ghosts.

The operation of  Hermitian  conjugation  transforms a  Faddeev-Popov ghost
to itself whereas an anti-ghost  is transformed to itself with the  opposite  
sign. Thus the rules of Hermitian
conjugation of gluon and W-boson ghosts are
\begin{eqnarray*}
(G.c)^+&=&G.c\\  
(G.C)^+&=&-G.C\\
 (\mbox{W+.c})^+&=& \mbox{W-.c} \\  
(\mbox{W+.C})^+&=&-\mbox{W-.C}
\end{eqnarray*}
Faddeev-Popov (anti)ghosts are scalar, anti-commutative  fields
\footnote{ The well-known spin-statistic relation is not valid for unphysical
fields.}.   The nonzero propagators for these fields are:
$$<0|T[\mbox{A+.c}(p_1), \; A.C(p_2)]|0> = <0|T[\mbox{A+.C}(p_1), \;
A.c(p_2)]|0>  = ScPr(p_1,p_2,M),$$
where  $M$ is a mass of the prototype particle. This equality of masses is
 a consequence of  the choice  of the t'Hooft-Feynman gauge.

The appearance of the  Faddeev-Popov ghosts in gauge theories can be
explained  in the following way. In the case of the t'Hooft-Feynman gauge
the quantum field of massless gauge particle has two unphysical 
components, because the  four-dimensional vector field 
describes a particle with two polarization states. The contributions of
Faddeev-Popov ghost and anti-ghost  compensate  the contribution of these 
unphysical polarizations.  In the case of massive  gauge boson  there are 
three physical polarizations and the compensation by means  of 
 Faddeev-Popov ghosts looks wrong from the viewpoint of naive arguments 
based on  counting  
degrees of freedom.  Indeed, in this case one additional 
{\it  Goldstone}  ghost appears \cite{BD}.

\paragraph{ Goldstone ghost.}

It is  generated for any  massive gauge vector particles.
  The name of this field is constructed by means of suffix {\it 'f'}.  
For example,  {\it W+.f, W-.f}  are the W-boson Goldstone ghosts.

This  ghost is scalar,  commutative,  and satisfies  the same  
conjugation rule as the prototype particle. For example, $(W+.f)^+= W-.f$.
The nonzero propagators for these fields are:

$$T[\mbox{A+.f}(p_1), \; A.f(p_2)]   = ScPr(p_1,p_2,M),$$
where  $M$ is a mass of the prototype particle. 

\paragraph{ Tensor ghost.}

  Whereas  the  Faddeev-Popov  and Goldstone ghosts are standard
elements of modern quantum field theory, the tensor ghost is an
original \CompHEP~  invention. This is an auxiliary field with a point-like
propagator which is used to construct  vertices with   
complicated color structure, for example, the  four-gluon vertex.

The tensor ghost is generated automatically for any vector particle with a
non-trivial SU(3) color group representation. Its name is constructed by
means of suffix {\it 't'}.
This  ghost is  commutative,  and satisfies  the same
conjugation rule as the prototype particles. It is  Lorentz-transformed 
like  a tensor field. The propagator is 
\begin{equation}
<0|T[\mbox{A+.t}^{m_1M_1}(p_1), \; A.t^{m_2M_2}(p_2)]|0> = \frac{1}{(2 \pi)^4i} \; 
\delta(p_1+p_2) \, g^{m_1m_2}\, g^{M_1M_2}\;.
\label{tensor_propagator}
\end{equation}
             \label{ghostFields}

  \subsubsection{Interaction  vertices}     
             \label{lagrangian}         
   The Table {\it Vertices} contains interaction vertices. 
The first four fields
$A1$, $A2$, $A3$, $A4$ include the names of the interacting particles.  
These fields
must contain particle names in \CompHEP~ notation.  $A4$ may be
empty.  The last two fields {\it 'Factor'} and {\it 'LorentzPart'} define a
vertex itself.  Let $S$ be the action, then a functional derivative of 
$S$ over fields is represented as  
\begin{eqnarray}
\label{eq1}
\frac{\delta S}{\delta A1_{[m1]}(p1)\, \delta A2_{[m2]}(p2)\, 
\delta A3_{[m3]}(p3)\,
 [\delta A4_{[m4]}(p4)]}
&  =  \nonumber   \\
(2 \pi)^4\delta(p1+p2+p3\, [+p4]) \, [\gamma_0] \, ColorStructure \cdot 
 Factor \cdot LorentzPart\;. &
\label{funcDeriv}
\end{eqnarray}   

  Here $p$ and $m$ denote 4-momenta and  Lorentz indices.  The brackets
\mbox{[\ ]} are used to mark the optional parts of  expression. Thus, 
 $A4$, $p4$, and  $m4$ appear only in the case of four particle 
vertex. In the case of anti-commuting fields the  right-side 
derivatives are assumed. The Fourier transformation  is defined by
\begin{equation}
       A(x) =  \int \! \exp(-i k x)  A(k) \,d^4k\;.
\label{Fourier}
\end{equation}
{\it 'Factor'} must be a rational  monomial constructed of the model
identifiers, integer numbers and imaginary unity.

   {\it 'LorentzPart'} must be  a tensor or  Dirac $\gamma$-matrix
expression. Coefficients of this expression are polynomials of the model
identifiers and scalar products of momenta.  The division '/' operator 
is forbidden in  {\it 'LorentzPart'}. It must be transferred to the {\it 'Factor'} field.

Similar to the \REDUCE~ notation, in order to  construct scalar products 
of momenta, momentum components, and metric tensors we use the  dot  symbol, for example, 
\begin{eqnarray*}
 p1.p2 & is & g_{\mu \nu}  p_1^\mu p_2^\nu\;; \\
 p1.m2 & is & p_1^{m_2}\;;\\
 m1.m2 & is & g_{m_1 m_2}\;.  
\end{eqnarray*}

  To  implement  the Dirac $\gamma$-matrix  with index $'m'$ we 
use a symbol $G(m)$, whereas  $G(p)$ denotes  $p_{\mu}\gamma^{\mu}$. Anti-commutation 
relations for $\gamma$  matrices should be written as
\begin{equation}
    G(v1)\, G(v2) + G(v2)\, G(v1) = 2 \;  v1.v2,
\label{anticommutation}
\end{equation}
where  $v1, v2$ are  momenta or  indices.

  The $\gamma_5$ matrix  is denoted by $G5$. It is defined by equation
   $$\gamma_5 = i\; \gamma_0\gamma_1\gamma_2\gamma_3$$

The number of fermion fields in one vertex must be two or zero. If you would like to 
implement a four-fermion interaction, use an auxiliary unphysical  field 
which may be constructed by means of the '*' symbol in the 'Aux' column 
of the particle table (see Section \ref{particles}).

 \CompHEP~ interprets the  anti-particle spinor  field as the Hermitian
conjugated particle field, rather than the Dirac conjugated one. Also it is
assumed that all spinor fields  are written in the
Majorana basis and  the matrix of C-conjugation is equal to
$(-\gamma_0)$ \footnote{ Above the Majorana spinor was defined as a
Hermitian self-conjugated one: $\psi=\psi^+$. In the same time 
it must be $C$-self-conjugated too:  $\psi=C \bar{\psi}^T$. Following 
these requests we get the phase of $C$ which is different from one 
chosen in the textbook \cite{BD} }. After substitution $C \to -\gamma_0$ all possible vertices could be
written as
$$  A1(p1) \, \gamma_0 \, G(v1) \, G(v2)  \ldots  G(vn) \, A2(p2) $$
where $A1$ and  $A2$ are  some spinor fields, each of them  corresponding to  particle 
or anti-particle.  This form of Lagrangian is assumed in the vertex
expression (\ref{eq1}).  $\gamma_0$ 
is substituted by \CompHEP~ automatically  in the case of a spinor 
particle vertex and is not expected in  {\it LorentzPart} of (\ref{funcDeriv}).

   Note that structures like $m1.m2$ and
$p1.m2$ are forbidden  for the vertex with fermions. In order to
implement these structures
use the equation (\ref{anticommutation}). 

   Let us note that by definition (\ref{funcDeriv}) the  {\it LorentzPart} 
has the  corresponding symmetry property in the case when  
identical particles appear in one vertex. This symmetry is not checked by
\CompHEP, but its absence will lead to  wrong results. The following 
equation may be used to check the symmetry in the case of fermion vertex:
\begin{eqnarray}
  A1(p1) \, \gamma_0 \, G(v1) \, G(v2)  \ldots [G5] \ldots  G(vn) \, A2(p2) = \nonumber \\
 A2(p2)\, \gamma_0  \, (-G(vn)) \ldots [G5] \ldots  (-G(v2)) \, (-G(v1)) \,
A1(p1) \;. & 
\label{reverse}
\end{eqnarray}
 It may be useful also to check the  Lagrangian self-conjugation property.
Note that the anticommutation  of {\it A1} and {\it A2}  is already taken into
account in (\ref{reverse}).

  {\it ColorStructure} is substituted by \CompHEP~ automatically. For a
colorless particle vertex it is equal to 1.  For $(3 \times \bar{3})$ and for
$(8 \times 8)$ vertices the unity tensor is substituted. If \CompHEP~ meets
a vertex
with three particles in the adjoint representation $(8 \times 8 \times 8)$, 
it substitutes
  $$ -\,i\, f(a1,a2,a3),$$ 
where $f^{\alpha_1}_{\alpha_2 \alpha_3}$  are the  structure constants of SU(3).
Color indices  $a1, a2, a3$  are taken in the same order as 
they appear in the  particle columns. 
For  the  $(3 \times \bar{3} \times 8)$  vertex  \CompHEP~ substitutes
        $$\frac{1}{2}\lambda(\bar{i},i,a),$$
where $\lambda(\bar{i},i,a)$ are the Gell-Mann  matrices.  
   More complicated color structures are not implemented yet, but it is
possible to  construct
them by means of  unphysical particles ({\it Aux}='*') or tensor ghosts (Section 
\ref{ghostFields}). In the case of tensor auxiliary 
field use the capital  {\it 'M'} for designation of the second Lorentz 
index of this field as it is shown in equation (\ref{tensor_propagator}).

 \subsection{Examples}  
 \label{examples}             
\subsubsection{Implementation of QCD Lagrangian}
\paragraph{ 3-gluon vertex.}
  Lagrangian (\ref{physLagQCD}) contains  the following  3-gluon vertex:
$$ S_{3G} =  -\frac{g}{2}\int(\partial_{\mu} G_{\nu}^{\alpha} - \partial_{\nu}
G_{\mu}^{\alpha})f^{\alpha}_{\beta \gamma} G^{\beta}_{\mu}
G^{\gamma}_{\nu}d^4x = 
-g\int \partial^{\mu_2} G_{\mu_1}^{\alpha_1} g^{\mu_1 \mu_3} 
 f_{\alpha_1 \alpha_2 \alpha_3} 
G^{\alpha_2}_{\mu_2} G^{\alpha_3}_{\mu_3} d^4x\;,    $$
where $G^{\alpha}_{\mu}$ is the gluon field, $g$ is the strong coupling
constant.  Applying the Fourier transformation (\ref{Fourier}) 
we get
 
$$ S_{3G} = (2\pi)^{4} g \int \delta(p_1+p_2+p_3)\;
i \; f_{\alpha_1 \alpha_2 \alpha_3} p^{\mu_2}_1  g^{\mu_1 \mu_3}
G_{\mu_1}^{\alpha_1}(p_1) G_{\mu_2}^{\alpha_2}(p_2)
G_{\mu_3}^{\alpha_3}(p_3)     d^4p_1 d^4p_2 d^4p_3\;. $$

This vertex contains three identical fields, so the  calculation of 
functional derivatives gives us six terms:
\begin{eqnarray*}
\lefteqn{\frac{\delta S_{3g}}{\delta G_{\mu_1}^{\alpha_1}(p_1)
\delta G_{\mu_2}^{\alpha_2}(p_2) \delta G_{\mu_3}^{\alpha_3}(p_3)}= 
 (2\pi)^4 \delta(p_1+p_2+p_3)\; i\; f_{\alpha_1 \alpha_2 \alpha_3}}\\
&&g 
(p_1^{\mu_2}g^{\mu_1 \mu_3} - p_3^{\mu_2}g^{\mu_1 \mu_3} 
+p_2^{\mu_3}g^{\mu_1 \mu_2} - p_1^{\mu_3}g^{\mu_1 \mu_2} 
+ p_3^{\mu_1}g^{\mu_2 \mu_3} -  p_2^{\mu_1}g^{\mu_2 \mu_3})\;.   
\end{eqnarray*}
Comparing this expression with the \CompHEP~  vertex  representation 
(\ref{funcDeriv}) where  {\it ColorFactor} is  $(-\,i\, f_{\alpha_1 \alpha_2
\alpha_3})$  we get  
$$Factor\cdot LorentzPart= -\,g \,
 \Bigl( (p_1^{\mu_2} - p_3^{\mu_2})g^{\mu_1 \mu_3}
+  (p_2^{\mu_3} - p_1^{\mu_3})g^{\mu_1 \mu_2}
+  (p_3^{\mu_1} -  p_2^{\mu_1})g^{\mu_2 \mu_3} \Bigr)\;.   $$

\CompHEP~ uses the notation $GG$ for the strong coupling constant $g$.
So for the 3-gluon vertex in the \CompHEP~ format we finally get
\vspace*{0.5cm}\\
{\footnotesize
\begin{tabular}{|l|l|l|l|l|l|}
\hline
A1&A2&A3&A4&Factor&Lorentz part\\
\hline
G&G&G& & GG & m1.m2*(p1-p2).m3+m2.m3*(p2-p3).m1+m3.m1*(p3-p1).m2 \\
\hline
\end{tabular}
}

\paragraph{ Quark-gluon interaction.}
 The interaction of a gluon with a quark is described by the 
following term of Lagrangian (\ref{physLagQCD}):
$$ 
S_{QqG}=  g \int G^{\alpha}_{\mu}(x) \bar{q}(x)\gamma^{\mu} 
\hat{t}_{\alpha}  q(x) \,  d^4x\;.
$$

Applying  the Fourier transformation and substituting $ \bar{q} =
q^+\gamma^0$   we get 
$$
S_{QqG}=  g  (2\pi)^4 \int \delta(p_1+p_2+p_3)  G^{\alpha}_{\mu}(p_3)
q^+(p_1)\gamma^0\gamma^{\mu} \hat{t}_{\alpha}  q(p_2) \,  d^4p_1  d^4p_2
d^4p_3\;;
$$

$$
\frac{\delta S_{QqG}}
{\delta q^+_i(p_1)\delta q^j(p_2) \delta G_{\mu}^{\alpha}(p_3)}=
 g  (2\pi)^4 \delta(p_1+p_2+p_3) (\hat{t}_{\alpha})^i_j
\gamma^0\gamma^{\mu_3}\;.
$$

The factor $(2\pi)^4 \delta(p_1+p_2+p_3)
(\hat{t}_{\alpha})^i_j \gamma^0 $ is substituted by \CompHEP~ 
automatically.  Thus, the quark-gluon  interaction is implemented in the 
\CompHEP~ {\it Vertex} table as the following record:
\vspace*{0.5cm}\\
\begin{tabular}{|l|l|l|l|l|l|}
\hline
A1&A2&A3&A4&Factor&Lorentz part\\   
\hline
Q&q&G& &  GG & G(m3) \\
\hline
\end{tabular}

\bigskip

\noindent
where $q$ and $Q$  are  designations for a quark  and 
a corresponding antiquark. 
\paragraph{ Interaction of ghosts with gluon.}
This interaction is described by the non-linear term of (\ref{fpLagQCD}),
namely 
$$
S_{\bar{c}cG}=  -g \int \bar{c}_{\alpha}(x) 
 \partial^{\mu}( f^{\alpha}_{\beta \gamma} G^{\beta}_{\mu}(x)
c^{\gamma}(x)) d^4x\;;
$$

Fourier transformation and    subsequent evaluation of functional
derivatives  gives us 
$$
S_{\bar{c}cG}= -g (2\pi)^4 \int \delta(p_1+p_2+p_3)  \bar{c}_{\alpha}(p_1)
(-i\,p_2 -i\,p_3)^{\mu_3}
 f^{\alpha}_{\beta \gamma} G^{\beta}_{\mu_3}(p_3)
c^{\gamma}(p_2) d^4p_1  d^4p_2  d^4p_3   
$$
$$
\frac{\delta S_{\bar{c}cG}}
{\delta \bar{c}_{\alpha_1}(p_1)
\delta c^{\alpha_2}(p_2)
 \delta G_{\mu_3}^{\alpha_3}(p_3)}=
 g  (2\pi)^4 \delta(p_1+p_2+p_3)
p_1^{\mu_3}i\, f^{\alpha_1}_{\alpha_2 \alpha_3} \;.
$$

The factor $(2\pi)^4 \delta(p_1+p_2+p_3) (-\,i\,
f_{ \alpha_1  \alpha_2  \alpha_3})$ is substituted by \CompHEP.
 Thus, this interaction may be implemented  as the  
 following record in the {\it Vertex} table.
\vspace*{0.5cm}\\
\begin{tabular}{|l|l|l|l|l|l|}
\hline
A1&A2&A3&A4&Factor&Lorentz part\\
\hline
G.C&G.c&G& & -GG &p1.m3 \\
\hline
\end{tabular}
\vspace*{0.5cm}\\
where $G.C$ and $G.c$ are the \CompHEP~ notations for 
the Faddeev-Popov ghosts $\bar{c}$ and $c$ respectively.

\paragraph{4-gluon interaction.}

Besides  the  3-gluon interaction term the Lagrangian (\ref{physLagQCD})
contains also the term of 4-gluon interaction:
$$
S_{4G}= -\,\frac{g^2}{4} 
g^{\mu\mu^\prime} g^{\nu\nu^\prime} \delta_{\alpha \alpha^\prime}  \int
f^{\alpha}_{\beta\gamma}G^{\beta}_{\mu}(x)G^{\gamma}_{\nu}(x)
f^{\alpha^\prime}_{\beta^\prime\gamma^\prime}
   G^{\beta^\prime}_{\mu^\prime}(x)
   G^{\gamma^\prime}_{\nu^\prime}(x) d^4x\;.
$$ 

The Fourier transformation and functional differentiation  lead us to the 
expression which contains three different $SU(3)$ color structures:

\begin{eqnarray}
\nonumber
&&\frac{\delta S_{4G}}{\delta G_{\mu_1}^{\alpha_1}(p_1)   
\delta G_{\mu_2}^{\alpha_2}(p_2) \delta G_{\mu_3}^{\alpha_3}(p_3)
 \delta G_{\mu_4}^{\alpha_4}(p_4)} =
 - g^2 (2\pi)^4 \delta(p_1+p_2+p_3+p_4) \delta_{\epsilon \epsilon^\prime
}\times\\ \nonumber
&&\Bigl( f^{\epsilon}_{\alpha_1 \alpha_2}f^{\epsilon^\prime}_{\alpha_3\alpha_4}
(g^{\mu_1\mu_3}g^{\mu_2\mu_4} - g^{\mu_1\mu_4}g^{\mu_2\mu_3})
+ f^{\epsilon}_{\alpha_1 \alpha_3}f^{\epsilon^\prime}_{\alpha_2\alpha_4}
(g^{\mu_1\mu_2}g^{\mu_3\mu_4} - g^{\mu_1\mu_4}g^{\mu_2\mu_3}) \\
&&+ 
f^{\epsilon}_{\alpha_1 \alpha_4}f^{\epsilon^\prime}_{\alpha_2\alpha_3}
(g^{\mu_1\mu_2}g^{\mu_3\mu_4} - g^{\mu_1\mu_3}g^{\mu_2\mu_4}) \Bigr)\;.
\label{4Gvertex}
\end{eqnarray}
The complicated color structure of 
this vertex  cannot be 
directly  written  down in the \CompHEP~ format. To implement this vertex 
we use the following trick. We introduce an auxiliary  tensor  field  
$t^{\alpha}_{\mu \nu}(x) $
and the Lagrangian of its interaction with
the gluon field:   
$$
   S_{aux}=  \int \left( \frac{i\,g}{\sqrt{2}} f^{\alpha}_{\beta\gamma} {t_{\alpha}}^{\mu \nu}(x)
 G^{\beta}_{\mu}(x)G^{\gamma}_{\nu}(x)
-\,\frac{1}{2}
  {t^{\alpha}}_{\mu \nu}(x) {t_{\alpha}}^{\mu \nu}(x)\right)d^4x\;.
$$

It is easy to notice that the  functional integration over the auxiliary field
  $t^{\alpha}_{\mu \nu}(x) $ reproduces  the term of 4-gluon
interaction in the partition function:

$$
 e^{i\,S_{4G}(G)}= \int e^{i\,S_{aux}(G,t)} \prod_{x,\alpha,\mu,\nu}
d\,t^{\alpha}_{\mu\nu}(x) \;.
$$  

For each colored  vector particle  \CompHEP~ 
adds     a tensor field with the same 
color to the  internal list of quantum fields.
The propagator of this field (\ref{tensor_propagator})  corresponds to the 
Lagrangian 
$( -\,\frac{1}{2} {t^{\alpha}}_{\mu \nu}(x) {t_{\alpha}}^{\mu\nu}(x))$. 
Consequently, in order to realize the 4-gluon
interaction  we must  introduce a vertex for the  interaction of the gluon  with 
this tensor field 
\begin{eqnarray*}
 S_{tGG}&=&\frac{i\, g}{\sqrt{2}}\int  f^{\alpha}_{\beta\gamma} {t_{\alpha}}^{\mu \nu}(x)
 G^{\beta}_{\mu}(x)G^{\gamma}_{\nu}(x) d^4x\;; \\
\frac{\delta S_{tGG}} { 
G^{\alpha_1}_{\mu_1}(p1)
G^{\alpha_2}_{\mu_2}(p2)
\delta t^{\alpha_3}_{\mu_3 {\mu_3}\prime}(p_3)   
} 
&=& (2\pi)^4\delta(p_1+p_2+p_3) (-\,i\, f_{\alpha_1 \alpha_2 \alpha_3})
\\&\times& \frac{g}{\sqrt{2}}
 (g^{\mu_2 \mu_3} g^{\mu_1{\mu_3}\prime} 
  - g^{\mu_1 \mu_3} g^{\mu_2{\mu_3}\prime} 
 )\;.
\end{eqnarray*}
 
In  \CompHEP~ notations this vertex looks like the following
\vspace*{0.5cm}\\
\begin{tabular}{|l|l|l|l|l|l|}
\hline
A1&A2&A3&A4&Factor&Lorentz part\\
\hline
G&G&G.t& & GG/Sqrt2 & m2.m3*m1.M3 -m1.m3*m2.M3   \\
\hline
\end{tabular}
\vspace*{0.5cm}\\
where $G.t$ is a \CompHEP~ notation for the auxiliary  tensor field 
$t^{\alpha}_{\mu\nu}$ associated
with the vector field $G$. Capital $M$  denotes the  second Lorentz index of
the tensor field. 

\begin{figure}

{\scriptsize    
 
\begin{picture}(80,80)(0,0)

\Text(10,70)[r]{$G_1$}
\Text(10,10)[r]{$G_2$}
\Text(70,70)[l]{$G_3$}
\Text(70,10)[l]{$G_4$}

\DashLine(10,70)(70,10){3.0} 
\DashLine(70,70)(10,10){3.0}
\Vertex(40,40){2} 
\end{picture} 
\begin{picture}(10,80)(0,0)
\Text(5,40)[]{=}
\end{picture}
\begin{picture}(80,80)(0,0)
\Text(10,70)[r]{$G_1$}
\Text(10,10)[r]{$G_2$}
\Text(70,70)[l]{$G_3$}
\Text(70,10)[l]{$G_4$}

\DashLine(10,70)(10,10){3.0} 
\DashLine(70,70)(70,10){3.0} 
\DashLine(10,40)(70,40){3.0} 

\Text(40,42.0)[b]{$G.t$}
\Vertex(10,40){2}
\Vertex(70,40){2}
\end{picture} 
\begin{picture}(10,80)(0,0)
\Text(5,40)[]{+}
\end{picture}
\begin{picture}(80,80)(0,0)

\Text(10,70)[r]{$G_1$}
\Text(10,10)[r]{$G_2$}
\Text(70,70)[l]{$G_3$}
\Text(70,10)[l]{$G_4$}

\DashLine(10,70)(70,70){3.0} 
\DashLine(10,10)(70,10){3.0} 
\DashLine(40,10)(40,70){3.0} 

\Text(38,40.0)[r]{$G.t$}
\Vertex(40,10){2}
\Vertex(40,70){2}
\end{picture} 
\begin{picture}(10,80)(0,0)
\Text(5,40)[]{+}
\end{picture}
\begin{picture}(80,80)(0,0)

\Text(10,70)[r]{$G_1$}
\Text(10,10)[r]{$G_2$}
\Text(70,70)[l]{$G_3$}
\Text(70,10)[l]{$G_4$}

\DashLine(10,70)(70,10){3.0} 
\DashLine(70,70)(10,10){3.0} 
\DashLine(55,55)(55,25){3.0} 
\Text(57,40.0)[l]{$G.t$}
\Vertex(55,55){2}
\Vertex(55,25){2}
\end{picture} 
}

\caption{ Splitting of four-gluon vertex}
\label{4Gsplitt}
\end{figure}
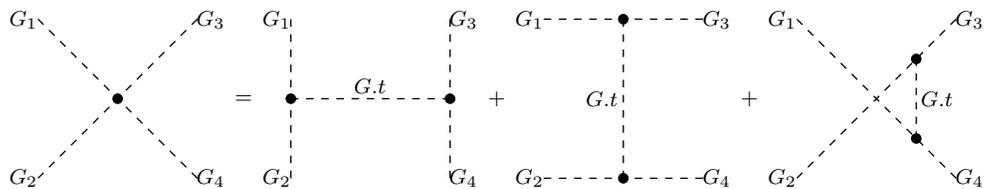

From the viewpoint of the  Feynman diagram technique such
realization of the 4-gluon interaction means that
instead of one 4-gluon vertex we substitute
three sub-diagrams presented  in Fig.\ref{4Gsplitt}.
Contribution of each of these diagrams corresponds to
one of the terms of expression (\ref{4Gvertex}).

\subsubsection{Neutrino as a Majorana fermion}

It is well known that  in the  Standard Model  only the left component 
of massless neutrino takes part in  interactions.
So one can describe neutrino by a Majorana  field which has the 
same number of degrees of freedom as a left Dirac one. To realize such a
particle  in the framework of \CompHEP one should add the  
following record to the table of particles:
\vspace*{0.5cm}\\
\begin{tabular}{|l|l|l|l|l|l|l|l|}
\hline
 Full  name    &A &A+ &2*spin &  mass    &width                  & color&aux \\
\hline
neutrino &MN&MN& 1      & 0 & 0 & 1 &  \\
\hline
\end{tabular}
\vspace*{0.5cm}\\

In terms of  Dirac  field  $\Psi_\nu$ a neutrino appears in the Standard Model
 Lagrangian  in the following way  (\ref{V-fermion})\footnote{where $Y=-1$,
$\Psi_1=\Psi_\nu$, $\Psi_2=\Psi_e$, $g_2=e/\sin{\Theta_w}$,
$g_1=e/\cos{\Theta_w}$ }:
\begin{eqnarray}
L_\nu &=& \frac{i}{2}(\bar{\Psi_\nu} \gamma_\mu \partial^\mu \Psi_\nu -
(\partial^\mu \bar{\Psi_\nu})\gamma_\mu \Psi_\nu) 
+ \frac{e}{4 \sin{\Theta_w} \cos{\Theta_w} } Z_\mu 
\bar{\Psi}_\nu \gamma^\mu ( 1 - \gamma^5) \Psi_\nu  \nonumber \\
&& + \frac{e}{2\sqrt{2}\sin{\Theta_w} }
\left(  W^-_\mu  \bar{\Psi}_e  \gamma^\mu (1 - \gamma^5)\Psi_\nu
+   W^+_\mu  \bar{\Psi}_\nu  \gamma^\mu (1 - \gamma^5)\Psi_e \right)\;, 
\label{neutrinoLag}
\end{eqnarray}
where  $\Psi_e$ is the  electron field. To rewrite it in terms of a Majorana
fermion let us perform the  substitution
$$ \Psi_\nu = \frac{1}{2}(1-\gamma^5) \psi_l + \frac{1}{2}(1+\gamma^5)
\psi_r\;,
$$
 where $\psi_l$ and  $\psi_r$ are   Majorana fermions.
Omitting  Lagrangian for  $\psi_r$  and applying the following identities for Majorana
fermions
\begin{eqnarray*}
\frac{i}{4}(\bar{\psi} \gamma_\mu \gamma^5 \partial^\mu \psi -
  (\partial^\mu \bar{\psi})\gamma_\mu \gamma^5 \psi) = \bar{\psi} \gamma^\mu  \psi &=&0
\end{eqnarray*}
which can be obtained  by means of (\ref{reverse}), we get  

\begin{eqnarray*}
L_\nu &=& \frac{i}{4}(\bar{\psi}_l \gamma_\mu  \partial^\mu \psi_l - 
   (\partial^\mu \bar{\psi}_l)\gamma_\mu \psi_l) 
- \frac{e}{4 \sin{\Theta_w} \cos{\Theta_w} } Z_\mu\bar{\psi_l} \gamma^\mu
\gamma^5  \psi_l  \\    
&+& \frac{e}{2\sqrt{2}\sin{\Theta_w} }\left(  W^-_\mu  \bar{\Psi}_e
\gamma^\mu (1 - \gamma^5)\psi_l    
+   W^+_\mu  \bar{\psi}_l  \gamma^\mu (1 - \gamma^5)\Psi_e \right)\;. \nonumber\\
\end{eqnarray*}
Here the first term is the  free Lagrangian for a massless  Majorana fermion.  
Other terms define the interaction. Using the definition 
(\ref{funcDeriv}) we can rewrite them in the \CompHEP~ notations:
\vspace*{0.5cm}\\
\begin{tabular}{|l|l|l|l|l|l|}
\hline
A1&A2&A3&A4&Factor&Lorentz part\\
\hline
MN &MN&Z& &-EE/(2*SW*CW) & G(m3)*G5   \\
\hline
E1&MN&W-& & EE/(2*Sqrt2*SW)& G(m3)*(1-G5)\\
\hline
MN&e1&W+& & EE/(2*Sqrt2*SW)& G(m3)*(1-G5)\\
\hline
\end{tabular}
\vspace*{0.5cm}\\

Let us emphasize  that there are two identical neutrino fields 
in that  term of the Lagrangian which describes the interaction of neutrinos
with a $Z$-boson. It leads to the additional factor  $2$  and to the symmetry
property
of the  corresponding  vertex. One of the typical mistakes in
realization of such a vertex is an introduction 
of  the G(m3)*(1-G5)  term which breaks the symmetry property.
Correct evaluation of the  functional derivative (\ref{funcDeriv})  with the 
help of the identity  (\ref{reverse}) never produces such a term.

\subsubsection{Leptoquarks}

In this section we  present an example of Lagrangian which contains 
the matrix of  {\it C}-conjugation. Such  matrix appears in interactions 
violating  the fermion number conservation. Let 
$\Psi_e$ and $\Psi_u$  be the fermion fields of electron and  {\it u}-quark, 
respectively, interacting   with the scalar complex leptoquark field
{\it F}
$$ L = \lambda  \bar{\Psi}_u^c (1+\gamma^5) \Psi_e F + \lambda  \bar{\Psi}_e
(1+\gamma^5) \Psi_u^c F^+   $$
where $\bar{\Psi}_u^c=\Psi_u^T C$ and  $\Psi_e^c=C \bar{\Psi}_e^T$.

In the Majorana basis which is used in \CompHEP~ the charge conjugation
operator  $C=-\gamma^0$. So the
Lagrangian  can be presented in the form
\begin{eqnarray*}
 L &=& - \lambda  \Psi_u^T \gamma^0 (1+\gamma^5) \Psi_e F - \lambda
\Psi_e^+\gamma_0 (1+\gamma^5) \gamma^0 (\gamma^0)^T \Psi_u^+ F^+ \\
&=&- \lambda  \Psi_u^T \gamma^0 (1+\gamma^5) \Psi_e F + \lambda \Psi_u^+  
\gamma^0  (1+\gamma^5)  \Psi_e^+ F^+\;.
\end{eqnarray*}

 Direct  implementation of the  definition (\ref{funcDeriv}) gives us 
the {\it Vertex} table 
\vspace*{0.5cm}\\
\begin{tabular}{|l|l|l|l|l|l|}
\hline
A1&A2&A3&A4&Factor&Lorentz part\\
\hline
u &e1&F& &-lambda & (1+G5)   \\
\hline
E1&U&F+& & lambda& (1+G5)\\
\hline
\end{tabular}
\vspace*{0.5cm}\\
By means of  equation (\ref{reverse}) we can rewrite this table  in the
equivalent form: 
\vspace*{0.5cm}\\
\begin{tabular}{|l|l|l|l|l|l|}
\hline
A1&A2&A3&A4&Factor&Lorentz part\\
\hline
e1 &u&F& &-lambda & (1+G5)   \\
\hline
U&E1&F+& & lambda& (1+G5)\\
\hline
\end{tabular}
\vspace*{0.5cm}

  \subsection{LanHEP and SUSY models}
             \label{LanHEP}             
There is a possibility of automatic conversion  of  the Lagrangian written in
the compact   form in the coordinate space  into the \CompHEP~ table format. It
can be performed  by means of  the  {\it LanHEP} program  written by A.~Semenov \cite{LanHEP}.
The input    is  expressed in terms of 
complex objects,  such as a covariant derivative and
a gauge field  tensor.  

The  {\it LanHEP} program was used \cite{MSSM}  to generate \CompHEP~ model files  
for the Minimal
Supersymmetric Model. Another result produced by {\it LanHEP}
is the realization the general two-Higgs-doublet model \cite{2Hmodel}   The CompHEP~  WWW page  
contains  references  to the LanHEP code  and to the  MSSM Lagrangian
model files.

\newpage
\section{\CompHEP~ output files}
             \label{output}
  \subsection{\LaTeX~ output}   
             \label{out-latex} 
 \CompHEP~ uses the {\it Axodraw} package by J.A.M.~Vermaseren  \cite{axodraw} 
to write diagrams and plots in \LaTeX~ format. To use this package the 
{\it Axodraw}
 style  should be included in the {\it documentstyle} statement. 
An example would be 

{\tt	{$\backslash$}documentstyle[axodraw]\{article\} }

Under  kind   permission of the author we put a copy of  {\it axodraw.sty} file 
in  the \$COMPHEP/ directory.
One bug in the axis drawing routine has been  corrected.    

    The {\it Axodraw} syntax is very easy and you can alter the \CompHEP~ 
output to get some local correction in picture if any.  The user can also 
change  line width,  scale of picture, and  size of characters. The
corresponding \LaTeX~ and {\it Axodraw} instructions are presented 
in the beginning of the output. For example,
\begin{verbatim}
{  \small              % letter  size control
   \SetWidth{0.7}      % line  width control
   \SetScale{1.0}      % line scale control
   \unitlength=1.0 pt  % text position control
   ............
}
\end{verbatim} 

Note that the  {\it {$\backslash$}SetScale}  instruction influences 
on  positions of lines, whereas the {\it {$\backslash$}unitlength} variable
is responsible for position if texts. Consequently, if  the user would like
to change the scale of picture, he must improve these instructions 
in the same manner.   For instance, to increase the picture by two times use 
\begin{verbatim}
   \SetScale{2.0}      % picture size control
   \unitlength=2.0 pt  % picture size control
\end{verbatim} 

In the case of Feynman diagram output \CompHEP~ substitutes \LaTeX~ 
names of particles as they are defined in the {\it Particle} table 
(see Section \ref{particles}).

  \subsection{Symbolic answer  in \REDUCE~ and \MATHEMATICA~ formats}
             \label{s-output}         \subsubsection{General structure}

The \CompHEP~ symbolic output can be used for  further 
manipulation with an answer which has been obtained by means of the 
built-in symbolic calculator. It  might be a summation of all diagrams to
a  common denominator expression, a symbolic integration of
answer, a representation of answer  as a function of special 
set of variables and so on. We have tried to present results in the form
which can be easily used for  different purposes.

All diagram contributions  for one  subprocess are stored in one 
file. The  subprocess ordering number is attached to  the file name. For
example, the {\it symb1.red} and {\it symb1.m}  files are generated  
to present the symbolic answer of the first subprocess in  the  \REDUCE~
\cite{REDUCE}  and \MATHEMATICA~ \cite{MATH}  format correspondingly. 

 The structure of the output  file can 
be described by the following scheme:
\begin{center}
{\small
\begin{tabular}{|c|} \hline
      Initial declarations; \\
      $initSum()$;\\
      Answer for the first diagram;\\
       $addToSum()$; \\
      Answer for the second diagram;\\
       $addToSum()$;\\   
      ~~~~...................\\
      ~~~~................... \\
      $finishSum()$; \\     
\hline
\end{tabular}
}
\end{center}

 {\it 'Initial declarations'} includes the  declaration of 
vector variables for momenta and the conservation law relations for them,
the declaration of independent parameters involved in calculation and
numerical values of them,
the declaration of constrained parameters and substitution rules for them,
and, at last, the  declaration of the process name.
 The momenta are named  
$p1,p2,p3,...$. They are  assigned to particles according 
to a particle sequence in the process name.  The signs  of momenta are defined 
in such a way that the sum of momenta of incoming particles is equal 
to the sum of  momenta of outgoing particles. The list of substitutions
of numerical values for  independent  parameters is written down in 
variable $parameters$. The list of substitutions
for the  constrained parameters   is stored in  variable $substitutions$.  
The lists of incoming and outgoing particles are stored in variables
$inParticles$ and $outParticles$ respectively.

\CompHEP~ writes down subsequently  expressions for diagram  contributions 
and after any record it calls a  summation procedure $addToSum()$.
Just before and after summation the procedures $initSum()$ and $finishSum()$
are called.  These three procedures  must  be written  by the user and loaded 
in advance. Such a construction of the output allows one  to use it for different 
purposes creating  appropriate  procedures.

Now we shall explain the structure of  diagram contribution.
It is started from a pseudo-graphic diagram image like in Fig.\ref{pseudo}.
After that  assignments for  $totFactor$, $numerator$, $denominator$
variables  follow: 
\begin{itemize}
\item[] $totFactor$ is a ration function depending on model parameters;
\item[] $numerator$ is a polynomial  of  model parameters and
momenta scalar products;
\item[] $denominator$ is presented as a  product of 
propagator denominators
$$ propDen(P,Mass,Width)\;, $$ 
where $P$, $Mass$, and $Width$ are the momentum, mass, and width of the 
corresponding
virtual particle. In the case  $Width=0$  $propDen$  must be defined as
$(Mass^2-P^2)$. The treatment of the $Width$ argument  can be arranged 
by  the user as he likes.

\end{itemize}
 In these terms the  diagram  contribution  to the squared matrix element 
may be expressed in the following way:

$$ totFactor \frac{numerator}{denominator}\;\;. $$

As it was mentioned above (Section \ref{sq_diagrams_menu}) the result obtained 
by summation of all diagrams 
must be symmetrized in the case  of identical outgoing particles. 
This work may be done by the $finishSum()$ procedure.

\subsubsection {Example: Summation of diagrams and symbolic integration 
by means of the  \REDUCE~  package}
\label{red-symb-int} 

 We have prepared some  summation programs for working with
\REDUCE~ output. They are: \\
\hspace*{5mm}$sum.red$ which presents the squared matrix element  as a common denominator
expression;\\
\hspace*{5mm}$sum2pole.red$  which presents the squared matrix element as a sum of
pole terms;\\ 
\hspace*{5mm}$sum2tot.red$  which presents a symbolical expression for the 
total cross
section.\\
The last two options are available only  for \verb|2->2|  processes.
These files are stored in the $COMPHEP/test$ directory.
 
Let you  prepare the symbolic output {\it symb1.red}  for the Compton scattering
{\it A,e1 \verb|->| A,e1}  in the 
framework of QED model \footnote{In other \CompHEP~ models the electron 
is massless.}.  Launch the \REDUCE~ system from
within the {\it result} directory. The possible \REDUCE~ sessions are:\\
{\small
\begin{verbatim}
%1
  in"$COMPHEP/test/sum.red"$     % loading the summation package
  in"symb1.red"$                 % reading contributions of diagrams 
  sum;                           % writing the answer 

(32*ee**4*(2*p1.p2**4-4*p1.p2**3*p1.p3+3*p1.p2**2*p1.p3**2-2*p1.p2**2*p1.p3
*me**2-p1.p2*p1.p3**3 + 2*p1.p2*p1.p3**2*me**2 + p1.p3**2*me**4))
/(propden(-p1-p2,me,0)**2*propden(p2-p3,me,0)**2)$

%2
  in"$COMPHEP/test/sum2pole.red"$  % loading the summation package  
  in"symb1.red"$                   % reading contributions of diagrams
  sum;                             % writing the answer 
2*ee**4*(4*sp(me)**2*me**4 + 8*sp(me)*up(me)*me**4 + 4*sp(me)*me**2 +
sp(me)*t+ 4*up(me)**2*me**4 + 5*up(me)*me**2 - up(me)*s + 1)$
\end{verbatim}
}
Here $s=(p1+p2)^2;\; t=(p1-p3)^2;$ the functions $sp,tp,up$ are defined by the 
following way
$$ sp(x)=1/(s-x^2);$$
$$ tp(x)=1/(t-x^2);$$
$$ up(x)=1/(1-u^2)\;\; where\;\; u=(p1-p4)^2 $$
{\small
\begin{verbatim}
%3
  in"$COMPHEP/test/sum2tot.red"$ % loading the summation package  
  in"symb1.red"$                 % reading contributions of diagrams
  sum;                           % writing answer for total cross section                   
(ee**4*(2*s**4*log(s/me**2) + s**4 - 12*s**3*log(s/me**2)*me**2 + 14*s**3*me**2 
- 6*s**2*log(s/me**2)*me**4 - 16*s**2*me**4 + 2*s*me**6 - me**8))/(16*s
**2*pi*(s**3 - 3*s**2*me**2 + 3*s*me**4 - me**6))$
\end{verbatim}
}

Sometimes the expression for total cross-section 
includes a cumbersome square root of kinematic variables
which appears as a result of evaluation of integrand limits.
In this case the integration routine  introduces 
a new variable \verb|be_|  for this square root to  express 
the total cross-section in a more compact form. 
The substitution for the  \verb|be_**2| is generated by the integration
routine. 

There are similar packages $sum.m$, $sum2pole.m$, and  $sum2tot.m$ 
for operation with the \MATHEMATICA~ output.


  \subsection {\REDUCE~ program}
               \label{r-code}       
The \REDUCE~  program  was the   first \CompHEP~ output 
which opened the possibility   to produce some physical results
by means of this package. Later on, 
when the \CompHEP~ built-in symbolic calculator was created,
the \REDUCE~ output  became unnecessary. But we still
keep it in the package for testing. 

The \CompHEP~ symbolic calculator  looks like a black box.
On the contrary the \REDUCE~ program is written in terms which
can be understood and checked. The comparison of result 
produced by the built-in symbolic calculator with that of
 \REDUCE~   evaluation of the generated code
is a good check of  \CompHEP~  software (see Section  \ref{symb-calc-check}).
  Below we shall describe 
the structure  of \REDUCE~ program.
  
\CompHEP~ generates a separate file for each squared diagram.
 The files are named as  {\it pNNN-MMM.red} where 
{\it NNN} is  the  subprocess number 
and {\it MMM} is the  diagram ordering number. 

The file begins with from the declaration of momenta and Lorentz indices.
For example,\\

\begin{verbatim}
% ----------- VARIABLES ------------------ 
 vector  A,p1,p2,p3,p4,p5,p6,p7,p8,p9,p10,p11,p12,ZERO_;
 vector  m1,m2,m3,m4,m5,m6,m7,m8,m9,m10,m11,m12,m13,m14,m15,m16;
%
%--------- Mass shell declarations -----------
 MASS  P1 = 0$  MSHELL P1$
 MASS  P2 = 0$  MSHELL P2$
 MASS  P3 = Mm$  MSHELL P3$
 MASS  P4 = Mm$  MSHELL P4$

%-------- Momentum substitutions --------
 Let  p4 = +p1+p2-p3$
 Let  p5 = +p1+p2$
 Let  p6 = -p1-p2$

\end{verbatim}

Vector {\it A} is used by the \REDUCE~ package  to construct 
the $\gamma^{5}$ matrix, $\gamma^{5}=G(ln,A)$. Vectors  whose names begin
 with {\tt 'p'} are used for designation of momenta. Vectors whose names begin
with {\tt 'm'} are reserved for Lorentz indices.
We see then the  mass-shell declarations for incoming and outgoing particles.
After that the file contains the substitutions of momenta of on-shell and 
virtual particles  which we have according to the  conservation law.

Then the diagram total factor  with self explanatory 
comments is written down. For example, in the case of $e^+ e^- \rightarrow \mu^+ \mu^-$
process we have 
\begin{verbatim}
%---------- Factors ---------------
 SymmFact:=1/1$    % Diagram symmetry factor
 AverFact:=1/4$       % Normalization factor of polarization average
 FermFact:=1$      % (-1)**(number of in-fermion particles)
 ColorFact:=1/1$    %  QCD color weight of diagram  
%
totFactor_:=EE**4$
totFactor_:=totFactor_*SymmFact*AverFact*FermFact*ColorFact$
\end{verbatim}

These declarations are self-explanatory except, maybe, of the $SymmFact$
variable. Generally   $SymmFact=N/D$  where  $N$ is equal to 1 in the 
case of   left-right    squared diagram   symmetry. Otherwise
the factor equals  2.  $D$ is a factorial connected with presence  of 
identical  outgoing particles and  partially reduced by a number of various 
possibilities to assign the momenta  of outgoing particles to the 
corresponding diagram lines.

The  program for evaluation of  one squared diagram actually  
includes  the codes for a set of diagrams which appear after
replacing of some physical particles by their ghosts according to the
existing interaction  vertices. 
We shall call all diagrams of the set as {\it ghost} diagrams.
Note that all these  diagrams have the same denominator. 
The evaluation is started from the initialization of
variable for the sum of numerators of the set:

{\tt numerator\_:=0\$}

The program for evaluating  each  diagram of the set is advanced by  the 
 pseudo-graphical image of diagram. (See, for example, Fig.\ref{pseudo}).
  The name of particle and corresponding momentum are written down near
the line. The Lorentz index is written just on the line.

The diagram code is started from the fermion loops evaluation.
The program moves along the  fermion line and multiplies 
the vertex and propagator terms. The    $nospur$ 
instruction is declared before each loop  evaluation to prevent
the default trace evaluation in the end of any instruction.

If  the result of multiplication contains
 Lorentz indices which can  be contracted, the program 
declares the corresponding  vectors as indices by means of 
the  {\it index  }  instruction.

As a rule the  fermion  vertices are multiplied 
in the order which corresponds to moving in the direction
opposite to the direction of fermion arrows. But if the  diagram  
contains a vertex with the {\it C-conjugate} operator or
Majorana particles the order of multiplication is chosen at random, 
because the propagator lines have different orientations or have no them at all. 
If the orientation of fermion line is chosen,  the  vertices 
are divided into two groups, normal and reverse. For the normal vertex the 
incoming fermion line is attached to the second fermion,
 whereas for the reverse  vertex it is attached to the  first one.
 \CompHEP~ transforms the reverse vertices to normal form by means 
of the rule (\ref{reverse}) and writes down the  corresponding comment.

After the last multiplication the state of the $spur$ switch is 
restored and the $\gamma$-matrix trace   is evaluated  with simultaneous 
multiplication by factor of '-4'. The '4' is needed because the \REDUCE~
trace evaluation omits this factor and the minus  appears from 
the Feynman rules.

The next step  is the  multiplication of vertices 
and contraction of   Lorentz indices . The indices for contraction 
are declared by the {\it Index } instruction before the 
multiplication. The fragment of the corresponding code looks like \\

\begin{verbatim}
 Index m2$
 Vrt_3:=Vrt_3*Vrt_5$
 RemInd  m2$
\end{verbatim}    

This code means that the vertex number 3 is multiplied by the  vertex number 5
with contraction of $index\; m2$. The result of multiplication is 
considered as a   generalized vertex  number 3.

  Here we would like to note that there is a bug 
in the \REDUCE~ package which  forbids to  convolute  
several indices in one operation. To bypass this bug we declare the '$m$'
vectors as  indices step by step and recalculate the expression on each step. 

If two vertices are connected by a propagator of massive vector particle
treated  in the physical gauge (\ref{mvPropagator})  the code are organized 
in the following way.
\CompHEP~ writes the code for the vertices product with contracted  indices
and the code for the  product of the  same vertices with the prior multiplication by 
 relevent momentum. After that  these contributions are  summarized.  
Below we present an example of such a code:

\begin{verbatim}
 Index m1$
 Vrt_0:=Vrt_1*Vrt_2$
 RemInd  m1$
 Vrt_L:=Vrt_1$   Vrt_R:=Vrt_2$
 Vrt_L:=(Vrt_L where m1=>(+P2+P3)/MZ)$
 Vrt_R:=(Vrt_R where m1=>(+(-P2)+(-P3))/MZ)$
 Vrt_0:=Vrt_0 + Vrt_L*Vrt_R$
\end{verbatim}    

The code for any diagram  evaluation  is terminated by instruction
 $$ numerator\_:=numerator\_ +DiagramFactor*GhostFact*Vrt\_1 $$
which  summarizes the ghost diagram contributions.
Here  {\it GhostFact} equals  $(-1)^{l+v}$ where the $l$ is a number 
 of loops of Faddeev-Popov ghosts and $v$ is a number of  vector
field lines in the diagram. The $(-1)^{v}$  factor  appears because 
the evaluations discribed above correspond to substitution of the 
$(g_{\mu\nu} - k_{\mu}k_{\nu}/M^2)$ factor for the propagator and the density 
matrix of vector field  whearas the correct
expression has the opposite sign.

The last step of the program is the assignment   of variable
$denominator\_$. It is expressed as a production of $propDen(P,Mass,Width)$
functions as it is explained in Section \ref{s-output}.

As a result the symbolic answer for the evaluated diagram
may be presented in the form

$$ totFactor\_ \frac{ numerator\_}{denominator\_}\;.  $$

  \subsection{\Fortran~ output files}    
             \label{out-fort}           
   \CompHEP~ \Fortran~ output is used  to compile  the squared matrix element
 for  subsequent 
 evaluation of cross sections and distributions.  The explanation
below is necessary  only if you would like to use the \CompHEP~ \Fortran~
output for other phase space integration program  or,  vice versa, to use
the \CompHEP~ 
routines  to integrate a phase space function  produced in some other
manner outside  \CompHEP.

Below we insert the  declaration of  parameter  type into the header of \Fortran~ 
routine. It is done for brevity only  and contradicts to the \Fortran~
style. We use also the {\it Real*X}  designation  for declaring  
the parameter of floating point type  which is {\it Real*8} or {\it Real*16} depending on 
the  type of   output.

   {\tt FUNCTION LENR()}\\ 
returns 8 in the case when \CompHEP~ produces a code with 
Real*8 (DOUBLE PRECISION) floating arithmetics or
16 in the case when \CompHEP~ produces a code with 
Real*16 (QUADRUPLE PRECISION) floating arithmetics.

   {\tt   FUNCTION NIN()}\\ 
returns a number of incoming particles.

   {\tt   FUNCTION NOUT()}\\
returns a number of outgoing particles.

   {\tt   FUNCTION NPRC()} \\
returns a total  number of subprocesses.

   {\tt   CHARACTER*6 FUNCTION PINF(NSUB,NPRTCL) } \\
returns a particle name for the
subprocess {\tt NSUB}. {\tt NPRTCL} is the particle ordering number in the subprocess.
The first {\tt NIN()} numbers numerate the incoming particles.

    {\tt  SUBROUTINE PMAS(NSUB,NPRTCL,REAL*X VAL)}\\ 
returns the particle mass. Incoming parameters are: {\tt NSUB} - the subprocess
number and {\tt NPRTCL} - the ordering number of particle.
 Returned parameter is VAL. In the case when LENR()=8 VAL must be a REAL*8
variable, otherwise it must be described as a REAL*16 one.

    {\tt  FUNCTION NVAR()}\\ 
returns a number of physical parameters involved in the calculation.
      
     {\tt SUBROUTINE VINF(NUMVAR,CHARACTER *6 NAME,REAL*8  VAL)}\\
provides the information about physical parameters involved into evaluations.
It returns the physical   parameter name NAME and its value VAL. 
The incoming {\tt NUMVAR}  is the parameter ordering number which
must be smaller than or equal to {\tt NVAR()}.
 
      {\tt SUBROUTINE ASGN(NUMVAR,REAL*8 VALNEW)}\\
assigns to the parameter {\tt NUMVAR} a new value {\tt VALNEW}.

      {\tt SUBROUTINE VINI}\\
makes the initialization of all physical parameters.  

{\tt      SUBROUTINE CPTH(CHARACTER*60 PATH, CHARACTER*1 F\_SLASH) }\\
assigns new values to all its parameters. {\it PATH}
is the path to the \CompHEP~ root directory (see Section \ref{u-install}). 
 {\tt F\_SLASH} is a symbol which separates file name in
the path. It is "/" or "\verb|\|" in the {\it UNIX} and {\it MS-Windows}  cases correspondingly.

  {\tt    REAL*8 FUNCTION SQME(NSUB)}\\ 
returns a numerical value of squared matrix element for the subprocess {\it NSUB}. 
Summation over out-particle polarization states and an averaging over
in-particle polarization states are carried out. Information about scalar
products of momenta must be passed on to the {\it SQME} through 
\begin{equation}
\label{SCLR} 
{\it COMMON/SCLR/ REAL*X~ PP(15)}
\end{equation}
 The assignments of scalar product values  to elements of the  {\it PP}
array,  like
$$PP(INDX(k,l)) = p_k.p_l\;\;\;  , $$
must be done before the {\it SQME} call. We assume that all four-momenta 
$p_i$ of particles have  positive energy components $( p_i^0 > 0)$.
 {\it PP} must be of the   {\it REAL*8} type in the case of double precision calculation 
and of the {\it REAL*16} type for the quadruple precision one.
      
  {\it SQME} is sensitive to the states of {\it LOGICAL} flags {\it
GWIDTH} and {\it RWIDTH} from
the {\it COMMON/WDTH/GWIDTH,RWIDTH}. They define the treatment of  particle 
finite width. The {\it GWIDTH} flag indicates that the widths do not break gauge
invariance. {\it RWIDTH} states that the running widths are substituted. See the
discussion in Section \ref{Breit-Wigner}.

   It is assumed that the same units ( for example,  powers of  GeV ) are used
 for momenta and for physical variables.  {\it SQME} produces an answer in
the
 $$[P]^{(2*(NIN()+NOUT()-4))}$$
 units, where {\tt [P]} is a unit of momentum.
Standard normalization for SQME is used. See 
 Eqs. (23.12) and (23.26) in \cite{ParticlesFields}.  When {\it n\_comphep\_f} 
 calls the {\it SQME} function it substitutes momenta in
the {\it GeV} units.
  
      {\tt FUNCTION INDX(K,L)}\\ 
returns the position of  scalar product $p_k.p_l$ in the 
{\it PP} array  (\ref{SCLR}). 

  \subsection{\C~ output files}   
             \label{out-c}              
   \CompHEP \C-output is  quite similar to the \Fortran~ one. 
Units and normalizations are also the same.
   Now we shall list the functions and the global constant parameters 
defined in this output and  called by the program {\it n\_comphep\_c}.   

  \subsubsection*{Parameter section}
 
\hspace*{1cm} {\tt const int  nvar\_}  is a number of independent physical parameters 
involved in  the evaluation of  squared matrix element.\\
\hspace*{1cm} {\tt const int nfunc\_}  is a number of constrained parameters involved 
in  the evaluation of  squared matrix element.\\
\hspace*{1cm}  {\tt int calcFunc(void)}  calculates all constrained parameters for current 
values of independent ones. It returns 0 in the case of success, otherwise 1.\\
\hspace*{1cm}{\tt int vinf\_(int numvar, char *name, double *val)}
 provides the  information about parameters. Here the incoming  
parameter   {\it numvar}  is a parameter number. The outgoing 
parameters {\it *name} and {\it *val} are the parameter name and  value, 
correspondingly.
If *name or {\it *val} is {\it NULL} no assignment for the  corresponding parameter 
will be done.

  The parameter  numbers in the range from 1 till {\it nvar\_} are 
reserved for independent parameters and the next  {\it nfunc\_} numbers are used  
for constraints.
  Equal-to-zero {\it numvar}  is associated with the center of mass energy.
If the {\tt numvar} value is out of this range, {\tt vinf\_(..)} returns 1.\\
\hspace*{1cm}  {\tt int asgn\_(int numvar, double valnew);}
assigns a new value {\it valnew}  to the   {\it numvar} parameter. 
It returns $0$ if $0 <= numvar <= nvar\_$, or $1$ otherwise.

\subsubsection*{Process section}

\hspace*{1cm}  {\tt const int nin\_}   is a number of incoming particles\\
\hspace*{1cm}  {\tt  const int nout\_}  is a number of outgoing particles\\
\hspace*{1cm}  {\tt const int nprc\_}  is a total number of subprocesses \\
\hspace*{1cm}  {\tt char processch[]}   contains the process name. It is
needed to be only  displayed  on the  screen  during the {\it n\_comphep\_c} session.\\
\hspace*{1cm} {\tt int pinf\_(int nsub, int nprtcl, char *name, double * mass)} 
   returns the {\it name}   and  the {\it mass} of particle   with the 
{\it nprtcl} 
ordering number  for the subprocess {\it nsub}. First {\it nin\_}~  numbers are 
reserved for 
incoming particles, while next {\it nout\_} are assigned to  outgoing ones.
One  can  substitute the {\it NULL} constant  instead  of any  outgoing 
parameter  if the corresponding information is not needed.
The   function returns $0$ if $nsub <= nprc\_$ and $nprtcl <= nin\_+nout\_$,
  otherwise 1 is returned.

\subsubsection*{The squared matrix element}
\hspace*{1cm} {\tt double sqme\_(int nsub,double * momenta, int * err)}
returns the numerical value of squared matrix element after   
summation over out-particle polarizations  and averaging over
in-particle polarizations.  Here 
 {\it nsub} is a  subprocess number,
 {\it momenta} is an array of particle momenta.   The $i^{th}$ Lorentz component 
of the particle number {\it N} corresponds to \mbox{\it momenta[4*N+i]},\\ 
 {\it *err} is a return parameter which does not equal  zero if   
{\it sqme\_} cannot be evaluated.

  We assume that the zero Lorentz components of momenta  are positive
and  the sum of incoming particle momenta is equal to the sum of 
outgoing particle momenta.

\subsubsection*{Widths  implementation parameters}
    
\hspace*{1cm}  {\tt int rwidth, gwidth} are the logical variables which  
define a particle width 
treatment for calculations of the squared matrix element.  They mean 
 {\it running width} and {\it gauge invariant width}, correspondingly . 
See Section \ref{Breit-Wigner}  for details. The value of this variables can
 be changed  by means of the {\it n\_comphep\_c} menu.

\newpage

\appendix
\addcontentsline{toc}{section}{Appendix}
\section*{Appendix}

\section{Self-check of the \CompHEP~ package} 
                   \label{symb-calc-check} 
 The   \CompHEP~ authors  have  invented some tools for  testing  the
program.
The positive result of these checks allows us to justify that \CompHEP~
works correctly. Our tests  relate to the symbolic result
level and have been realized with the help of the  \REDUCE~ \cite{REDUCE} 
symbolic manipulation system. All test routines are stored in the 
 {\it \$COMPHEP/test}  directory.  If the user is going to repeat
them he has to  copy the corresponding files  into his 
working directory.
All check commands must be started from the same directory.

\subsection {Check  of the   built-in symbolic calculator}

   The first check is a comparison of results 
produced by  the  \CompHEP~ symbolic calculator (Section \ref{s-output})
 with those by   \REDUCE~ 
evaluation of the corresponding code (Section \ref{r-code}). 
The  positive result of this
comparison    means that our   built-in  symbolic calculator works 
correctly.
  Note that the 
\REDUCE~ code may be viewed through by the user and  the 
\CompHEP~ algorithms for  evaluation of the squared matrix element 
 can be verified in this  way. 

The check is realized by means of the program {\it check.red} which must be 
started from within the \REDUCE~ session by instruction\\
\hspace*{2cm}{\tt in"check.red";}\\
It is assumed that the \REDUCE~ code for diagrams and the corresponding 
expressions evaluated by \CompHEP~ are stored in the {\it results}
directory in advance.  The results of this check are saved in the  
{\it message} file. It consists of a list of diagram  numbers  accompanied 
by the  labels {\it OK} or {\it Error} depending on the  result of comparison.

\subsection{Comparison of results produced  in two different gauges}

The comparison of results produced in  the unitary gauge and in the  
t'Hooft-Feynman gauge 
is a very impressive test of the package. In this way  we
check not only  the \CompHEP~ code for symbolic 
evaluation  but also the  correctness of model implementation. 

To perform this check we evaluate  the symbolic sum of  
diagrams in different versions of the Standard Model and compare 
results. The non-zero 
difference is a signal of mistake.
Symbolic summation is performed by  \REDUCE. This summation    is the
most difficult step of  comparison because the sum of diagrams
can be extremely cumbersome.

The \REDUCE~ program {\it cmp.red}  carries out a  summation of  symbolic 
answers written down in {\it results/symb1.red}  as well as   in  
{\it results\_/symb1.red}  and a comparison of  the sums.
In the case of zero difference  {\it cmp.red} puts down
 {\it OK} into the {\it message} file, otherwise the word {\it Error} 
appears.

\subsection{Automatic check for a set of processes} 

We have created the unix script commands {\tt cycle\_check} and
{\tt cycle\_cmp}  which perform the above checks automatically 
for some set of processes. The {\tt cycle\_check} needs one 
numerical parameter which indicates the number of model which is tested.
 The {\tt cycle\_cmp} needs two 
numerical parameters which indicate the numbers of models which are compared.

In both cases the  list of processes is read  from the standard 
input and  the generated {\it message} files are directed to the 
standard output device.

To simplify these tests we open a possibility to restrict the generated
diagrams by excluding   particles of the  third generation.
To realize this possibility just pass one addition parameter to 
{\tt cycle\_check} and  {\tt cycle\_cmp}. The value of this parameter is
 unessential.     
   
We have created various lists of Standard Model processes for testing. They are

\begin{tabular}{l    l  l}
 22\_0gen.prc  &  {\it Higgs and vector particles sector} &               \\
               &  {\it for 2\verb|->|2} processes         & 11 processes,  \\
 22\_2gen.prc  & {\it 2\verb|->|2   for two generations} & 142~processes, \\
 22\_3gen.prc  &  {\it 2\verb|->|2  for three generations}& 294~processes, \\
 23\_2gen.prc  &  {\it 2\verb|->|3  for two generations}  & 455~processes. \\
\end{tabular}

Each list contains only one representative of the  cross-symmetrical 
processes.

The following tests have been carried out for the current version:\\
\begin{verbatim}
    cycle_check 4      < 22_3gen.prc
    cycle_check 3      < 22_3gen.prc
    cycle_check 4   2g < 23_2gen.prc
    cycle_cmp   3 4    < 22_3gen.prc
\end{verbatim}

  \section{Ghost fields and the squared diagram technique for the
t'Hooft-Feynman gauge}
             \label{squaring}           
\subsection{The problem}

Every time  when we are  trying to create a model containing a massive 
vector particle we meet a problem caused by  
bad asymptotics of its propagator

\begin{equation}
   \frac{i}{  (2\pi)^4}\,  \frac{g_{\mu\nu} -k_\mu k_\nu/m^2}{m^2-k^2}\;.
\label{mass_vect_phys_prop}
\end{equation}

The $( g_{\mu\nu} - k_\mu k_\nu/m^2  )$ factor gives a projection on
 physical degrees of freedom in the polarization space. This term appears
because a 4-component vector
field is used to describe a particle with 3 degrees of freedom.  The $k_\mu
k_\nu/m^2$ term  leads to a  fast  growth of  amplitudes
at high energies, what breaks unitarity and  is not compatible with 
the renormalizability of  theory.

The problem mentioned above is  solved 
in the framework of  gauge field theories where the  gauge symmetry is 
responsible for  mutual cancellation of rapidly growing  contributions
of separate diagrams \cite{BD}.  
 Let  our model of particle interaction be based on a gauge
theory.  Then on a step of 
numerical evaluation  we  expect a  cancellation  of
contributions which come from various Feynman diagrams. Consequently,
finite precision numerical calculations may lead to wrong results.
   Accompanying problem
is a  cumbersome  expression for  each diagram as a result 
of appearance of mutually canceling  terms. 

At the same time  there is a freedom in formulation of  Feynman rules
for gauge theories caused by an  ambiguity of  gauge fixing terms \cite{BD}. 
 These terms   modify the quadratic 
part of the Lagrangian and consequently may improve the vector 
particle propagator.  	Indeed, in the case of t'Hooft-Feynman  gauge the 
propagator  of vector particle takes the form
\begin{equation}
   \frac{i}{ (2\pi)^4}  \frac{g_{\mu\nu}}{m^2 - k^2}\;.
\label{hf_ propagator}
\end{equation}
that  provides a formulation of the theory
where the problem of vector particle propagator is solved explicitly.

A price for this solution  is an appearance  of three 
additional unphysical particles in the model. They are 
a couple of Faddeev-Popov ghosts and one Goldstone ghost.
All of them have  scalar type propagators with the same mass $m$.
Opposite to (\ref{mass_vect_phys_prop}) the propagator (\ref{hf_ propagator}) 
does not vanish when it projected onto  the temporal polarization state
\begin{equation}
 e^0=k/m\;\;,
\label{temporal}
\end{equation}
that also leads to the  appearance of  additional unphysical state.
The  main principles of gauge invariance guarantee \cite{BD} that
an  expression for the amplitude should  be  the same for any  gauge if 
only physical incoming and outgoing states are considered.

Generally the t'Hooft-Feynman  gauge solves the problem of 
cancellations.  But while  calculating  
processes with incoming or outgoing massive vector particles, 
we meet a similar problem. Indeed, we need to multiply 
the diagram contributions  by polarization vectors. The polarization 
vectors $(e^1,e^2,e^3)$ constitute an orthonormal basis 
in the sub-space orthogonal to a momentum $k$.  Due to the relation
\begin{equation}
  e^1_{\mu} e^1_{\nu} + e^2_{\mu} e^2_{\nu} + e^3_{\mu} e^3_{\nu} =
 k_\mu k_\nu/m^2  -g_{\mu\nu}
\label{polar_sum} 
\end{equation}      
at  least one of the polarization vectors  has large (of order of $k$)
components for any 
choice of  polarization  basis.  Let  vector  $k$ have the 
components
\begin{equation}
   k=(\sqrt {m^2 +  p^2}, 0,0,p)\;.
\end{equation}
Then the polarization vectors can  be chosen as
\begin{eqnarray}
e^1=(0,1,0,0)\;;\\
e^2=(0,0,1,0)\;;\\
\label{longitudinal}
e^3=(p/m,0,0,\sqrt {1 +  p^2/m^2)}\;.
\end{eqnarray}

The first two vectors correspond to transversal polarizations and the third
one corresponds to  a longitudinal one. We see that the
longitudinal vector has large ( of order of  $k$) components.
It may imply a fast  increase of cross-sections of processes with the
longitudinal
polarizations at high energies or an appearance of cancellations between various 
diagrams. Indeed the second case is realized and,
hence, we have got a problem with diagram  cancellations.   
When evaluating  squared diagrams  we  get 
 cumbersome  expressions again as a result of  substitution of the projector
(\ref{polar_sum}).

\subsection{Incoming and outgoing ghosts}
A solution of the above problem   looks as a chess sacrifice. The 
main idea is to include  incoming and outgoing unphysical states
into the  consideration.
 In the  t'Hooft-Feynman  gauge the   masses of ghost partners
equal  the mass of  parent vector particle. Let us consider
ghost states as  unphysical polarizations alike the temporal one (\ref{temporal}).  
Note that the Faddeev-Popov ghost states and the  temporal
state have a negative norm, whereas the Goldstone state has a
positive norm \cite{BD}. So  the unphysical  polarizations  can give
 a positive as well as a negative  contribution to the  polarization sum.

The main statement is that a contribution of 
all unphysical polarizations to the sum of  squared matrix element over 
 polarizations   equals  zero \cite{LB}. As a result
\begin{equation}
\sum_{i \in S_{phys}} A_i A_i^{*} = \sum_{i \in S_{all}} \sigma(i)  A_i
A_i^{*}\;,
\label{phys.eq.full}
\end{equation}
where  $i$ is a multi-index for  polarization states;
$A_i$ is an amplitude of the process; $S_{phys}$ is a set of physical
polarization states; $S_{all}$ is a full set of polarizations including 
unphysical ones; $\sigma(i)=\pm 1 $ 
depending on a signature of the Hilbert space norm of the polarization 
state $i$.
  
A drawback  due to enlarging a set of polarization states 
is clear:  we have much more matrix element terms  for evaluation
and subsequent summation. To see an  advantage of this trick
let us sum the temporal (\ref{temporal})  and  
longitudinal (\ref{longitudinal}) polarization contributions. 
Note that both of them have components  of order of $k$, but for calculation 
of the corresponding sum we can  alter  the basis of polarization states
\begin{equation}
  e^0_{\mu} e^0_{\nu} - e^3_{\mu} e^3_{\nu} =   e'^0_{\mu} e'^0_{\nu} - e'^3_{\mu}
e'^3_{\nu}\;, 
\end{equation}
and in such a way to have new basis vectors of order
of unity  in spite of possible large value of the momentum $k$
\begin{eqnarray*}
e'^0=(1,0,0,0)\;;\\
e'^3=(0,0,0,1)\;.
\end{eqnarray*}
   In  other words, the  inclusion of the 
temporal polarization replaces the normalization condition (\ref{polar_sum})  
by
\begin{equation}
e^0_{\mu} e^0_{\nu}- e^1_{\mu} e^1_{\nu} - e^2_{\mu} e^2_{\nu} - e^3_{\mu} e^3_{\nu} =
g_{\mu\nu}\;.
\label{polar_sum_0}
\end{equation}
which does not lead to the  fatal requirement on a value of polarization
vectors.
In the case of unpolarized calculation  by the 
squared diagram technique we  substitute $g_{\mu\nu}$ for 
the polarization sum  according to (\ref{polar_sum_0}).  
But now we have to add   incoming and outgoing ghost states to the
polarization sum in order to compensate  the contribution of 
temporal polarization state. 


\subsection{Massless vector-particle case}

We have discussed above the massive vector particle case.
The Lagrangian of free  massless vector field  has got 
a gauge symmetry.  So we need anyway to ensure a gauge symmetry for the 
 interaction  model to support a compatibility with  the free field model.
In this case the Feynman gauge  leads to the 
propagator (\ref{hf_ propagator}) with  $m=0$  and  
to the appearance of  Faddeev-Popov ghosts \cite{BD}. However a Goldstone 
ghost does not appear and the longitudinal polarization becomes
unphysical like temporal one. Summation over physical
polarization states can be replaced by that over an
extended set of polarizations like (\ref{phys.eq.full}). See 
the corresponding proof in \cite{CL}.

 In  the framework of  amplitude technique there is no reason to include the
incoming and outgoing   ghost partners of massless vector particle  into 
consideration. 
Longitudinal polarization  becomes unphysical, but the transversal
polarization vectors may be  chosen of the order of unity.  On the contrary,
 in the case of squared diagram technique the extension of polarization states
 is  very useful and has been used in numerous calculations. 

If only physical polarization states are taken into account, then
for the squared diagram evaluation  we must convolute  free Lorentz 
indices, which appear after evaluation of the left and right parts of 
squared diagram, with the projector on physical sub-space. In the
massless  case this projector equals  \cite{CL}
$$  g_{\mu\nu}  - \frac{k_\mu \eta_\nu + k_\nu \eta_\mu}{(k.\eta)} \;,$$
where $\eta$ is an auxiliary vector with the zero Lorentz norm.
Due to the gauge invariance the  sum over  all diagrams does not depend on 
this  $\eta$,
but each squared diagram contribution contains it. This leads 
to cumbersome expressions of squared diagram contributions 
in comparison with  the case when the  unphysical  polarizations
are included into the sum.

\subsection{Summation of ghost diagrams  in \CompHEP}

\CompHEP~ uses the  squared diagram technique with summation over
polarizations. Basically  one squared diagram corresponds to the
\begin{equation}
\sum_{i \in S_{all}} A_i^k A_i^{*l}
\end{equation}  
contribution in a squared matrix element, where $A^k$ and $A^l$ 
are the amplitudes related to some Feynman diagrams.

\begin{figure}
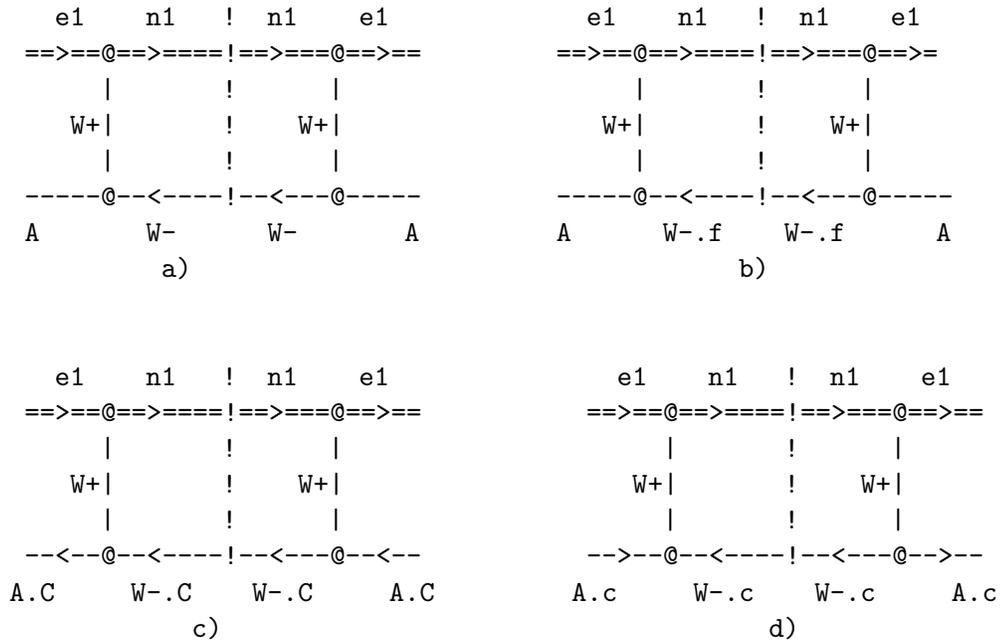

{\small
\begin{verbatim}
       e1    n1   !  n1    e1             e1    n1   !  n1    e1
     ==>==@==>====!==>===@==>==         ==>==@==>====!==>===@==>=
          |       !      |                   |       !      |
        W+|       !    W+|                 W+|       !    W+|
          |       !      |                   |       !      |
     -----@--<----!--<---@-----         -----@--<----!--<---@-----
     A       W-      W-       A         A      W-.f    W-.f      A
              a)                                    b)


       e1    n1   !  n1    e1               e1    n1   !  n1    e1
     ==>==@==>====!==>===@==>==           ==>==@==>====!==>===@==>==
          |       !      |                     |       !      |
        W+|       !    W+|                   W+|       !    W+|
          |       !      |                     |       !      |
     --<--@--<----!--<---@--<--           -->--@--<----!--<---@-->--
    A.C     W-.C    W-.C     A.C         A.c     W-.c    W-.c     A.c
                c)                                    d)
\end{verbatim}
}

\caption{Ghost diagrams}

\label{ghostDiagrams}
\end{figure}

If we follow an idea of the previous section and take into account 
the ghost incoming particles,  a number of squared diagrams 
increases significantly. For example, in  the simplest case of 
$e^-,\gamma \rightarrow  n_e, W^-$  process one squared diagram of
Fig.\ref{ghostDiagrams}(a) is accompanied by three ghost diagrams with 
the similar topology.  

Let us note that all diagrams of Fig.\ref{ghostDiagrams} have the same 
denominators. The  numerators of these diagrams are 
 polynomials of scalar products of momenta. The powers of these polynomials 
are the same for all diagrams, so one might expect that the  symbolic 
sum of all these diagrams has approximately the same size as one term. 

Following  this note \CompHEP~ calculates the  symbolic sum of all sets of 
diagrams which become identical after replacing the ghost particles by their
parents. 

\subsection{Gauge symmetry and cancellations} 
                      \label{cancellations} 

Cancellation of diagram contributions 
is an essential point  both  for  symbolic and  numerical processing,
because a relatively small variation  of one diagram contribution may lead  to 
a significant error. Such variation can be caused either by finite
precision of floating operations or by correction of Feynman rules, for
instance, by including particle widths into consideration, or  by 
removal of  some diagram subset. 
We would like to stress  again these obstacles  
to warn the user.
  
 There are two well known  examples of gauge cancellations.
The first one is the ultraviolet  cancellation of terms originating from
the propagators of massive vector particles.
This problem could  be resolved  by the calculation of squared matrix element
in the t'Hooft - Feynman gauge. 

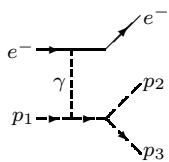
\begin{figure}[thb]
{\scriptsize
\linethickness{0.5pt}
\begin{center}
\begin{picture}(62,77)(0,0)
\put(17.5,57.1){\vector(1,0){2}}
\put(9.9,57.1){\makebox(0,0)[r]{$e^-$}}
\put(10.9,57.1){\line(1,0){13.0}}
\put(24.0,57.1){\line(1,0){13.0}}
\put(43.5,63.6){\vector(1,1){2}}
\put(51.1,70.2){\makebox(0,0)[l]{$e^-$}}
\put(37.0,57.1){\line(1,1){13.0}}
\put(21.9,44.0){\makebox(0,0)[r]{$\gamma$}}
\multiput(23.7,57.1)(0.0,-0.3){8}{\rule[-0.25pt]{0.5pt}{0.5pt}}
\multiput(23.7,53.1)(0.0,-0.3){8}{\rule[-0.25pt]{0.5pt}{0.5pt}}
\multiput(23.7,49.1)(0.0,-0.3){8}{\rule[-0.25pt]{0.5pt}{0.5pt}}
\multiput(23.7,45.1)(0.0,-0.3){8}{\rule[-0.25pt]{0.5pt}{0.5pt}}
\multiput(23.7,41.1)(0.0,-0.3){8}{\rule[-0.25pt]{0.5pt}{0.5pt}}
\multiput(23.7,37.1)(0.0,-0.3){8}{\rule[-0.25pt]{0.5pt}{0.5pt}}
\multiput(23.7,33.1)(0.0,-0.3){8}{\rule[-0.25pt]{0.5pt}{0.5pt}}
\put(17.5,30.9){\vector(1,0){2}}
\put(9.9,30.9){\makebox(0,0)[r]{$p_1$}}
\multiput(10.7,30.9)(0.3,0.0){12}{\rule[-0.25pt]{0.5pt}{0.5pt}}
\multiput(15.6,30.9)(0.3,0.0){12}{\rule[-0.25pt]{0.5pt}{0.5pt}}
\multiput(20.5,30.9)(0.3,0.0){12}{\rule[-0.25pt]{0.5pt}{0.5pt}}
\put(30.5,30.9){\vector(1,0){2}}
\multiput(23.7,30.9)(0.3,0.0){12}{\rule[-0.25pt]{0.5pt}{0.5pt}}
\multiput(28.6,30.9)(0.3,0.0){12}{\rule[-0.25pt]{0.5pt}{0.5pt}}
\multiput(33.5,30.9)(0.3,0.0){12}{\rule[-0.25pt]{0.5pt}{0.5pt}}
\put(51.1,44.0){\makebox(0,0)[l]{$p_2$}}
\multiput(36.7,30.9)(0.3,0.3){12}{\rule[-0.25pt]{0.5pt}{0.5pt}}
\multiput(41.6,35.8)(0.3,0.3){12}{\rule[-0.25pt]{0.5pt}{0.5pt}}
\multiput(46.5,40.7)(0.3,0.3){12}{\rule[-0.25pt]{0.5pt}{0.5pt}}
\put(43.5,24.4){\vector(1,-1){2}}
\put(51.1,17.8){\makebox(0,0)[l]{$p_3$}}
\multiput(36.7,30.9)(0.3,-0.3){12}{\rule[-0.25pt]{0.5pt}{0.5pt}}
\multiput(41.6,26.0)(0.3,-0.3){12}{\rule[-0.25pt]{0.5pt}{0.5pt}}
\multiput(46.5,21.1)(0.3,-0.3){12}{\rule[-0.25pt]{0.5pt}{0.5pt}}
\end{picture} \ 
\end{center}
}
\caption{Example of process with the $1/t^2$ pole cancellation.}
\label{t^2_cancel}
\end{figure}

The second example is the cancelation  of double pole $(t^{-2})$ 
terms  of  t-channel photon 
propagator. There is a wide class of processes where the  incoming 
electron goes out in the forward direction emitting a virtual  photon
like in Fig.\ref{t^2_cancel}. The corresponding 
diagrams have got  the $1/t$ pole, where $t$ is the squared momentum of the virtual
photon.  For the above kinematics   the photon appears very close to its
mass shell ($ t \approx 0$), hence
this configuration gives a large contribution to the cross section. 

For the squared  matrix element  we expect the $1/t^2$ pole, 
but it appears to be reduced 
up to $1/t$ pole  \cite{strfunWW} in the zero-electron-mass limit. 
This fact is caused by electro-magnetic U(1) gauge  invariance. 
If diagrams of Fig.\ref{t^2_cancel} type contribute to your process, we
 strongly recommend to 
 to set the {\it 'Gauge invariance'} switch 
{\it ON} (see Section \ref{Breit-Wigner}) to prevent the  gauge  symmetry
breaking by width terms.
Another way to solve this problem is the usage of  the Weizsaecker-Williams
approximation (see Section \ref{WW-approx}).

\section{Distribution functions and beam spectra}  
             \label{str-fun}            \subsection{\CTEq~  distribution functions}

There are several sets of  disribution functions   presented by CTEQ
collaboration which describe the densities of quarks and gluons in
(anti)proton. In \CompHEP~ we  have implemented the  
\CTEq\cite{strfunCTEQ} set. 
Apart from  the Feynman variable $x$  they  depends on  
the QCD-scale parameter $Q$.
  The available  range is  
\[ 
\begin{array}{lcl}
       10^{-5} & <  x  < & 1; \\ 
       1.6     & <  Q  < & 10000  \; \mbox{GeV}. 
\end{array} \]
\noindent 
For  points outside  these ranges  the  extrapolation is used and the
corresponding warning is  directed to the standard output file.

  The \CTEq~ set of functions is produced in the framework of the
next-to-leading calculations in the  standard $\overline{MS}$ scheme 
with $\alpha_s(M_Z) = 0.116$.
  See further explanations in \cite{strfunCTEQ}.

\subsection{\MRS~  structure functions}

 The  \MRS~ group also  presents several sets of proton structure
functions. Two of them are implemented in \CompHEP. They are
\MRS($A^\prime$) and \MRS($G$) \cite{strfunMRS}.
 
   In the $A^\prime$ case it is supposed that the sea quark and gluon
densities are parameterized at small $x$ by the $x^{-\lambda}$
function  where $\lambda$ is the same  for the gluon and for the sea
quarks. This suggestion is motivated by the QCD theory.
  The G set uses two different  $\lambda$ parameters  for the sea quarks
and for the gluon densities.  A fit of experimental data  for the G set is 
better.
     
  The next-to-leading order evaluation  and the $\overline{MS}$
factorization scheme
are  used for evaluation  of the $Q^2$ dependence of structure  function.
$\alpha_s(M_Z) = 0.113 \; [0.114]$ in the $A^\prime$ \ [$G$] case, correspondingly.

The available  range is
\[
\begin{array}{lcl}
                  10^{-5} & <  x   < & 1;  \\ 
                  5      & <  Q^2 < & 1310720  \;\mbox{GeV}^2. 
\end{array} \]
See complete explanation in \cite{strfunMRS}.

\subsection{Backscattered  photon spectrum}
    
This function describes the spectrum of photons scattered backward
from the interaction of laser light with the  high energy  electron
beam:

\[ f(x) = \left\{ 
\begin{array}{ll}
0,                                             & \mbox{for~} x>x_{max} \\
N(1-x+1/(1-x) \,(1-4x/x_0 \, (1-x/(x_0 (1-x)) \,))), & \mbox{for~} 0<x<x_{max}
\end{array} \right. \]
\noindent
where  $x_0=4.82, \; x_{max}=x_0/(1+x_0), \; N$ is a normalization factor.

The above spectrum  corresponds to the special initial condition  when 
unpolarized photons are created.
See \cite{strfunLaser} for more details.

\subsection{Weizsaecker-Williams approximation}
\label{WW-approx}
  Weizsaecker-Williams approximation is used to describe  processes 
of electro-production in the case of  small angle of charged particle 
  scattering.
 In this case the  virtual photon emitted by the scattering particle  
appears near to the mass shell (see Fig.\ref{t^2_cancel}). It gives a 
possibility to reduce the process of
electro-production to the photo-production one with an appropriate photon 
spectrum:
$$ f(x) = (q^2\,\alpha/(2 \pi))
     (\log((1-x)/(x^2 \delta)) (1+(1-x)^2)/x -2(1-x-\delta \, x^2)/x),$$
where $\alpha$ is the  fine structure constant,
$q$ is a charge of incoming particle,
$m$ is its mass,  
$\delta = (m/Q_{max})^2.$
$Q_{max}$  sets out the  region of  photon virtuality $(P^2 >-Q_{max}^2)$ 
 which  contributes to  the process.
It is assumed  that region of large virtuality can be 
taken into account by direct calculation of electro-production.
As a rule this contribution is small enough.  

Parameters $q$, $m$, and $Q_{max}$ are defined by the user. 
See \cite{strfunWW} for the further explanations. In the case of 
  \CompHEP~  the Weizsaecker-Williams photon spectrum is 
available for charged leptons only.

\subsection{ISR and Beamstrahlung}
ISR (Initial State Radiation) is a process of photon radiation  by the
incoming electron due to its interaction with other collision particle.
 The resulting spectrum of  electron has been
calculated by Kuraev and Fadin \cite{strfunISR1}.
 In \CompHEP~ we realize the similar expression by Jadach, Skrzypek, and Ward
\cite{strfunISR2}:
\[
\begin{array}{ll}
F(x) = & \exp(\beta (3/4-Euler)) \beta (1-x)^{\beta-1}  \\ 
 & ((1+x^2)-\beta ((1+3 x^2) \ln(x)/2 + (1-x)^2)/2) /(2 \Gamma(1+\beta)),
\end{array} \] 
\noindent
where 
\[
 \begin{array}{lcl}
\alpha &=&1/137.0359895 \;\;\mbox{is the fine structure constant;} \\
\beta&=& \alpha (2 \ln(SCALE/m)-1) / pi   \\
  m    &=&0.00051099906 \;\;\;\;\;\;\mbox{is the electron mass;} \\
Euler  &=&0.5772156649 \;\;\;\;\;\;\;\;\mbox{is the Euler constant;} \\
\Gamma()  & \mbox{is} &        \mbox{the gamma function;} \\
SCALE   & \mbox{is} &          \mbox{the energy scale of reaction.}
\end{array} \] 
In the Kuraev and Fadin article  the parameter  SCALE equals to  the total 
energy of the process because  they 
considered the process of direct $e^+ e^-$ annihilation. In order to 
apply this structure function to another  processes we 
provide the user with a possibility  to define this parameter. 

Beamstrahlung is a  process of energy loss  by the
incoming electron due to its interaction with the  electron (positron)
bunch  moving in the opposite direction.    
The effective  energy spectrum of electron can be  described by 
the following function  \cite{strfunBeam}
$$
F(x)=\frac{1}{N_{cl}} [(1-E^{-N_{cl}})\,  \delta(1-x) + 
     \frac{\exp(-\eta(x))}{1-x}\,  h(\eta(x)^{1/3}, N_{cl}) ]\;\;\;,                  
$$
where 
\begin{eqnarray*}
\eta(x) &=& 2 /(3 \Upsilon) \, (1/x -1) \;\;\;, \\
h(z, N_{cl}) &=& \sum_{n \geq 1} \frac{z^n}{n! \Gamma(n/3)} \gamma(n+1, N_{cl})
\;\;\;,
\end{eqnarray*} 
and $\gamma$ is the incomplete  gamma function.

Function $F(x)$ depends on two parameters, $N_{cl}$ and $\Upsilon$,
which in their turn are determined by a  bunch  design:
\begin{eqnarray*}
 N_{cl}& =&  \frac{25\, \alpha^2 N}{12\, m  (\sigma_x+\sigma_y)}\;\;\;,\\
 \Upsilon &=& \frac{5\,\alpha\, N\, E}{6\, m^3 \sigma_z (\sigma_x+\sigma_y)\;,}
\end{eqnarray*}
where 
\[
\begin{array}{lcl}
   N & & \mbox{is number particles in the bunch,} \\
   \sigma_x,\sigma_y,\sigma_z & & \mbox{are sizes of bunch,} \\
   E & & \mbox{is a center-of-mass momentum.}
\end{array} \]

The Beamstrahlung spectrum  cannot be integrated by the current \CompHEP~ 
version  because it contains a $\delta$-function. 
Instead of it we provide the user with a possibility to integrate the squared 
matrix element with a convolution of Beamstrahlung and ISR  spectra.

 \section{Monte Carlo phase space integration}
             \label{Monte-Carlo}
  \subsection{Adaptive Monte Carlo integration package \VEGAS}
             \label{vegas}            
This section contains a short description of the adaptive Monte Carlo program
VEGAS. See for details \cite{Lepage, NumRecFORT}.


 The Monte Carlo method reduces a task of integral evaluation to 
the task of  mean value  calculation.  Let $g(x)$ is a density 
function satisfying  

      $$ \int \! g(x) \,dx = 1,$$
then 
      $$ \int \! f(x) \, dx = \int \! f(x)/g(x) \, \,g(x) \, dx = \;
 <\!\!f/g\!\!> \; = \lim_{ N \to \infty} \sum (f(x_i)/g(x_i)) / N, $$
where points $x_i$ are sampled with the  probability density $g(x) \, dx.$

 The uncertainty $\sigma_N$  of $<\!f/g\!>$ estimation  by $N$ sample points 
is  proportional to  square root of function's  variance divided over $N$:  
   $$\sigma_N =  \sqrt{ (<\!\!(f/g)^2\!\!> - <\!\!f/g\!\!>^2)  / N }\;. $$
VEGAS uses two techniques which allow to decrease the uncertainty of Monte
Carlo calculation, namely the {\it importance sampling}  and the 
{\it stratified sampling}.

\subsubsection{Importance sampling}

The idea of importance sampling technique is based on diminution   of variance by  
a proper choice of the density function  g(x). The general solution of this
problem could be in choosing 
      $$ g(x) = |f(x)| \; / \int \! |f(x)| \,dx.$$
      
However this solution is useless because  it returns us to the problem
of evaluation of $f(x)$  integral   and requires a generation of sampling
points for complicated density function.  
  
  To bypass these problems  VEGAS seeks  this function 
in the factored form 
     $$g(x_1, x_2, \ldots ,x_n) = g_1(x_1) \, g_2(x_2)  \ldots g_n(x_n).$$
The optimal functions $g_i(x)$ could be easily evaluated in terms of $f(x)$
\cite{Lepage, NumRecFORT}.
VEGAS is an adaptive program. For the first iteration it puts $g_i(x)=1.$
The information about $f(x)$ which VEGAS gets during the iteration 
is used to refine the density function. Generally VEGAS performs several
iterations improving the density function after each of them.

 The following parameters manage VEGAS work:
 \begin{enumerate}
\item {\it Itmx}  is a  number of iterations;
\item {\it Ncall} is a number of integrand calls for one iteration.
\end{enumerate}

\subsubsection{Stratified sampling}
 
The idea of stratified sampling method is to divide a volume of 
integration into a large number of sub-volumes and calculate integrals
separately in each sub-volume. This method produces a smaller uncertainty
comparing with  the  direct Monte Carlo method  because here the uncertainty 
is  caused only by a function  variance in the  sub-volumes, while 
the integrand variation from one sub-volume to another  does not 
contribute to the uncertainty. 

  The stratified sampling method is used to estimate the integral for 
any VEGAS iteration. The larger number Ncall is chosen, the smaller
size of sub-volume becomes available and, consequently, the more
successfully the stratified sampling works.


  \subsection{Parameterization of multi-particle phase space} 
             \label{kinematics}       
\subsubsection{Parameterization via  decay scheme}

The element of phase space volume  for a $n$-particle  state
 is equal to   \cite{ParticlesFields}
\begin{equation}
d\Gamma_n(q) = (2\pi)^4
\delta^4(q-p_1-p_2-p_3-...-p_n)\prod_{i=1}^{n}\frac{\delta(p_i^2-m_i^2)}
{(2\pi)^3}d^4p_i\;.
\label{phaseSpace}
\end{equation}
The same expression is valid for both  the decay of unstable particle with
momentum $q$ and the interaction of two particles with momenta $q_1$ and 
$q_2$  such that $q_1+q_2=q$.
For further discussion we need a designation for a 
phase space volume of some subset $S$ of the full $n$-particle set.
According to (\ref{phaseSpace}) 
\begin{equation}
d\Gamma(q,S)= (2\pi)^4
\delta^4(q-  \sum_{i \in S}p_i)\prod_{i \in S}\frac{\delta(p_i^2-m_i^2)}
{(2\pi)^3}d^4p_i\;.
\end{equation}

Let $S_1$ and $S_2$ be two disjoint particle subsets, then 

\begin{eqnarray}
\lefteqn{d\Gamma(q,S_1\cup S_2) =} \nonumber\\ &&\int ds_1ds_2\Bigl((2\pi)^4
\delta^4(q- q_1 -q_2 )  \frac{\delta(q_1^2-s_1)}{(2\pi)^3}d^4q_1
\frac{\delta(q_2^2-s_2)}{(2\pi)^3}d^4q_2\Bigr)  \nonumber \\ &&
\times \frac{d\Gamma(q,S_1)}{2\pi}\times\frac{ d\Gamma(q,S_2)}{2\pi}\;.
\label{recursion}
\end{eqnarray}

The above formula  expresses a   multi-particle volume 
in terms of  two-particle one,  the volumes $d\Gamma(q_1,S_1)$ and
$d\Gamma(q_2,S_2)$  with a  reduced 
number of particles, and the virtual squared masses  $s_1,\;s_2$ of clusters 
$ S_1,\;S_2$. 

Recursive application of this formula
allows one to   express the  multi-particle phase space in terms 
of two-particle phase space. In its turn the 
two-particle  phase space is explicitly  described by 
spherical angle $\Omega$ of motion  of the first decaying particle in the rest 
frame of initial state \cite{ParticlesFields}. 
\begin{equation}
\frac{d\Gamma(q,[1,2])}{2\pi}=\frac{k d\Omega}{4(2\pi)^3 \sqrt{q^2}}\;,
\label{1_2}
\end{equation}
where  $k$ is the absolute  value of three-dimensional 
momentum of outgoing particles in the rest frame.
Thus,  applying  recursively (\ref{recursion}) and (\ref{1_2})  to
(\ref{phaseSpace}) we obtain an explicit expression for the phase space
volume  in terms of the  squared  masses  $s_j$ of virtual  clusters and
the two-dimensional
spherical angles  $\Omega_j$, where $j$ is an ordinal number of decay:
\begin{equation}
d\Gamma _n(q) =   
\frac{k_1 d^2\Omega_1 }{4(2\pi)^2 \sqrt{q^2}}
\prod_{j=2}^{n-1} \frac{k_j d^2\Omega_j  }{4(2\pi)^3 \sqrt{s_j}}\;.
\prod_{j=2}^{n-1} ds_j 
\label{phaseSpace2}
\end{equation}
Here  $k_j$ is a  momentum of  outgoing clusters 
produced by decay of the $j^{th}$ cluster in its 
 center-of-mass.

The expression (\ref{phaseSpace2}) means some sequential \verb|1->2| decay 
scheme which starts from   incoming state  and finishes with outgoing 
particles of the process. For example, the integration domain for 
 $s_j$ parameters  depends on this scheme. Below we present 
two such schemes for a process with four outgoing particles: 

\begin{picture}(120,120)
\put(0,60){q} 
\put(5,60){\line(1,0){10}}
\put(22,55){$\Omega_1,k_1$}
\put(15,60){\line(1,1){30}} 
\put(18,75){$s_2$}  
\put(52,85){$\Omega_2,k_2$}
\put(45,90){\line(1,1){15}} \put(63,103){$p_1$}
\put(45,90){\line(1,-1){15}} \put(63,73){$p_2$}

\put(15,60){\line(1,-1){30}}
\put(20,40){$s_3$}
\put(52,25){$\Omega_3,k_3$}
\put(45,30){\line(1,-1){15}}  \put(63,43){$p_3$}
\put(45,30){\line(1,1){15}}    \put(63,13){$p_4$}

\end{picture}
~~~~~~~~
\begin{picture}(120,120)
\put(0,90){q} 
\put(5,90){\line(1,0){10}}  \put(20,87){$\Omega_1,k_1$}

\put(15,90){\line(1,1){15}}  \put(33,103){$p_1$}

\put(15,90){\line(1,-2){15}} \put(35,57){$\Omega_2,k_2$}

\put(10,75){$s_2$} 
\put(30,60){\line(1,1){15}}  \put(48,73){$p_2$}

\put(30,60){\line(1,-2){15}} \put(50,27){$\Omega_3,k_3$}

\put(25,45){$s_3$} 
\put(45,30){\line(1,1){15}}   \put(63,43){$p_3$}

\put(45,30){\line(1,-1){15}}   \put(63,13){$p_4$}

\end{picture}

  In the case of \CompHEP~ project  such  decay scheme
is defined by the user via the `Kinematics' menu (see Section \ref{Kinematics}).

\subsubsection{Polar vectors}

To complete  phase space parameterization we must 
fix a polar coordinate system  choosing the polar and
the azimuthal  angles for each of decays
\begin{equation}
 d^2\Omega_j=d\cos{\Theta_j}d\Phi_j
 \label{polarCoordinate} 
\end{equation}

We have  an ambiguity in the choice of  polar coordinate.  Let us remind 
that  our goal is not only  parameterization of phase space but also 
 regularization of the squared matrix element in  the 
phase space manifold.  The main idea of such regularization is 
a  cancellation of integrand sharp peaks  by the  phase space measure.
Originally the  phase space measure (\ref{phaseSpace2})  has no 
 cancellation factors, but we can create them by  means of a Jacobian of
transformed variables. To get an appropriate Jacobian  we need to have
the initial phase space variables  related  to   poles  of the
 squared matrix element. 

   In their  turn the poles of  squared matrix element are
caused by virtual particle propagators and  generally  have  
one of the   forms (\ref{pole_1}), (\ref{pole_2}) or (\ref{pole_w})
 (Section \ref{regul_menu})
 depending on  a  squared sum of momenta. Variables $s_j$ in
(\ref{phaseSpace2}) are also equal  to  squared sums of momenta. 
So, the parameterization (\ref{phaseSpace2}) allows us to 
 smooth some peaks of the matrix element.

It  appears to be that the  polar coordinates can be chosen in such a way 
that all  $cos{\Theta_j}$  have  simple linear 
relations to the  squared sums of momenta \cite{kinemat, AIHENP5}.
 The polar angle $\Theta_j$ can  be  unambiguously fixed by  the
{\it polar vector} $Pole_j$  whose space components 
in the rest frame of decay  correspond to the 
$\Theta_j=0$ direction. 
Let $q_{j1}$  and $q_{j2}$ be the   momenta
of the first and the second clusters produced by the $j^{th}$ decay. 
Then 
\begin{eqnarray*}
(Pole_j+q_{j1})^2= (Pole_j^0+q_{j1}^0)^2 - \mid \overline{Pole}_j\mid ^2 -
  \mid\overline{q}_{j1}\mid ^2 - 2cos{\Theta_j} \mid
\overline{Pole}_j  
\mid \mid \overline{q}_{j1}  \mid  \nonumber \\   
(Pole_j+q_{j2})^2= (Pole_j^0+q_{j2}^0)^2 -  \mid \overline{Pole}_j\mid ^2 -
  \mid\overline{q}_{j2}\mid ^2 + 2cos{\Theta_j} \mid
\overline{Pole}_j 
\mid \mid \overline{q}_{j2}  \mid 
\end{eqnarray*}

Thus, in order to get    $cos{\Theta_j}$
 related to  a squared sum of some particle momenta we may 
construct the polar vector as a sum of particle   momenta
\cite{kinemat,AIHENP5}.

For the  non-contradictory construction  
we need to  set  the decays  in some order with a natural requirement 
that the sub-decays of clusters produced by the $j^{th}$ decay  have the ordinal 
numbers  larger than $j$. In giving such ordering we can 
  construct a polar vector  for each decay 
based on the incoming momenta and on those of particles produced by 
decays possessing smaller ordinal numbers.

The following statements can be proved.   
{\it
In the framework of any ordered  scheme of decays and for any sum $P$ 
of particle  momenta one can find the decay  number  $j$ such that 
either  $P^2=s_j$   or  $P$ might  be represented as  $Pole_j + q_j$, 
where   $q_j$ is the momentum of one of the clusters  in the  $j^{th}$ 
decay and   $Pole_j$ is a polar vector constructed according to the above
rule.
}
In other words, any of poles (\ref{pole_1}), (\ref{pole_2}), (\ref{pole_w}) 
can be expressed   either in terms  of  $s_j$ parameters or in terms some of  $cos{\Theta_j}$ for
an  appropriate choice of the polar vector \cite{kinemat,AIHENP5}.

In  \CompHEP~  the  ordering is arranged automatically, so
that   all sub-decays of the first cluster 
have smaller numbers than those of the second cluster.
Polar vectors are also constructed automatically  according to the
list of peaks prepared by the user.

\subsubsection{Smoothing}

The general idea of the integrand smoothing is trivial.
Let us need to evaluate 
\begin{equation}
\int_a^b F(x)dx \;\;\;,
\label{integral} 
\end{equation}
and let $F(x)$ have a peak like  $f(x)$,  where $f(x)$    is a simple
symbolically integrable function  in contrast to 
 $F(x)$:
\begin{equation}
g(x)=\int_a^x f(x') dx'\;.
\end{equation} 

Now we may represent the integral (\ref{integral}) as 
\begin{equation}
\int_a^b F(x)dx = \int_0^{g(b)}dy\frac{F(g^{-1}(y))}{f(g^{-1}(y))},  
\end{equation}
where $g^{-1}(y)$ is the inverse  function for   $g(x)$. The 
integrand is a smooth function now.

We face very often  squared matrix elements which have several 
poles in one of variables. For example, the
$\gamma \rightarrow b,\bar{b}$, $Z \rightarrow b,\bar{b}$ and
$H \rightarrow b,\bar{b}$ virtual  subprocesses may contribute just to 
the same amplitude.
Although in this case we can evaluate  the  integral function  $g(x)$
symbolically,  the inverse  function $g^{-1}(y)$ can be computed
only as a  numerical solution of the corresponding equation.
To bypass   the calculation of inverse function  
\CompHEP~  uses the   multi-channel  Monte Carlo  
(branching) method to  smooth a sum of peaks.

   The idea of the branching method is the following. Let  
$F(x)$ have two peaks, one is similar to $f_1(x)$ and
another to $f_2(x)$.
 $f_1(x)$ and  $f_2(x)$  are singular but elementary functions.
Then, instead of 
one integration (\ref{integral}),  we could  perform two ones:
\begin{equation}
\int F(x)dx=
 \int \frac{F(x)f_1(x)}{f_1(x)+f_2(x)} dx + 
\int \frac{F(x)f_2(x)}{f_1(x)+f_2(x)} dx\;, 
\label{branching}
\end{equation}
 but now each integration  has only a single peak! It is easy to
extend this method for an arbitrary number of peaks. 

The branching method was used in \cite{excalibur} to separate  peaks
which came from various diagrams. In that paper there was also  proposed
to use the expression (\ref{branching}) where  $f_i(x)$ is replaced by 
$\alpha_i f_i(x)$  with a subsequent 
search for optimal coefficients $\alpha_i$. 
\CompHEP~ passes on this weight optimization to \VEGAS,
combining two integrals in one \VEGAS~ hypercube.

As was mentioned above, \CompHEP~ automatically  searches for a  polar 
vector for each angle integration in order to  reach a  linear relation
between   $cos\Theta$  and one of the squared sum of
momenta which is responsible for the peak. 
It could happen that various peaks  need different polar vectors 
for the same decay. In this case \CompHEP~ uses the branching method again,
but now for the whole two-dimension sphere integration.  In other words,
we use the branching equation (\ref{branching})  where  $x$ is the
two dimensional sphere angle \cite{kinemat,AIHENP5}.

 \section{Lagrangian of the Standard Model}
 \label{theory}  \subsection{Gauge theories}
\paragraph{Group.} Group is defined by its structure constants 
$ f^{\alpha}_{\beta \gamma} $ which are anti-symmetric 
$( f^{\alpha}_{\beta \gamma} = -f^{\alpha}_{\gamma \beta})$
and obey the Jacoby  identity:
$$  f^{\delta}_{\alpha \epsilon}f^{\epsilon}_{\beta \gamma} +
f^{\delta}_{\gamma \epsilon}f^{\epsilon}_{\alpha \beta } +
f^{\delta}_{\beta \epsilon}f^{\epsilon}_{\gamma \alpha} = 0\;. $$

Group generators are Hermitian matrices   $\hat\tau_{\alpha}$ which 
satisfy the commutation relations:
$$\hat\tau_{\alpha}\hat\tau_{\beta} - \hat\tau_{\beta}\hat\tau_{\alpha}=
 i\,f^{\gamma}_{\alpha \beta} \hat\tau_{\gamma}\;. $$ 
In particular the generators in the   adjoint representation are
$$ (\hat\tau_{\alpha})^i_j = -i\,f^i_{j\alpha}\;. $$  
Group transformation may be represented with the help of group 
generators as
   $$ \hat g(w)=exp(i\hat\tau_{\alpha}\omega^{\alpha})\;. $$
We   assume  that the Killing  metric is orthonormal: 
$$ -\, \frac{1}{2} f^{\epsilon}_{\gamma \alpha} f^{\gamma}_{\epsilon
\beta} = \delta_{\alpha \beta}\;. $$
This metric allows one to raise and lower the group indices. 
In the case of  orthonormal Killing metric the structure constants
are fully antisymmetric under interchange of any pair of indices.

\paragraph{Local gauge invariance.} It is an invariance of Lagrangian
under group     transformations $g(x)$ which depend on a point in the space-time 
manifold. Fields in such theories are divided into two classes: 
matter fields and gauge  fields. The gauge fields $A^{\alpha}_{\mu}(x)$
are  vector ones. The number of such fields is equal 
to the number of group generators. The matter fields $\psi^i(x)$  can  have 
an arbitrary 
Lorentz  structure. Their internal components are transformed
according to some representation of the group. Let $D(\omega)$ be an 
operator  which performs an  infinitesimal transformation of fields under 
the local gauge transformation. For the matter fields we have
$$ \hat D(\omega)\psi(x)= i\,\omega^{\alpha}(x){\hat \tau}_
{\alpha}\psi(x)\;. $$
For  gauge fields the local gauge transformations  are defined by 
$$ (\hat D(\omega) A_{\mu})^{\alpha}(x)= 
f^{\alpha}_{\beta \gamma} A^{\beta}_{\mu}(x) \omega^{\gamma}(x) 
 + \partial_{\mu} \omega^{\alpha}(x)\;.$$
The following expressions, namely a  covariant derivative and a gauge field 
tension, are used to construct a local invariant Lagrangian:
$$  \nabla_{\mu}\psi(x) = \partial_{\mu} \psi(x) - \hat D(A_{\mu})\psi(x) =
 \partial_{\mu} \psi(x) - iA^{\alpha}_{\mu}(x){\hat \tau}_{\alpha}\psi(x)\;; $$
$$ F^{\alpha}_{\mu \nu}(x) = \partial_{\mu}A^{\alpha}_{\nu}(x) 
-\partial_{\nu}A^{\alpha}_{\mu}(x) + 
f^{\alpha}_{\beta\gamma}A^{\beta}_{\mu}(x)A^{\gamma}_{\nu}(x)\;. $$
It can be proved that 
$$ \hat D(\omega)[\nabla_{\mu}\psi(x)]= \hat \tau_{\alpha}\omega^{\alpha}(x)
\nabla_{\mu}\psi(x)\;; $$
$$ (\hat D(\omega) F_{\mu \nu})^{\alpha}(x)= f^{\alpha}_{\beta \gamma}
F^{\beta}_{\mu \nu}(x)\omega^{\gamma}(x)\;.$$
In these terms the Lagrangian of gauge theory is defined by the following 
expression \cite{BD}
\begin{equation} 
 L = -\, \frac{1}{4g^2} {F^{\alpha}}_{\mu \nu} {F_{\alpha}}^{\mu \nu} 
+ L_m( \nabla_{\mu} \psi, \psi)
\label{physLag}
\end{equation}
where $g$ is the coupling constant and $L_m(\partial_{\mu}\psi,\psi)$ is some 
Lagrangian of the matter fields which is invariant under the global gauge transformations.

\paragraph{Gauge fixing terms.}
In order to quantize the gauge theory  one must add  to 
 (\ref{physLag})   gauge fixing term  and the corresponding Faddeev-Popov 
term. The first term breaks the gauge symmetry and in this way removes the 
divergence of the functional  integral. The second term improves the integration 
measure to provide  correct predictions for gauge invariant
observables.

The general form of the gauge fixing term is
\begin{equation}
 L_{FG}(x)= -\, \frac{1}{2}\sum_{\alpha} (\Phi^{\alpha}(x))^2\;. 
\label{gfLag}
\end{equation}
The corresponding Faddeev-Popov term is
\begin{equation}
 L_{FP}= - \bar{c}_{\alpha}(x) (D(c)[\Phi^{\alpha}])(x) \;,
\label{fpLag}
\end{equation}
where $c^{\alpha}(x)$ and $ \bar{c}_{\alpha}(x)$ are the auxiliary  
anti-commutative fields. They are called the Faddeev-Popov ghosts.
Note that we may multiply  (\ref{fpLag}) by an  arbitrary factor which 
can be hidden in the definition of ghost fields. As a rule
it is chosen in such a way to provide a convenient form of the  ghost propagator.

The well-known   choice of the gauge fixing terms, the Feynman-like gauge,  is 
$$ L_{GF}= -\,\frac{1}{2g^2}\sum_{\alpha} 
(\partial^{\mu} A^{\alpha}_{\mu} +...)^2\;.$$   
In this case the quadratic  part of gauge field  Lagrangian takes  
the simplest form
\begin{equation}
  -\,\frac{1}{2g^2} \partial_{\mu}A_{\nu}\partial^{\mu}A^{\nu}\;. 
\label{FeynmQ} 
\end{equation} 
The corresponding Faddeev-Popov Lagrangian is

$$ L_{FP}= - \bar{c}_{\alpha}(x) (\Box c^{\alpha}(x) +
 \partial^{\mu}( f^{\alpha}_{\beta \gamma} A^{\beta}_{\mu}(x)
c^{\gamma}(x)) + ...)\;. $$
The normalization of $L_{FP}$ is chosen  to have
the Faddeev-Popov ghost propagator equal to the propagator of scalar
particle:
$$T(\bar{c}_{\alpha}(p),c^{\beta}(q))= \delta_{\alpha}^{\beta} \delta(p+q)
\frac{1}{i(2\pi)^4}\;\frac{1}{-p^2}\;. $$

\paragraph{Normalization.}
For the purposes of applying the  perturbation theory  
the  gauge field is rescaled by substitution  
$$ A  \rightarrow A/g\;. $$
In this way the coupling constant $g$ leaves the quadratic part of 
the Lagrangian (\ref{physLag}) and appears in the  interaction terms. 

\subsection{QCD Lagrangian}
QCD is a gauge theory based on the  $SU(3)$  group. The corresponding 
 gauge field  $G^{\alpha}_{\mu}(x)$ is called a {\it gluon}.  
The matter fields $q_k(x)$ are {\it quarks}.
They  are triplets and are transformed  according to the fundamental 
representation. Index $k$  enumerates a sort of quarks. 

QCD Lagrangian in the Feynman  gauge is written down 
following  the rules (\ref{physLag}), (\ref{gfLag}), (\ref{fpLag}).

\begin{equation} 
 L_{phys} = -\, \frac{1}{4} {F^{\alpha}}_{\mu \nu}(x) {F_{\alpha}}^{\mu
\nu}(x) +  \sum_k \frac{i}{2}( \bar{q}_k(x)  \gamma^{\mu}\nabla_{\mu}q_k(x) -
\nabla_{\mu}\bar{q}_k(x)\gamma^{\mu}q_k(x) )\;.
\label{physLagQCD}
\end{equation}
where 
$$  \nabla_{\mu}q(x) = \partial_{\mu} q(x) - i\,g\, G^{\alpha}_{\mu}(x) 
\hat{t}_{\alpha}q(x)\;;  $$
$$ F^{\alpha}_{\mu \nu}(x) = \partial_{\mu}G^{\alpha}_{\nu}(x)
-\partial_{\nu}G^{\alpha}_{\mu}(x) +
g\,f^{\alpha}_{\beta\gamma}G^{\beta}_{\mu}(x)G^{\gamma}_{\nu}(x)\;; $$

\begin{equation}
L_{GF} = -\,\frac{1}{2} \sum_{\alpha}(\partial^{\mu} G^{\alpha}_{\mu})^2 \;;
\label{gfLagQCD}
\end{equation}

\begin{equation}
 L_{FP}= - \bar{c}_{\alpha}(x) (\Box c^{\alpha}(x) + g\,
 \partial^{\mu}( f^{\alpha}_{\beta \gamma} G^{\beta}_{\mu}(x)
c^{\gamma}(x)))\;,
\label{fpLagQCD}
\end{equation}
$ c^{\alpha}(x)$ are Faddeev-Popov ghosts,  $g$ is coupling constant, 
$ f^{\alpha}_{\beta \gamma} $ are the  $SU(3)$ structure constants,
$t_{\alpha}$ are the   generators in the fundamental representation\footnote{
$ t_{\alpha}$ are equal to   $ \lambda_{\alpha}/2$ where  $
\lambda_{\alpha}$ are the Gell-Mann matrices.}.

\subsection{Lagrangian of electroweak interactions}

\subsubsection{Vector bosons}
Gauge theory of electroweak interactions is based on the $SU(2)\times U(1)$ group.
So, we have a triplet of $SU(2)$ vector fields $W^{\alpha}_\mu(x)$ and a single 
vector field $B_\mu(x)$. The $SU(2)$ structure constants are
presented by the absolutely antisymmetric tensor
$\epsilon^{\alpha\beta\gamma}$. The  Lagrangian of gauge fields is 
written down    according to (\ref{physLag}):
\begin{equation}
\label{SU2U1}
 L_{GF} = -\, \frac{1}{4} {FW^{\alpha}}_{\mu \nu}(x) {FW_{\alpha}}^{\mu \nu}(x)
     -\, \frac{1}{4} FB_{\mu \nu}(x) FB^{\mu \nu}(x)\;,
\end{equation}
where  
\begin{eqnarray*}
FW^{\alpha}_{\mu \nu}(x)& =& \partial_{\mu}W^{\alpha}_{\nu}(x)
-\partial_{\nu}W^{\alpha}_{\mu}(x) + g_2
\epsilon^{\alpha\beta\gamma}W^{\beta}_{\mu}(x)W^{\gamma}_{\nu}(x)
\;;\\
 FB_{\mu \nu}(x)& =& \partial_{\mu}B_{\nu}(x) - \partial_{\nu}B_{\mu}(x)
\;;
\end{eqnarray*}
$g_2$ is a coupling constant for the $SU(2)$ gauge interaction.

 Infinitesimal local gauge transformations are defined as follows
\begin{eqnarray}
(\hat D(w,b) W_{\mu})^{\alpha}(x) &=& 
g_2 \epsilon^{\alpha\beta \gamma} W^{\beta}_{\mu}(x) w^{\gamma}(x) 
 + \partial_{\mu} w^{\alpha}(x) \;;\nonumber \\
\label{SU2U1tranf}
  D(w,b) B_{\mu}(x)&=&\partial_{\mu} b(x) \;.
\end{eqnarray}

Let us express $W^1_\mu$ and $W^2_\mu$ 
in terms of a mutually conjugated couple  $W^+_\mu$ and $W^-_\mu$ 
\begin{eqnarray*}
 W^1_\mu&=&(W^-_\mu+\,W^+_\mu)/\sqrt{2}\;;  \\
 W^2_\mu&=& (W^-_\mu-\,W^+_\mu)/(i \sqrt{2})\;.
\end{eqnarray*}
Thus, the Lagrangian of  self-interaction for the  $SU(2)$ gauge fields  
in term of $W^+$ and $W^-$  has the form:
\begin{eqnarray}
i\,g_2 \left(\partial_\mu W^3_\nu(W^+_\mu W^-_\nu - W^-_\mu W^+_\nu)
              +\partial_\mu W^+_\nu(W^-_\mu W^3_\nu - W^3_\mu W^-_\nu) \right.
&&\nonumber \\
    \left.     -\partial_\mu W^-_\nu(W^+_\mu W^3_\nu - W^3_\mu W^+_\nu) \right)
&&\nonumber \\
    +g_2^2 \left( W^3_\mu W^+_\nu(W^-_\mu W^3_\nu - W^3_\mu W^-_\nu) 
    +\frac{1}{2}  W^+_\mu W^-_\nu  (W^+_\mu W^-_\nu - W^-_\mu W^+_\nu) \right)\;.
&&\label{3V4V}
\end{eqnarray}

All matter fields in the electroweak theory are either  $SU(2)$
 invariant singlets or  belong  to its fundamental 
representation. In the latter case  they form  doublets. 
Generators for these doublets   are expressed via the Pauli $\sigma$-matrices
$$
 \hat \tau_{\alpha} = \hat \sigma_{\alpha}/2 \;.          
$$
 Thus the infinitesimal local gauge transformations for doublets take a
form:

$$
    D(w,b) \psi(x)= \frac{i\,g_2}{2}w^{\alpha}(x)\hat \sigma_{\alpha}\psi(x) + 
    \frac{i\, g_1}{2} Y\, b(x) \psi(x) \;.
$$   
Here $g_1$ is the coupling constant of $U(1)$ gauge interaction.
The constant $Y$ depends on a type of the doublet. It is called a {\it hypercharge}.

In the gauge theory of electroweak interaction the gauge fields  
interact  with a  scalar ({\it Higgs})  doublet  which  has a nonzero 
vacuum state. 
Without loss of generality one can put  $Y=1$ for the Higgs doublet.
By means of the gauge transformation  the vacuum state of this field 
may be presented in the form:
$$
   \Phi_{\Omega}= \left( \begin{array}{c} 0 \\ \phi_0/\sqrt{2} 
 \end{array} \right)\;,
$$     
where $\phi_0$ is a  real constant.

As a result of spontaneous symmetry breaking the $ W^{\alpha}_{\mu}$ and 
$B_{\mu}$ fields do not correspond to physical particles. Physical 
particles in this model are the photon ($A_\mu$), W-bosons ($W^+_{\mu}\,,
W^-_{\mu}$) and Z-boson ($Z_{\mu}$).
 The photon field $A_{\mu}$ is a combination of  gauge fields 
responsible for the local gauge transformations which save  
the Higgs vacuum $\Phi_{\Omega}$: 
$$
     A_{\mu} = B_{\mu} \cos{\Theta_w} + W^3_{\mu} \sin{\Theta_w}\;,
$$
where the mixing angle $\Theta_w = \arctan(g_2/g_1)$. To complete $W^+$,  $W^-$,
 and $A$ up to the orthonormal  basis of gauge fields we introduce   
$$  Z_{\mu} = -B_{\mu} \sin{\Theta_w} + W^3_{\mu} \cos{\Theta_w}\;. $$

Let  $w^+(x)$, $w^-(x)$, $a(x)$, and $z(x)$ be parameters of the local
gauge transformation corresponding to the fields $W^+_\mu$,  $W^-_\mu$,
$A_\mu$, and  $Z_\mu$. Then  for a matter  doublet  with a hypercharge $Y$
the gauge transformation is given  by the following expression:
\begin{eqnarray}
\lefteqn{\hat D(w^+,w^-,a,z) \Psi=}\nonumber \\ & &\frac{i\,g_2}{2} \left[\sqrt{2} 
\left( \begin{array}{cc} 0 & w^+ \\ w^-&0 \end{array} \right) +
\sin\Theta_w\,a\, \left( \begin{array}{cc} 1+Y & 0 \\ 0&Y-1 \end{array} \right) \right.
\nonumber \\
\label{DPsi}
&+&  \cos\Theta_w  \,z\,
\left.\left( 
\begin{array}{cc} 1 -Y\tan^2\Theta_w &  0 \\ 0& -1  -Y\tan^2\Theta_w \end{array}
\right)   \right] \Psi\;.
\end{eqnarray}

\subsubsection{Lagrangian of Higgs field}

In the framework of renormalizable field theory  the general 
expression for the  gauge invariant  Higgs  Lagrangian is:  
$$
 L_{Higgs}= (\nabla_\mu \Phi)^+(\nabla^\mu \Phi) -\, \frac{\lambda}{2}
(\Phi^+ \Phi - \frac{1}{2} \phi_0^2)^2 \;, 
$$
where $\lambda$ is a new coupling constant.

 Some perturbations of the Higgs vacuum  can be realized by 
means of the local gauge transformation which do not correspond to physical
degrees of freedom. To separate physical and gauge degrees of freedom for a
small perturbation of the Higgs vacuum we present $\Phi$ in the
following form: 
\begin{equation}
 \Phi= \left(\begin{array}{c} i\,W^+_f \\(\phi_0 + H -i\,Z_f)/\sqrt{2} \end{array}
\right)\;.
\label{HiigsField}
\end{equation}
Here $W^+_f$ and $Z_f$ are the unphysical {\it Goldstone} fields
corresponding
to variation of the Higgs vacuum caused by the  $D(w^+,0,0,0)$ and 
$D(0,0,0,z)$ gauge transformations. The real field  $H$ corresponds to 
a physical degree of freedom which is assoshated with the Higgs patricle.  

The  term $(-\, \frac{\lambda}{2}( \Phi^+ \Phi - \frac{1}{2} \phi_0^2)^2)$ 
 contains  the mass term for the $H$ field:
$$
     -\, \frac{\lambda}{2} (\phi_0)^2  \, H^2 \;,
$$
and the following terms of self-interaction for  $H$ and 
the Goldstone fields:
\begin{equation}
-\, \frac{\lambda}{2} \left( (W^+_f W^-_f + ( H^2 + Z_f^2)/2)^2 
   + 2\phi_0 H(W^+_f W^-_f + ( H^2 + Z_f^2)/2) \right)\;.
\label{HiggsSelfInt}
\end{equation}
Usually   $\lambda$ is expressed via the  mass of $H$-boson $M_H$:
$$
   \lambda=\left(\frac{M_H}{\phi_0}\right)^2\;.
$$
Let us remind that 
$$\nabla_\mu \Phi = \partial_\mu \Phi - \hat D(W+_\mu,W-_\mu,A_\mu,Z_\mu)
\Phi \;\;,$$
where due to (\ref{DPsi})
\begin{eqnarray}
\lefteqn {\hat D(W+_\mu,W-_\mu,A_\mu,Z_\mu) \Phi=}&&  \nonumber \\
&& \frac{i\,g_2}{2} 
\left( \begin{array}{c}  
W^+_\mu (\phi_0 + H -i\,Z_f) + i\,(2 \sin{\Theta_w} A_\mu + \cos{\Theta_w}
(1-\tan^2{\Theta_w}) Z_\mu) \,W^+_f
\\
 ( 2 i W^-_\mu   W^+_f    - Z_\mu(\phi_0 + H -i\,Z_f)/\cos{\Theta_w} )/\sqrt{2}
\end{array} \right)\;. \nonumber \\
\label{DPhi}
\end{eqnarray}

The term $(\hat D_\mu \Phi)^+(\hat D^\mu \Phi)$  gives us the
mass terms for the $W$ and $Z$-bosons: 
\begin{equation}
\left( \frac{g_2 \phi_0}{2} \right)^2 W^+_\mu {W^-}^\mu + \frac{1}{2}
\left( \frac{g_2 \phi_0}{2 \cos{\Theta_w} } \right)^2 Z_\mu Z^\mu \;,
\label{WZ-mass-terms}
\end{equation}
and the following terms  describing the interaction of a couple of 
vector bosons with Higgs and Goldstones:
\begin{eqnarray}
\frac{g_2^2}{4}(
&& W^+_\mu {W^-}^\mu ( H^2 +Z_f^2 +2\phi_0 H +2 W^+_f W^-_f) \nonumber \\
&& + 4 A_\mu A^\mu \sin^2{\Theta_w} W^+_f W^-_f \nonumber \\
&& + \frac{Z_\mu Z^\mu}{\cos^2{\Theta_w}}(  (H^2 + Z_f^2)/2 + \phi_0 H + (1-2
\sin^2{\Theta_w})^2  W^+_f W^-_f) \nonumber  \\ 
&& + 4  A_\mu  Z^\mu \tan{\Theta_w}(1 -2 \sin^2{\Theta_w}) W^+_f W^-_f \nonumber\\
&& -2i\tan{\Theta_w} W^+_\mu (\cos{\Theta_w} A^\mu - \sin{\Theta_w} Z^\mu) 
           (\phi_0 + H -i\,Z_f) W^-_f\nonumber \\
&& + 2i\tan{\Theta_w} W^-_\mu (\cos{\Theta_w} A^\mu - \sin{\Theta_w} Z^\mu)
           (\phi_0 + H +i\,Z_f) W^+_f\nonumber\\
 )\;.&&  \label{VVH}
\end{eqnarray} 

The constant $\phi_0$ now can be expressed in terms of the $W$-boson mass
$M_W$:
$$
 \phi_0= 2\,M_W/g_2\;.
$$
The  $Z$-boson mass is related to  the $W$-boson mass by means of
constraint:
$$ M_Z=M_W/\cos{\Theta_w}\;.$$
 
The term $( - (\hat D_\mu \Phi)\partial^\mu \Phi^+  - (\hat D_\mu
\Phi)^+ \partial^\mu \Phi) $  gives us  off-diagonal quadratic terms:
\begin{equation} \label{offDiagonal}
- g_2 \phi_0 \left( \frac{1}{2}( W^+_\mu \partial^\mu W^-_f + 
W^-_\mu \partial^\mu W^+_f )  
+\frac{1}{\cos{\Theta_w}}Z_\mu \partial^\mu Z_f \right)  
\end{equation}
and the following  terms of interaction:
\begin{eqnarray}
\frac{g_2}{2}\left(\right. 
&&-H ( W^-_\mu \partial^\mu W^+_f  + W^+_\mu \partial^\mu W^-_f ) 
+  (\partial^\mu H)(W^-_\mu W^+_f +W^+_\mu W^-_f) \nonumber \\
&&-i(Z_f (  W^-_\mu \partial^\mu W^+_f  -W^+_\mu \partial^\mu W^-_f ) 
+ (\partial^\mu Z_f)  (W^+_\mu W^-_f - W^-_\mu W^+_f)) \nonumber \\
&&+i( 2\sin{\Theta_w} A_\mu + \cos{\Theta_w}(1-\tan^2{\Theta_w}) Z_\mu)
( W^-_f \partial^\mu W^+_f  -  W^+_f \partial^\mu W^-_f )\nonumber \\
&&+  (\partial^\mu H) Z_\mu Z_f/\cos{\Theta_w} 
 - (\partial^\mu Z_f) Z_\mu H/cos{\Theta_w} \nonumber \\
\left. \right)\;. \label{VHH}
\end{eqnarray}

The off-diagonal quadratic terms are canceled by the gauge fixing terms.
See below.

\subsubsection{Gauge fixing and ghost terms for the t'Hooft-Feynman gauge}
In the case of t'Hooft-Feynman gauge the   gauge fixing terms are
$$
   -\frac{1}{2} (\partial^\mu A_\mu)^2  
-\frac{1}{2} (\partial^\mu Z_\mu  + M_Z\, Z_f)^2
-| \partial^\mu W^+_\mu + M_W\, W^+_f|^2\;.
$$
The squared  divergences  of fields  transform the  quadratic part of Lagrangian 
for vector 
bosons to a diagonal form like (\ref{FeynmQ}). The squared Goldstone field 
terms  give a mass to the
Goldstone particle   equal to the mass of the corresponding 
vector boson field. The off-diagonal quadratic terms, which follow from
the gauge fixing Lagrangian, cancel the off-diagonal terms (\ref{offDiagonal})
 up to  complete  divergency terms. 

 According to the general rule (\ref{fpLag})  the  Faddeev-Popov Lagrangian is
\begin{eqnarray*}
 &&  -A_{\bar{c}} D(W^+c,W^-_c,A_c,Z_c)(\partial^\mu A_\mu) \\  
 &&  -Z_{\bar{c}} D(W^+c,W^-_c,A_c,Z_c) (\partial^\mu Z_\mu  + M_Z\, Z_f)\\
 && -W^-_{\bar{c}} D(W^+c,W^-_c,A_c Z_c)(\partial^\mu W^+_\mu +
M_W\,W^+_f)\\ 
 && -W^+_{\bar{c}} D(W^+c,W^-_c,A_c,Z_c)(\partial^\mu W^-_\mu + 
M_W\,W^-_f)\;.
\end{eqnarray*}
Note that due to (\ref{SU2U1tranf}) 
\begin{eqnarray*}
\hat D W^+_{\mu}& =&  i\,g_2\, ( W^+_\mu( \sin{\Theta_w} A_c+ \cos{\Theta_w} Z_c)
-W^+_c(\sin{\Theta_w}A_\mu+\cos{\Theta_w} Z_\mu) ) +\partial_\mu W^+_c   \;;\\
\hat D A_{\mu}& =& -i\,g_2 \sin{\Theta_w}(W^+_\mu W^-_c - W^-_\mu W^+_c) +\partial_\mu
A_c \;;\\
\hat D Z_{\mu }& =& -i\,g_2 \cos{\Theta_w} (W^+_\mu W^-_c - W^-_\mu W^+_c)
+\partial_\mu Z_c\;,
\end{eqnarray*}
and according to (\ref{DPhi})
\begin{eqnarray*}
 D W^+_f &=& W^+_c M_W +  \frac{g_2}{2} 
( W^+_c ( H -i\,Z_f) + i\,(2 \sin{\Theta_w} A_c + \cos{\Theta_w}
(1-\tan^2{\Theta_w}) Z_c) \,W^+_f)\;;
\\
 D Z_f & =& M_Z Z_c +\frac{g_2}{2}( \frac{Z_c H} 
{\cos{\Theta_w}} - i(W^-_c W^+_f - W^+_c W^-_f) 
)\;.
\end{eqnarray*}
After substitution of these derivatives to the Faddeev-Popov Lagrangian 
  we see  that it contains the quadratic part:
$$ -A_{\bar{c}}\Box A_c -Z_{\bar{c}}\Box Z_c - W^-_{\bar{c}}(\Box
W^+_c +M_W W^+_c) - W^+_{\bar{c}}(\Box W^-_c +M_W W^-_c)\;,
$$
and the following  vertices of interaction:
\begin{eqnarray}
-g_2(&&  i \,\sin{\Theta_w} (\partial^\mu A_{\bar{c}})  (W^+_\mu W^-_c - W^-_\mu W^+_c)
\nonumber \\
&&+ i\,\cos{\Theta_w}  (\partial^\mu Z_{\bar{c}}) (W^+_\mu W^-_c - W^-_\mu W^+_c)
\nonumber\\ 
&&+  \frac{M_Z Z_{\bar{c}}}{2}\left(\frac{Z_c H}{\cos{\Theta_w}} - 
i(W^-_c W^+_f - W^+_c W^-_f) \right) \nonumber\\
&&- i(\partial^\mu  W^-_{\bar{c}}) ( W^+_\mu( \sin{\Theta_w} A_c+ \cos{\Theta_w} Z_c)
-W^+_c(\sin{\Theta_w}A_\mu+\cos{\Theta_w} Z_\mu)) \nonumber \\
&&+ \frac{M_W W^-_{\bar{c}}}{2} 
( W^+_c ( H -i\,Z_f) + i\,(2 \sin{\Theta_w} A_c + \cos{\Theta_w}
(1-\tan^2{\Theta_w}) Z_c) \,W^+_f)  )  )\nonumber  \\
&&+ i(\partial^\mu  W^+_{\bar{c}}) ( W^-_\mu( \sin{\Theta_w} A_c+ \cos{\Theta_w} Z_c)
-W^-_c(\sin{\Theta_w}A_\mu+\cos{\Theta_w} Z_\mu)) \nonumber \\
&&+ \frac{M_W W^+_{\bar{c}}}{2} 
( W^-_c ( H + i\,Z_f) - i\,(2 \sin{\Theta_w} A_c + \cos{\Theta_w}
(1-\tan^2{\Theta_w}) Z_c) \,W^-_f)  )  ) \nonumber \\
)\;.&& \label{FP-interaction}
\end{eqnarray}

\subsubsection{Unitary gauge}

The {\it unitary} gauge  may be considered as a limit of $\xi \rightarrow
\infty$ of the following  gauge fixing Lagrangian:
$$
 -\frac{1}{2} (\partial^\mu A_\mu)^2  -\frac{M_Z^2}{2\xi} (Z_f)^2
-\frac{M_W^2}{\xi} (W^+_f W^-_f)\;. 
$$
In this limit one gets $Z_f=0$ and $W_f=0$, what decreases a number
of vertices. Under these constraints the Faddeev-Popov Lagrangian
(\ref{fpLag}) takes the form:  

\begin{eqnarray*}
  -A_{\bar{c}} \partial^\mu (-i\,g_2 \sin{\Theta_w}(W^+_\mu W^-_c - W^-_\mu W^+_c) +\partial_\mu
A_c ) 
  -Z_{\bar{c}}  M_Z\ ( M_Z  + \frac{g_2 H} {2\cos{\Theta_w}} ) Z_c &&  \\
 -W^-_{\bar{c}}  M_W ( M_W +  \frac{g_2}{2}   H )  W^+_c 
 -W^+_{\bar{c}} M_W ( M_W +  \frac{g_2}{2}   H  )  W^-_c \;. &&
\end{eqnarray*}
Integration over $A_c$ and $A_{\bar{c}}$ can be performed explicitly 
and gives a result which does not depend on other fields. So  $A_c$ and
$A_{\bar{c}}$ ghosts may be omitted.  

In the unitary gauge only  physical polarization states of the 
incoming and outgoing $W^\pm$ and $Z$ bosons are considered. So 
$W^\pm_c$ and $Z_c$ are not needed in the external lines of Feynman diagrams
and may be omitted also in tree level calculations.
Consequently, in the unitary gauge all ghost and Goldstone fields may be omitted. 

\subsubsection{Summary of vertices for the boson sector}
In the case of t'Hooft-Feynman gauge the full set of vertices is 
described by  expressions (\ref{3V4V}), (\ref{HiggsSelfInt}), (\ref{VVH}),
(\ref{VHH}), and (\ref{FP-interaction}), where 
\begin{eqnarray*}
 W^3_\mu &=& \sin{\Theta_w}  A_\mu + \cos{\Theta_w} Z_\mu \;; \\
\lambda&=&\left(\frac{g_2 M_H}{2 M_w}\right)^2 \;.
\end{eqnarray*}
The coupling constants $g_2$ and $g_1$ may be expressed in terms of the electromagnetic
coupling constant: $g_2 = e \sin{\Theta_w}$ and  $g_1 = e \cos{\Theta_w}$ .

In the case of unitary gauge the interaction is defined by a
 subset of vertices which appears after removing the Faddeev-Popov ghosts and the 
Goldstone  fields.

\subsubsection{Interaction of vector bosons  with fermions}
Experiments in particle physics show that the  $W$-bosons 
interact with left-handed components of fermions
$$ \Psi^L = \frac{1 - \gamma^5}{2} \Psi\;. $$
Thus, the $SU(2)$ group  must transform only the left-handed fermion components.
The initial Lagrangian of  fermion field has the corresponding  
global  symmetry only if all fermions are massless. Indeed, the
Lagrangian of free massless fermion field  splits into two independent
parts: 
\begin{eqnarray*}
 L &=& \frac{i}{2}(\bar{\Psi} \gamma_\mu \partial^\mu \Psi -
(\partial^\mu \bar{\Psi})\gamma_\mu \Psi)\\ 
&=&  \frac{i}{2}(\bar{\Psi}^L \gamma_\mu \partial^\mu \Psi^L -
(\partial^\mu \bar{\Psi}^L)\gamma_\mu \Psi^L) +  
  \frac{i}{2}(\bar{\Psi}^R \gamma_\mu \partial^\mu \Psi -
(\partial^\mu \bar{\Psi}^R)\gamma_\mu \Psi^R)\;,
\end{eqnarray*} 
which allows to  apply  the $SU(2)$ gauge transformations to the left-handed 
components of fermion doublets. In the same time the mass term contains a 
product of left-handed and right-handed  fermion components:
$$ m \bar{\Psi} \Psi = m ( \bar{\Psi}^L \Psi^R + \bar{\Psi}^R \Psi^L)\;,$$
what forbids an appearance of such terms in the  invariant Lagrangian.
Later on we shall show how such fermion particles will  acquire masses in  
result  of the gauge invariant 
interaction  of formerly massless fermion fields with the Higgs doublet.

Whereas $SU(2)$ transforms only the left-handed components of doublets,
$U(1)$ interacts with both. Left-handed  and right-handed components of fermions 
must have the same electric charge. It allows to find the hypercharge 
of right-handed components if a hypercharge of left-handed doublet is known.
The  $U(1)$ gauge field $B_\mu$ is equal to $ \cos{\Theta_w} A_\mu
- \sin{\Theta_w} Z_\mu$. So, the electromagnetic  coupling constant for 
right-handed fermion with a hypercharge $Y^R$ is $(g_1 Y^R \cos{\Theta_w})/2$.
Comparing it with the expression (\ref{DPsi}) we see that the hypercharges of
right-handed components  of doublet are 
\begin{eqnarray*}
Y^R_1&=&1+Y\;, \\
Y^R_2&=&Y-1\;.
\end{eqnarray*}
Thus, we may unambiguously write down the vertices of interactions 
associated with the covariant derivative:
\begin{eqnarray*}
&&\frac{g_2}{2} \left( \begin{array}{cc} \bar{\Psi}_1 & \bar{\Psi}_2 \end{array} \right)
\left[    \sqrt{2} 
\left( \begin{array}{cc} 0 & W^+_\mu \\ W^-_\mu&0 \end{array} \right) +
\sin\Theta_w\,A_\mu\, \left( \begin{array}{cc} 1+Y & 0 \\ 0&Y-1 \end{array} \right) \right.\\
&+&  \cos\Theta_w  \,Z_\mu
\left.\left( 
\begin{array}{cc} 1 -Y\tan^2\Theta_w &  0 \\ 0& -1  -Y\tan^2\Theta_w \end{array}
\right)        \right]
\left( \begin{array}{c} \gamma^\mu \Psi^L_1 \\ \gamma^\mu \Psi^L_2 \end{array} \right) \\
&+& \frac{g_1}{2} (\cos{\Theta_w} A_\mu- \sin{\Theta_w} Z_\mu )  
\left( \begin{array}{cc} \bar{\Psi}_1 & \bar{\Psi}_2 \end{array} \right)
 \left( \begin{array}{cc} Y+1 &0\\0&Y-1 \end{array}  \right)  
\left( \begin{array}{c}
 \gamma^\mu \Psi^R_1 \\  \gamma^\mu  \Psi^R_2 \end{array} \right)\;. 
\end{eqnarray*}

After  matrix multiplication we obtain:
\begin{eqnarray}
&& \frac{g_2}{2\sqrt{2}} 
\left(  W^-_\mu  \bar{\Psi}_2  \gamma^\mu (1 - \gamma^5)\Psi_1 
+   W^+_\mu  \bar{\Psi}_1  \gamma^\mu (1 - \gamma^5)\Psi_2 \right) \nonumber \\
&+& \frac{g_1}{4 \sin{\Theta_w}} Z_\mu \left(\;
\bar{\Psi}_1 \gamma^\mu ( 1 - \gamma^5 -2(Y+1) \sin^2{\Theta_w} ) \Psi_1 \right.
\nonumber \\
&& \;\;\;\;\;\;\;\;\;\;\;\;\;\;\;\;\;\left.
- \bar{\Psi}_2 \gamma^\mu (1-\gamma^5 +2(Y-1)\sin^2{\Theta_w} ) \Psi_2 \;\right) 
\nonumber \\
&+& \frac{g_1}{2}\cos{\Theta_w} A_\mu \left( (Y+1) \bar{\Psi}_1 \gamma^\mu
\Psi_1 + (Y-1) \bar{\Psi}_2 \gamma^\mu \Psi_2 \right)\;.
\label{V-fermion} 
\end{eqnarray}

\subsubsection{Interaction of the Higgs doublet with fermions and generation
of fermion masses}
The mass terms for fermions  in the electroweak theory are generated via their
$SU(2)\times U(1)$ invariant Yukawa interaction with Higgs doublet. 
Namely, the  Yukawa Lagrangian
$$
 - \frac{m_1 \sqrt{2}}{\phi_0} \left( \bar{\Psi}^L_i \epsilon^{ij} \stackrel{*}{\Phi^{j}} \Psi^R_1 -
\bar{\Psi}^R_1 \Phi^{i} \epsilon^{ij} \Psi^L_j \right)
-\,\frac{m_2 \sqrt{2}}{\phi_0} \left( \bar{\Psi}^L_i \Phi^i \Psi^R_2 +
\bar{\Psi}^R_2 \stackrel{*}{\Phi^{i}} \Psi^L_i \right) 
$$
produces the mass terms for doublet components
$$  -m_1 \bar{\Psi}_1 \Psi_1 -m_2 \bar{\Psi}_2 \Psi_2\;, $$ 
which are accompanied by  vertices of interactions of fermions 
with Goldstone fields: 
\begin{eqnarray}
&-&  \frac{ m_1\, g_2}{2 M_W} \left ( \frac{i}{\sqrt{2}}(W^-_f \bar{\Psi}_2 
(1 + \gamma^5) \Psi_1 - W^+_f \bar{\Psi}_1 (1-\gamma^5) \Psi_2) 
+  H \bar{\Psi}_1 \Psi_1 + i Z_f \bar{\Psi}_1 \gamma^5 \Psi_1 \right)
\nonumber\\
&-&  \frac{ m_2\, g_2}{2 M_W} \left ( \frac{i}{\sqrt{2}}(W^+_f \bar{\Psi}_1 
(1 + \gamma^5) \Psi_2 - W^-_f \bar{\Psi}_2 (1-\gamma^5) \Psi_1) 
+  H \bar{\Psi}_2 \Psi_2 -i Z_f \bar{\Psi}_2 \gamma^5 \Psi_2 \right)\;.
\nonumber\\ \label{H-fermion}
\end{eqnarray}

If there are several doublets with the same hypercharge, then 
a general form of Yukawa   Lagrangian  contains a  product of terms from
different doublets. Such terms form  two  mass matrices: one  for
upper   and another for lower fermions. Each of these matrices can be diagonalized 
by means of the unitary transformation of doublets, but this cannot be done for both 
of them at the
same time.  In this case the basis  of doublets is chosen in such a way 
to  present one of these matrices, for example,  for upper fermions,
in the diagonal form. 
Then   the   physical particles correspond to linear combinations of
lower doublet fields   realized by some unitary matrix which is called 
a {\it mixing matrix}.    

Generally the Lagrangian is written down in terms of fermion fields which 
directly correspond to particles.  Interactions of such  fields
with $A$, $Z$, $H$, and $Z_f$  are the same as defined by (\ref{V-fermion}) 
and (\ref{H-fermion}), whereas interactions with $W^\pm$ and $W^\pm_f$
contain elements of the  mixing matrix.

\subsubsection{Quarks and leptons}

The Standard Model contains three doublets of leptons with a hypercharge
$Y=-1$:  
$$
\left( \begin{array}{c} \nu_e \\ e^-  \end{array} \right)
\left( \begin{array}{c} \nu_\mu   \\ \mu^-  \end{array} \right)
\left( \begin{array}{c} \nu_\tau \\ \tau^-  \end{array} \right)\;,
$$ 
and three  doublets of quarks with a hypercharge $Y=\frac{1}{3}$:
$$
\left( \begin{array}{c} u \\d^\prime  \end{array} \right)
\left( \begin{array}{c} c \\s^\prime  \end{array} \right)
\left( \begin{array}{c} t \\b^\prime  \end{array} \right)\;.
$$ 

Neutrinos, the upper components of lepton doublets, are so far massless as 
a result of experimental measurements. So the mixing matrix for 
leptons is not needed. Meantime both the  upper and down components of  quark
doublets have nonzero   masses. The corresponding mixing matrix is
named the Cabibbo-Kobayashi-Maskawa matrix. It expresses the lower components of 
quark doublets in terms of the real quark fields:
$$
\left(\begin{array}{c} d^\prime \\ s^\prime \\ b^\prime \end{array}\right) =
\left(\begin{array}{c} Kobayashi \\ Moskawa \\ matrix \end{array}\right)
\left(\begin{array}{c} d \\ s \\ b \end{array}\right)\;.
$$

\addcontentsline{toc}{section}{Pictures and figures}

\newpage
\section*{Pictures and figures}

\begin{figure}[htb] 
{\scriptsize
\begin{center}
\begin{verbatim}
******************************************************************
*                            M A I N  M E N U                    *
*    CompHEP numerical module    u  ,D   -> b  ,B  ,W+           *
****************************************************************** 
*     x: Exit   hN: Help (N-menu position)  m: MAIN menu         *
*                                                                *
*    1: Subprocess                        2: IN state            *
*    3: Model parameters                  4: QCD SCALE           *
*    5: Breit-Wigner                      6: Cuts                *
*    7: Kinematics                        8: Regularization      *
*    9: Vegas                            10: Simpson             *
*   11: Batch                            12: View result         *
 *****************************************************************
   Type number of menu position and press ENTER:_ 
\end{verbatim}
\end{center}
}
\caption{ Example of the \FORTRAN~  menu}
\label{screen_menu_f}
\vspace*{-5cm}
\end{figure}
\vspace*{-5cm}


\begin{figure}
{\small
\begin{verbatim}
               e1          E2   !  E2          e1
             ==>==\     /==<====!==<==\     /==>==
               P1 |     |  P4   !  P4 |     |  P1
                  |     |       !     |     |
               E1 |  A  |  e2   !  e2 |  A  |  E1
             ==<==@-1---@==>====!==>==@---2-@==<==
               P2    P5    P3   !  P3   -P6    P2
\end{verbatim}
}
\caption{Example of  pseudo-graphic diagram  image  }

\label{pseudo}

\end{figure}

\end{document}